\documentclass[letterpaper,12pt]{report}

\usepackage[letterpaper,textwidth=17cm,textheight=20cm]{geometry} 
\usepackage[english]{babel}
\usepackage[utf8]{inputenc} 
\usepackage[T1]{fontenc}
\usepackage{lipsum}
\usepackage{rotating}
\usepackage{fancyhdr}               
\usepackage{parskip}
\usepackage[font=small]{caption}  
\usepackage{lettrine}
\usepackage{subcaption} 
\usepackage{csquotes} 
\usepackage{epigraph} 
\usepackage{slashed}
\usepackage{amssymb}
\usepackage{amsmath}
\usepackage{amsthm}         
\usepackage{mathtools}

\usepackage{graphicx}
\usepackage[dvipsnames]{xcolor}         
\usepackage{listings}          
\usepackage{hyperref}     
\usepackage[normalem]{ulem}

\fancypagestyle{plain}{
  \fancyfoot{}
  \fancyhead{}
  
}

\begin{document}

\begin{titlepage}
\vspace*{-3cm}
\begin{figure}[!htb]
    \centering
    \includegraphics[keepaspectratio=true,scale=0.1]{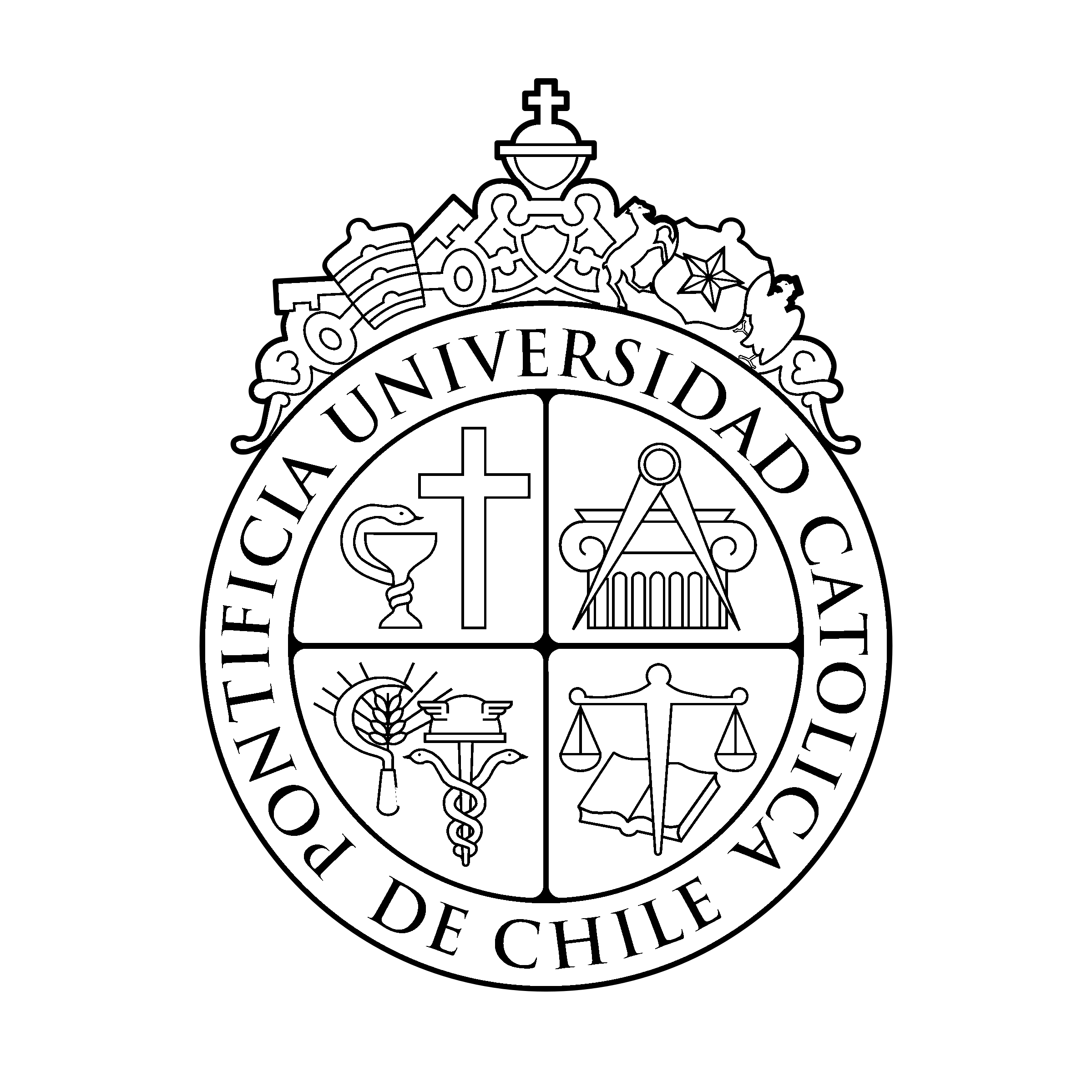}
\end{figure}

\begin{center}
    \LARGE{PONTIFICIA UNIVERSIDAD CATÓLICA DE CHILE}
    \vspace{5mm}
    \\ \Large{INSTITUTE OF PHYSICS}
    \vspace{5mm}
    \\ \large{Submitted to the Faculty of Physics for the attainment of the Master's degree in Physics}
    \vspace{3mm}
    \\ \large {\scshape\today}
\end{center}

\vspace{4mm}
\begin{center}
\hrulefill

\vspace*{1mm}
    {\Large{\bf Thermal and quantum phase transitions in a holographic anisotropic Dirac semimetal}}\\
    
\hrulefill    

\end{center}
\vspace{5mm}

\begin{minipage}[t]{0.47\textwidth}
	{\large{Advisor:}{\normalsize\vspace{3mm}
	\bf\\ \large{Rodrigo Soto-Garrido PhD}}}\\[8mm]
	{\large{Committee:}{\normalsize\vspace{3mm}}\bf\\
	\large{Rodrigo Soto-Garrido PhD} \\[2mm]
	\large{Enrique Muñoz PhD} \\[2mm]
	\large{Alberto Faraggi PhD}
	}
\end{minipage}
\hfill
\begin{minipage}[t]{0.47\textwidth}\raggedleft
	{\large{Candidate:}{\normalsize\vspace{3mm} \bf\\ \large{Sebastián Bahamondes BSc}}}
\end{minipage}

\end{titlepage}

\thispagestyle{plain}

\begin{center}
    {\LARGE\bf  Thermal and quantum phase transitions in a holographic anisotropic Dirac semimetal}\\[5mm]
    {\Large Thesis dissertation for the title of Master of Science (MSc)}\\[4mm]
    {\Large\bf Sebastián Bahamondes BSc}\\[1cm]
    
    {\large\bf Abstract}
\end{center}
    In this thesis we build a phenomenological, strongly
    coupled quantum field theory in $2+1$-dimensions through AdS/CFT holography, 
    by building a $3+1$-dimensional, negatively curved gravity theory with a $SU(2)$ gauge field,
    and a scalar field in the adjoint of $SU(2)$. We locate a phase transition between two distinct phases at zero and finite temperature, which 
    are characterized through the dispersion relation of quasi-normal modes of probe fermions in the bulk,
    and correspond either to a Dirac semimetal or a band insulator. These phases are separated by a
    critical phase/critical point (depending if $T>0$ or $T=0$, respectively) where the band structure
    of boundary fermions exhibits semi-Dirac anisotropy. We 
    characterize each phase at $T=0$ by explicit solutions to the bulk equations of motion in the infra-red,
    and determine that the critical point's spacetime is a Lifshitz geometry, whose dynamical critical exponent is
    approximately equal to $2$. We also find that this anisotropy induces a non-trivial
    scaling of the shear viscosity-entropy density ratio with respect to temperature in the $T\to 0$ limit, and find evidence 
    that the anisotropic phase of the system corresponds to a finite-temperature quantum critical phase.

\tableofcontents

\chapter*{}
\begin{center}
    \textit{To Lady Gaga.}
\end{center}
\chapter{Introduction}\label{chap:introduction}

\lettrine{T}{heoretical physics} has seen great progress in the last few decades regarding the physical realization of various condensed matter systems with exciting properties. It can be argued that the most recent breakthrough which inspired a new wave of scientific productivity in the Condensed Matter Theory (CMT) community was in 2007 with the discovery of graphene \cite{Novoselov2007:Nature}. Indeed, the electronic and transport properties of this simple yet rich material has sparked a wave of research into novel ways two-dimensional materials can be combined in the laboratory; from magic-angle bilayer graphene \cite{Cao2018:Nature,Oh2021:Nature} to quantum Hall bi-layers \cite{Liu2022:Science}. Out of the different solid systems whose research has seen great development since graphene, Dirac semimetals are one of the most important. These are solid-state systems that feature low-energy quasiparticle excitations (QPEs) around the Fermi energy at discrete points of intersection between conduction and valence bands, which behave as Dirac fermions from high-energy physics. These quasiparticles have been measured in various experimental settings, from $\mathrm{Au_2Pb}$ through ARPES \cite{SanchezBarriga2023}, and in $\mathrm{VO_2}-\mathrm{TiO_2}$ heterostructures \cite{banerjee2009:PhysRevLett,Link:2017ora}. One can modify a Dirac semimetal by instead making it a semi-Dirac, or anisotropic, semimetal. This is characterized by low-energy QPEs around special points in the Brillouin zone with a linear dispersion along one direction in crystal momentum space, and quadratic along the other \cite{Uryszek:2019joy}.

 On the other hand, one of the most intriguing aspects of modern many-body systems, regardless of their band structure classification, is the role that strong interactions and correlations play in their emergent quantum characteristics. For example, when fermionic degrees of freedom are involved, these theories usually escape standard Fermi-liquid analysis. As \cite{Iqbal2011:Proceeding} indicates, examples of these challenging systems are the quark-gluon plasma (QGP) manufactured at RHIC from heavy-ion collisions, as well as different ultra-cold atom systems. The Fermi liquid approach to CMT is one of the most successful developments in modern quantum theory, and it relies in the fact that different bulk systems in the thermodynamic limit can be understood in the context of interactions between degrees of freedom that collectively have particle-like behavior: quasi-particles \cite{Iqbal2011:Proceeding,Sachdev2023:book}. However, not all many-body systems will feature quasi-particles in their spectrum, and as such they require non-perturbative methods for their theoretical understanding since standard methods of many-body physics break down, specially near criticality \cite{Zaanen2015:BookChptr2}.
  
  The breakdown of standard many-body physics methods usually implies one can't implement perturbation theory approaches to measure correlation functions in strongly interacting quantum systems. One successful theoretical approach to bypass this limitation is the AdS/CFT holographic correspondence, first proposed in \cite{Maldacena1999:goatPaper} and further enhanced for its application to CMT. Holography applied to CMT (usually dubbed AdS/CMT) works by translating the measurement of observables and correlation functions in a strongly coupled $d+1$-dimensional Quantum Field Theory (QFT) into a weakly coupled, classical gravitational problem in a negatively-curved (AdS) $d+2$-dimensional spacetime \cite{hartnoll2018:MITpressChapter,Zaanen2015:Book}. Operators in the QFT are mapped to classical fields in the bulk spacetime following a standard holographic dictionary, and the UV of the QFT is interpreted as located in the conformally flat boundary of the bulk \cite{Witten1998:TheorMatPhys,Maldacena1999:goatPaper}. Even though it was proposed originally in the context of string theory, the AdS/CMT approach to holography for building so-called bottom-up models that dualize strongly interacting CMT systems has exposed a wide range of qualitative predictions for such theories. In AdS/CMT one usually breaks conformal invariance in the IR by deforming the holographic theory in the bulk, so that the spacetime's geometry reflects departure from conformal invariance. Given AdS/CFT's conjectural status, this allows for the engineering of different geometries ruled by General Relativity (GR) that dualize phenomenological CMT models classified by symmetry, with the UV of the theory corresponding to a fixed point in the renormalization group (RG) flow. This allows not only for the dualization of effective field theories at $T=0$, but also of thermal QFTs and thermodynamic phase transitions.

  Given all of the above, the goal of the present work is to further expand the literature of strongly coupled CMT holographic theories. Specifically, we aim to describe a phase transition, both at finite and zero temperature, of a $2+1$-dimensional, strongly coupled, CMT system between a Dirac semimetal and a band insulator, through a critical semi-Dirac point. As indicated in \cite{Bahamondes2024:JHEP}, this type of phase transition is predicted to occur in black phosphorus \cite{Kim2015:Science}, in  TiO$_2$/VO$_2$ nanostructures under confinement \cite{Pardo2009:PhysRevLett}, and photonic metamaterials \cite{Wu2014:Optics}. Such a phase transition, both at finite and zero temperature (in which case we talk about a Quantum Phase Transition (QPT)), in a strongly coupled many-body system is engineered in this thesis using holographic tools. In particular, we obtain the system's phase diagram, and measure its transport coefficients through linear-response theory applied to holography. Finally, we 
  also corroborate that parameters of the model have a well-defined 
  critical behavior near the $T=0$ critical point, and measure the non-relativistic time scaling 
  of the critical spacetime through its characteristic Lifshitz dynamical exponent. Finally, we give numerical 
  evidence for the characterization of the finite temperature, semi-Dirac region above the $T=0$ critical point 
  as a Quantum Critical Region. Motivation for this type of project stems from the recent development of similar bottom-up metallic phase transitions, like Weyl and multi-Weyl semimetals \cite{Landsteiner2015:PhysLettB,Landsteiner2019:SciChinaPhysMechAstron,Juricic2020:JHEP,Juricic2024:arXiv} and holographic fermionic flat bands \cite{Grandi:2021bsp,Grandi2024:JHEP}. 

This report is organized as follows: in chapter \ref{chap:theoretical_background} we introduce the main background required for understandinf the results of this thesis; mainly concepts related to solid-state physics and AdS/CMT in general. In chapter \ref{chap:results} we show the main results on the location of the three distinct phases that the holographic construction reproduces, and characterize their bulk geometry both at finite and zero temperature. We conclude in chapter \ref{chap:conclusions} and subsequent appendices.

\chapter{Theoretical background and model construction}\label{chap:theoretical_background}
In this chapter we show the mathematical formalism of both the toy model that inspires the holographic construction that is the main object of study of this work, as well as the minimal ingredients of AdS/CMT that will be applied to measure phase transitions and transport coefficients on such system. In all subsequent expressions, both in this chapter and in following ones, natural units will be used: $\hbar = c = k_B = 1$. Also, when relativistic covariant notation is involved, we will use the mostly-plus sign convention for the metric: $\eta_{\mu\nu} = \mathrm{diag}(-1,1,\ldots,1)$, where the first index is always associated to the time coordinate.

\section{Tight-binding graphene toy model}\label{sec:toy_model}
\begin{figure}[!htb]
    \centering
    \includegraphics[width=\linewidth]{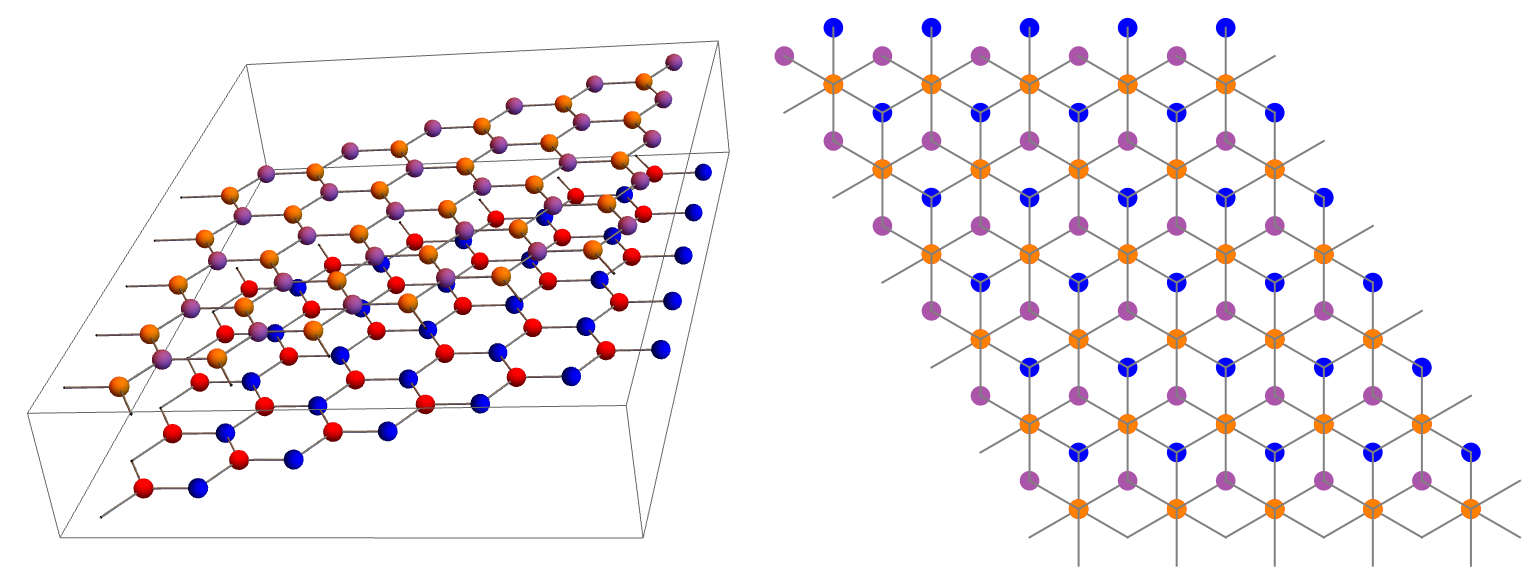}
    \caption{Sketch of a graphene bilayer with Bernal stacking. Each graphene layer is rotated by $60^\circ$ with respect to each other. The Hamiltonian in eq.~\eqref{eq:bilayer_hamiltonian} could represent this system by introducing the interaction strengths $\Delta_1$ and $\Delta_2$ between fermionic excitations in each layer, among other types of systems.}
    \label{fig:graphene_stacked_lattice}
\end{figure}

As was described in chapter \ref{chap:introduction}, a semimetal is a specific type of solid-state system whose band structure in the first Briollouin zone features fermionic quasiparticles at specific points of symmetry where conduction and valence bands intersect, while a band insulator features a gap between said conduction and valence bands \cite{Bahamondes2024:JHEP}. A specific system that realizes both of these structures is the following Hamiltonian, which represents a free model in $2+1$ dimensions corresponding to two fermionic quasi-particles coupled to each other through a pair of parameters that we call $\Delta_1$ and $\Delta_2$ (see \cite{Grandi:2021bsp,Grandi2022:PhysRevD} for similar constructions):
\begin{equation}\label{eq:bilayer_hamiltonian}
    H= H_D \otimes 1_{2\times 2}+ \left( \Delta_1\sigma_1 \otimes \sigma_1+ \Delta_2\sigma_3 \otimes \sigma_3  \right).
\end{equation}
In \eqref{eq:bilayer_hamiltonian} we have the free-Dirac Hamiltonian $H_D$ in momentum space, defined by $H_D = -\gamma^t(\gamma^xk_x+\gamma^yk_y)$ with the $2+1$-dimensional gamma matrices $\gamma^a = (\sigma_3,-i\sigma_2,i\sigma_1)$ (here $\sigma_j$ are the Pauli matrices, with $j=1,2,3$). This Hamiltonian can represent a variety of different solid-state systems in the tight-binding approximation, one of them being Bernal stacked bilayer graphene \cite{Katsnelson2020:BookChptr14}, sketched in Figure \ref{fig:graphene_stacked_lattice}.

\begin{figure}[!htb]
    \centering
    \includegraphics[width=\linewidth]{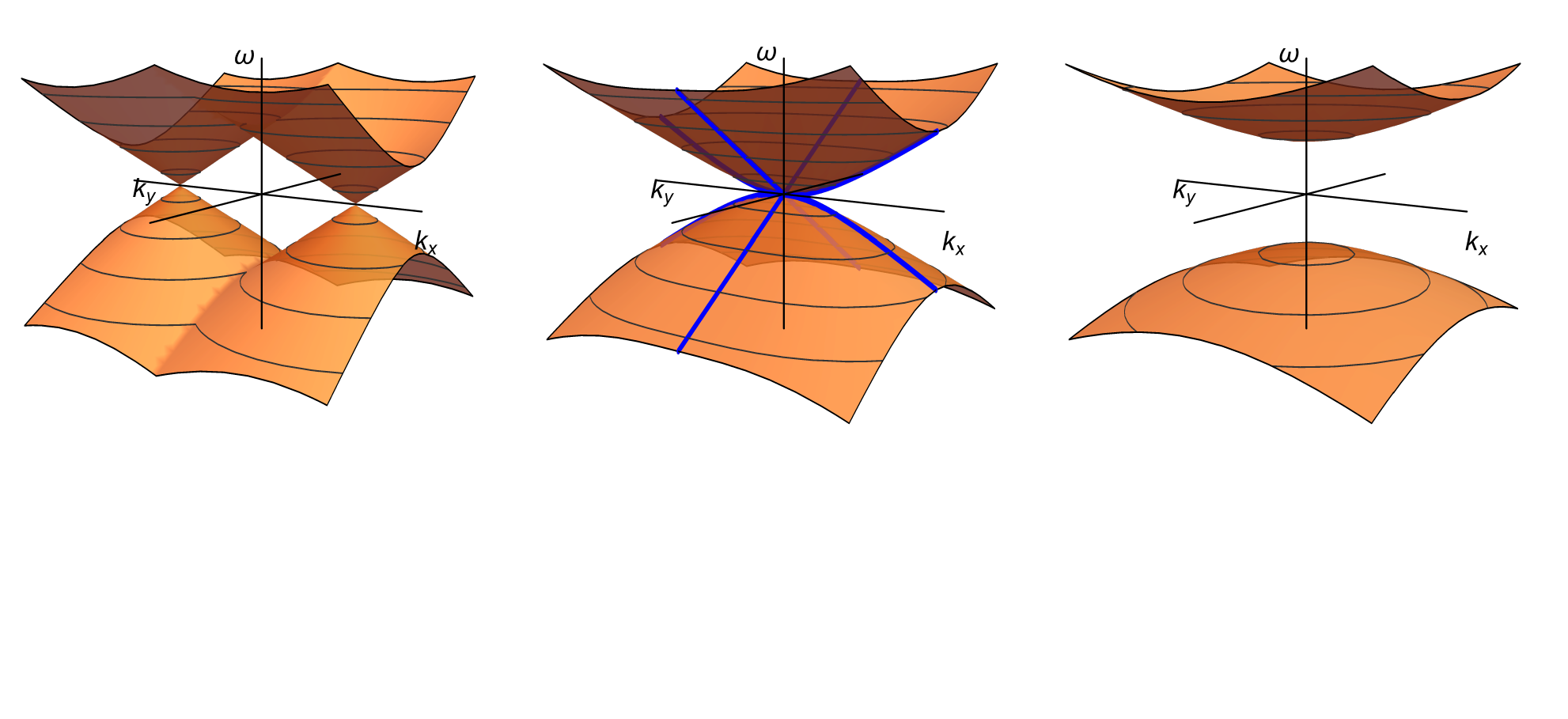}
    \vspace*{-3.1cm}
    \caption{Dispersion relations for three distinct cases of relative values between $\Delta_1$ and $\Delta_2$. If $\Delta_1>\Delta_2$, the band structure features two Dirac cones separated in the $k_x$ direction of momentum space (left plot). If $\Delta_1 = \Delta_2$, the dispersion relation becomes anisotropic (center plot), and if $\Delta_1<\Delta_2$ a gap forms between the conduction and valence bands (right plot). Image taken from \cite{Bahamondes2024:JHEP}.}
    \label{fig:prety_dispersions}
\end{figure}

We refer to Hamiltonian \eqref{eq:bilayer_hamiltonian} as a toy model because its band structure is exactly solvable by direct diagonalization in momentum space. Doing so yields the following dispersion relation around the Fermi energy (from here on set equal to zero):
\begin{equation}\label{eq:anysotropic_relation}
    \omega(k_x,k_y\,;\,\Delta_1,\Delta_2)=\pm \sqrt{k_x^2+k_y^2+ \Delta_1^2+\Delta_2^2- 2\Delta_1\sqrt{(k_x^2 +\Delta_2^2)}},
\end{equation}
where the $\pm$ correspond to the conduction and valence bands, respectively. Depending on if $\Delta_1 >\Delta_2$, $\Delta_2 = \Delta_1$ or $\Delta_1<\Delta_2$, the dispersion relation of fermions that satisfy Dirac's equation $i\partial_t\Psi = H\Psi$ (where $\Psi$ is a four-tuple of a pair of two-duple Dirac spinors) will have either semimetallic, insulating, or anisotropic (semi-Dirac) dispersion relations, as is shown in Figure \ref{fig:prety_dispersions}. By fine tuning the parameters $\Delta_{1,2}$ the Dirac cones of the semimetallic regimen become increasingly closer to each other by approaching the origin along the $k_x$ direction. When $\Delta_1 = \Delta_2$ they merge into a semi-Dirac point in a quantum phase transition, and when $\Delta_1<\Delta_2$, the system becomes a band insulator. We know that the quantum critical point is a semi-Dirac semimetal by approximating eq.~\eqref{eq:anysotropic_relation} for small momenta when $\Delta_1=\Delta_2=\Delta$: 
\begin{equation}\label{eq:approximate_anisotropy}
\omega(k_x,k_y\,;\,\Delta,\Delta)\approx\pm\sqrt{k_y^2+\frac{k_x^2}{4\Delta^2}}.
\end{equation}
As was outlined above, the Dirac equation for these quasi-particle excitations is written as $i\partial_t\Psi = H\Psi$, where $\Psi$ can be expressed as a combination of two separate Dirac spinors $\Psi = \begin{bmatrix}
    \psi\\\xi
\end{bmatrix}$, which are both two-component spinors. These spinors would represent fermions that belong to the Hilbert space associated to each graphene layer, if Hamiltonian \eqref{eq:bilayer_hamiltonian} represented the graphene bilayer sketched in Figure \ref{fig:graphene_stacked_lattice}. The introduction of these spinors explicitly incorporates the graphene layer that each particle belongs to as an aditional flavor index.

With all of the above, the solutions to the equation of motion $i\partial_t\Psi = H\Psi$ can be equivalently be stated as the mean-field approximation of an effective field theory described by the following classical action \cite{Bahamondes2024:JHEP}: 
\begin{align}
    S&=\int d^3x \left(i\bar \psi \gamma^a\partial_a \psi+i\bar \xi \gamma^a\partial_a \xi+\Delta_2 \left( \bar\psi\psi-\bar\xi\xi \right)-\Delta_1\left( \bar\psi\gamma^1\xi+\bar\xi\gamma^1\psi \right)  \right)\nonumber\\
    &=i\int d^3x\;\bar\Psi\slashed{\partial}\Psi + \int d^3x\;\bar\Psi\Phi\Psi - \int d^3x\;\bar\Psi\slashed{B}\Psi.\label{eq:effective_toy_action}
\end{align}
 Two new fields have been introduced in eq.~\eqref{eq:effective_toy_action}: non-abelian $SU(2)$ uniform scalar and gauge fields, which are denoted by $\Phi$ and $B$, respectively. In \eqref{eq:effective_toy_action} the scalar field is in the adjoint representation of $SU(2)$, and is given by $\Phi = \Delta_2\sigma_3$, while the gauge field is given by $B = \Delta_1\sigma_1\mathrm{d}x$. This shows that the tight-binding theory described by the original Hamiltonian \eqref{eq:bilayer_hamiltonian} has a $U(2)$ symmetry that is explicitly broken down to $U(1)$ by the presence of the fields $\Phi$ and $B$. This is indication that the quantum phase transition is not driven by the standard Landau scheme of spontaneous symmetry-breaking. It also indicates that the semi-Dirac nature of the transition between the semimetallic and insulating phases is not necesarily bound to a specific lattice structure, since the dispersion relation \eqref{eq:anysotropic_relation} could be equally deduced from the action \eqref{eq:effective_toy_action} without ever resorting to solid-state physics constructions. This leads to the hypothesis that semi-Dirac anisotropy is closely related to an underlying symmetry-breaking mechanism associated to the $U(2)$ group, and as such should be present in a whole universality class of effective field theories.

 The above gives the necessary justification to attempt at reproducing this phase transition through a bottom-up holographic construction, which is what we will show in section \ref{sec:bulk_action}. Note that the holographic model that will be built is not a dual of the field theory represented by the action \eqref{eq:effective_toy_action}. Rather, it will represent a whole universality class of strongly coupled QFTs that have an explicitly broken $U(2)$ symmetry by $SU(2)$ scalar and gauge sources. This matter content will act as a strongly coupled background on top of which probe fermions will be placed, and their dispersion relations calculated through holographic means for different values of these operator sources.

\section{Elements from AdS/CFT}\label{sec:elements_from_adscft}
\subsection*{Bulk spacetime construction and statement of the correspondence}
In order to create a holographic dual for a CMT system that transitions between a semimetal and an insulator through a semi-Dirac phase, we need some understanding on the use of AdS/CFT for bottom-up constructions. In essence, the AdS/CFT correspondence creates a dual spacetime from a conformal field theory (CFT) by treating the energy scale of such theory as a geometric quantity. Therefore, the renormalization group scale of the CFT is taken as an additional spacetime coordinate in an equal footing to any of the other spacetime dimensions \cite{Hartnoll2009:ClassQuantGrav_lectures}. This additional spacetime coordinate results in a $d+2$-dimensional spacetime, built from the $d+1$ dimensions of the original CFT plus the energy scale dimension.

\begin{figure}[!htb]
    \centering
    \includegraphics[width=0.8\linewidth]{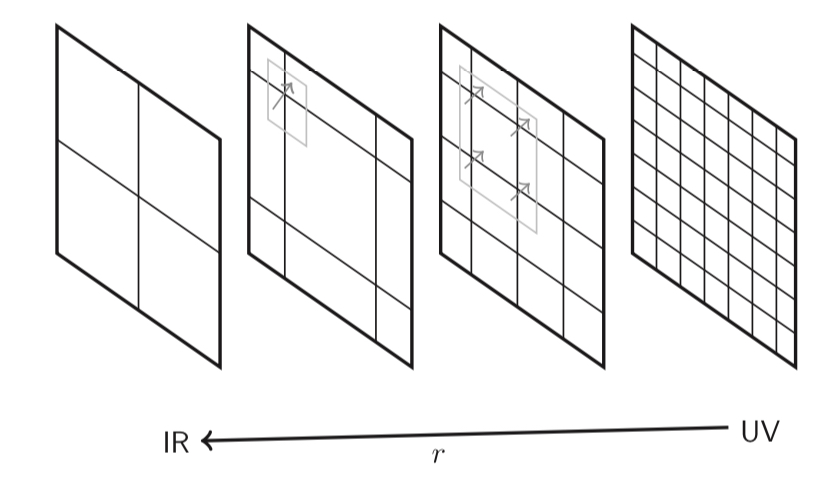}
    \caption{Stacking of different scaled versions of the $d+1$-dimensional spacetime of the dual CFT of interest at different values of the radial coordinate $r$. The interpretation of $r$ is as the renormalization group scale, with the UV and IR located at $r = 0$ or $r=\infty$, respectively, depending on the choice of coordinates. Image taken from \cite{Zaanen2015:Book}.}
    \label{fig:spacetime_stacking}
\end{figure}

The exact geometry of the dual spacetime is derived from symmetry principles by demanding the bulk spacetime isometries correspond to the symmetries of the CFT. For the simplest CFT these are the elements of the conformal group $SO(d+1,2)$, which translate into the following possible metric:
\begin{equation}\label{eq:pure_AdS_metric}
    \mathrm{d}s^2=\frac{r^2}{L^2}\eta_{ab}\mathrm{d}x^a\mathrm{d}x^b+\frac{L^2}{r^2}\mathrm{d}r^2.
\end{equation}
The latin indices $a$ and $b$ range from $0$ to $d$, and correspond to the CFT spacetime coordinates, while the radial coordinate $r$ is the geometrized energy scale of the theory \cite{Zaanen2015:Book}, with the ultraviolet (UV) of the renormalization group (RG) flow located at $r=\infty$ and the infrared (IR) at $r = 0$. The metric \eqref{eq:pure_AdS_metric} is the metric of pure anti-de Sitter spacetime in $d+2$ dimensions ($\mathrm{AdS}_{d+2}$) in Poincaré patch coordinates \cite{Erdmenger2015:book}, where $x^a\in\mathbb{R}^{d,1}$ and $r\in(0,\infty)$. The fixed constant $L$ is the anti-de Sitter radius, and fixes the scale of intrinsic curvature of the spacetime. The coordinate choice that will be used to describe the geometry of $\mathrm{AdS}_{d+2}$ spacetime will be slightly different by making the energy scale a more explicit "length scale" coordinate through the transformation $r\mapsto L^2/r$, resulting in the following line element \cite{Zaanen2015:BookChptr2}: 
\begin{equation}\label{eq:pure_AdS_metric_oficial}
 \mathrm{d}s^2=\frac{L^2}{r^2}\left(\eta_{ab}\mathrm{d}x^a\mathrm{d}x^b+\mathrm{d}r^2\right),   
\end{equation}
where the UV of the RG flow is now located at $r=0$ and the IR at $r=\infty$. The UV of the dual field theory at $r=0$ is of particular importance, since it is the conformal boundary of $\mathrm{AdS}_{d+2}$ spacetime. From here onwards, unless otherwise stated, this is the choice of Poincaré patch coordinates that we will always use when dealing with an $\mathrm{AdS}_{d+2}$ geometry.

It is clear from the metric \eqref{eq:pure_AdS_metric_oficial} that apart from a global $r$-dependent factor that diverges as $r\to 0$, the geometry tends towards a flat Minkowski spacetime; precisely the kind of geometry any QFT lives on. This is why we say that the dual field theory is located at the boundary of the $\mathrm{AdS}_{d+2}$ bulk, even though this boundary does not actually have a real physical location in the spacetime itself. The statement that a field theory is "located" at the boundary of this type of bulk should not be taken literally, yet it is common language in most of the literature to refer to the field theory of interest as a "boundary theory" \cite{Zaanen2015:BookChptr2}, and so we stick to this tradition.

In this work we deviate from exact AdS/CFT in the sense that the type of field theories that we describe are not necesarily conformal. Given the conjectural status of AdS/CFT we can get away with this deformation of the original correspondence as long as the geometry of the spacetime that is built from the geometrization of the RG flow respects the symmetries of the theory that is being dualized. Indeed, one of the foundations of AdS/CFT is that spacetime isometries in the bulk correspond to the symmetries of the dual field theory. This is usualy done by allowing the metric to be deformed away from pure $\mathrm{AdS}_{d+2}$ geometry in the IR of the RG scale, while maintaining it in the UV. This reflects the fact that different symmetries of a QFT can either appear or vanish when probing it at different energy scales, regardless of its microscopic details. This concept is called universality \cite{Spencer2010}, and therefore a given bulk gravitational theory will describe a whole universality class of QFTs linked together by their emergent symmetries. 

The only other requirement in this approach to AdS/CFT is that the UV of the theory remain a fixed point of the RG flow \cite{Hartnoll2009:ClassQuantGrav_lectures}. This is indeed accomplished if the geometry of the bulk is asymptoticaly $\mathrm{AdS}_{d+2}$ in the $r\to 0$ limit, since the metric \eqref{eq:pure_AdS_metric_oficial} is scale invariant under $r\mapsto \lambda r$, $x^a\mapsto \lambda x^a$. Recalling that bulk metric isometries correspond to QFT symmetries, this means that the $\beta$ functions of the dual QFT are invariant under scalings of the energy scale (which is taken to be the coordinate $r$). Therefore the Callan-Symanzyk equations imply that the $\beta$ functions of the QFT vanish in the UV, making it a fixed point of the RG flow \cite{hollowood20096lecturesqftrg}.

Finally, we must address how strongly coupled QFTs in particular are dualized by this type of spacetime construction; specifically by a weakly coupled, classical field theory that lives in such a bulk spacetime. The starting point of this work is to create a classical field theory that is governed by the laws of GR from the symmetries and field content of the QFT we are interested in. The metric \eqref{eq:pure_AdS_metric_oficial} is the simplest case where the only bulk field is the metric tensor $g_{\mu\nu}$. This classical tensor field is dual to the quantum field that any sensible QFT should posses: the energy-momemtum tensor $T_{ab}$ \cite{Balasubramanian1999:CommMathPhys}. However, more fields will be present in the bulk if bosonic and/or fermionic operators are also present in the boundary QFT. The most general way of stating the interactions and field content of the bulk theory is to write down its action, given in simplest terms by \cite{Zaanen2015:BookChptr1}:
\begin{equation}\label{eq:generic_bulk_action}
    S = \frac{1}{2\kappa^2}\int\!\mathrm{d}^{d+2}x\;\sqrt{-g}\left(R-2\Lambda+\cdots\right),
\end{equation}
where $R$ is the Ricci scalar, $\Lambda = -\frac{d(d+1)}{2L^2}$ is the negative cosmological constant, and $\kappa$ is the graviational coupling constant usually related to Newton's constant: $\kappa^2 = 8\pi G$. The terms implicit in the $\cdots$ include both the matter content of the bulk theory as well as higher derivative corrections to gravity. Pure $\mathrm{AdS}_{d+2}$ is a solution to the mean-field equations derived from the condition $\delta S = 0$ when no other fields are included in \eqref{eq:generic_bulk_action} and no corrections to the Einstein-Hilbert Lagrangian, $\mathcal{L}_{EH}=\frac{1}{2\kappa^2}(R-2\Lambda)$, are included.

 For the bulk theory to be a classical field theory determined by GR, corrections to the Einstein-Hilbert action should be ignored in \eqref{eq:generic_bulk_action}. In order to do this consistently, as is shown in \cite{Zaanen2015:BookChptr1}, the AdS/CFT dictionary requires that both the boundary degrees of freedom, $N$, and the QFT's coupling constant, $\lambda_{QFT}$, be arbitrarily large. This is because the ratio of the bulk's curvature to its Planck length, $L/\ell_{P}$ is proportional to $N$, and the ratio of $L$ to the length of strings from string theory in the bulk, $L/\ell_s$, is proportional to $\lambda_{QFT}$. To ignore quantum corrections to classical GR, and also ignore the effects of string theory so that the bulk fields can be classical, then $L/\ell_{P},L/\ell_{s}\to\infty$, which means taking $N,\lambda_{QFT}\to\infty$. This is known as the t'Hooft limit \cite{Zaanen2015:BookChptr1}. In this regimen, the gravitational sector of the bulk theory is accurately determined by GR, and all fields are classical since the presence of strings is negligible. On the other hand, the boundary theory is strongly coupled and has an arbitrarily large number of degrees of freedom. A small caveat is that, strictly speaking, the t'Hooft limit applies when dualizing an $SU(N)$ gauge CFT with matrix-valued trace operators \cite{hartnoll2018:MITpressChapter}, which is where the association of $N$ and $\lambda_{QFT}$ to $\ell_P$ and $\ell_s$ can be appropriately made using the AdS/CFT dictionary. However, several different top-down duals of AdS/CFT have shown the same type of correspondence, where a strong coupling and large number of degrees of freedom in the boundary theory leads to classical gravity in the bulk \cite{Maldacena1999:goatPaper,Aharony2008:JHEP}. This is why, from the bottom-up approach, a theory that is described by classical Einstein gravity is taken, at face value, to represent a dual QFT in the strong coupling and large $N$ limit, despite not knowing the explicit form of its Hamiltonian or Lagrangian.
 The particular coupling constant(s) of the boundary theory that is(are) arbitrarily large may not be known in bottom-up approaches, which does not mean that a large amount of observables can not be measured through the methods that will be outlined below. 
 
With all of the above, the statement of the AdS/CFT correspondence applied to strongly coupled QFTs is the following. The matter content and symmetries of a strongly coupled QFT, with an arbitrarily large number of degrees of freedom, is mapped into a bulk spacetime with negative curvature whose isometries are the symmetries of theory, and which has an asymptotic $\mathrm{AdS}$ geometry when approaching its conformal boundary. Any operator present in the QFT is mapped to a classical field with the same quantum numbers; a boundary scalar operator is mapped to a bulk scalar field, a vector boundary operator is mapped to a bulk vector field, etc... The interactions and dynamics of the bulk fields is determined by a bulk action $S_b$ with an Einstein-Hilbert term, and aditional matter content that depends on the specific interactions between matter fields that need to be reproduced:
\begin{equation}
    S_{\mathrm{b}} = \frac{1}{2\kappa^2}\int\!\mathrm{d}^{d+2}x\;\sqrt{-g}\left(R-2\Lambda\right)+S_{\mathrm{matter}}
\end{equation}

\subsection*{The GKPW formula}
The last paragraph of the previous subsection stated the AdS/CFT in a broader, qualitative footing. However, in order to make the duality concrete in the sense of being able to compute correlation functions of a strongly coupled QFT from quantities in the bulk, an explicit formula linking both sides of the duality is needed. This is achieved through the Gubser-Klebanov-Polyakov-Witten (GKPW) formula, formulated in \cite{Gubser1998:PhysLettB,Witten1998:TheorMatPhys}, which states that the partition functions of both sides of the duality coincide:
\begin{equation}\label{eq:GKPW_Rule}
   \int\!\mathcal{D}\left[\,\left\{\Xi_\alpha(x),\Xi^*_\alpha(x)\right\}_{\alpha\in I}\,\right] e^{iNS_{\mathrm{b}}[\{\Xi_\alpha\}_{\alpha\in I}]} = Z_{QFT}\left[\left\{J_{\alpha}=\Xi_{\alpha,(l)}^{\mathrm{sol}}(r\to 0)\right\}_{\alpha\in I}\right].
\end{equation}
In this formula, the left-hand side is the path integral over all bulk field configurations, with the latter indexed as $\{\Xi_\alpha(x)\}_{\alpha\in I}$ for some set of indices $I$. Each of these bulk fields could be a scalar, vector, tensor, or spinorial field, depending on the specific dual operator they are built to dualize. The spacetime coordinate $x$ that these fields depend on include the full bulk coordinates: $x^\mu = (t,\mathbf{x},r)$, with $x^a = (t,\mathbf{x})$ being the boundary coordinates. The constant $N$ represents the number of degrees of freedom of the boundary theory. The right-hand side is the partition function of the boundary QFT. In the $T=0$ case it corresponds to the generating functional of correlation functions \cite{Meert2022:PhDthesis}. The fields $J_\alpha \equiv J_\alpha(t,\mathbf{x})$ correspond to sources in the action of the dual QFT that couple to the operators that the bulk fields dualize. These source fields are taken to be the leading coefficient of the solution to the bulk equations of motion (EOMs) evaluated on the conformal boundary \cite{Erdmenger2015:bookChptr5}, which are labeled as $\Xi_{\alpha,(l)}^{\mathrm{sol}}\equiv\Xi_{\alpha,(l)}^{\mathrm{sol}}(t,\mathbf{x},r)$. 

For example, if the dual QFT's action (whatever it may be) has a scalar complex operator $\widehat{\phi}_\mathrm{s}\equiv\widehat{\phi}_\mathrm{s}(t,\mathbf{x})$, then $Z_{QFT}$ is a functional of two c-number valued sources $\{J,J^*\}$ that couple to $\widehat{\phi}\,,\,\widehat{\phi}^*$ as in:
\begin{equation}\label{eq:example_functional}
    Z_{QFT}\left[J\,,\,J^*\right]=\left\langle \exp \left[i\int\!\mathrm{d}t\mathrm{d}^d\mathbf{x}\;\left(J(t,\mathbf{x})^*\phi(t,\mathbf{x})+J(t,\mathbf{x})\phi^*(t,\mathbf{x})\right)\right]\right\rangle_{QFT}.
\end{equation}
The connection between this expression and the AdS spacetime is that both $J$ and $J^*$ are evaluated as a set of boundary conditions supplied to the leading solution of the dual scalar bulk field's EOMs\footnote{For all QFT path integrals, it is assumed that the vacuum to vacuum amplitude $Z_{QFT}[J_\alpha=0]$ is normalized to $1$ to simplify notation.}.

The GKPW rule as stated in \eqref{eq:GKPW_Rule} is the strong version of the correspondence. When dealing with strongly coupled QFTs in the large $N$ limit, the left-hand side of \eqref{eq:GKPW_Rule} can be approximated by a saddle point located at the stationary point of the bulk action \cite{Erdmenger2015:bookChptr5}. This weaker version of GKPW is the one that will be employed in this work:
\begin{equation}\label{eq:GKPW_weaker}
    e^{iS_{\mathrm{b}}[\{\Xi^{\mathrm{sol}}_\alpha\}_{\alpha\in I}]} = Z_{QFT}\left[\left\{J_{\alpha}=\Xi_{\alpha,(l)}^{\mathrm{sol}}(r\to 0)\right\}_{\alpha\in I}\right].
\end{equation}

In order to compute any $n$-point correlation function in the QFT as vacuum expectation values (VEV) of operator products, functional derivatives with respect to the operator sources are taken in $Z_{QFT}$, which are subsequently set equal to zero \cite{Erdmenger2015:bookChptr1}:
\begin{align}
\left\langle\widehat{\mathcal{O}}_1(t_1,\mathbf{x}_1)\cdots\widehat{\mathcal{O}}_n(t_n,\mathbf{x}_n)\right\rangle_{QFT} &= \int\!\mathcal{D}\left[\mathcal{O}_\alpha(t,\mathbf{x})\right]\,\mathcal{O}_1(t_1,\mathbf{x}_1)\cdots\mathcal{O}_n(t_n,\mathbf{x}_n)\,e^{iS_{QFT}}\nonumber\\
&=\left.(-i)^n\frac{\delta^{\,n}Z_{QFT}[\{J_\alpha\}_{\alpha\in I}]}{\delta\,J_1(t_1,\mathbf{x}_1)\cdots\delta\,J_n(t_n,\mathbf{x}_n)}\right|_{\{J_\alpha=0\}_{\alpha\in I}},\label{eq:n-point_function}
\end{align}
where time-ordering of the operators in the left-hand side is assumed. By using the GKPW formula in eq.~\eqref{eq:GKPW_weaker} we have the following way of calculating QFT correlation functions from quantities in the bulk:
\begin{equation}\label{eq:n-point_function_GKPW}
\left\langle\widehat{\mathcal{O}}_1(t_1,\mathbf{x}_1)\cdots\widehat{\mathcal{O}}_n(t_n,\mathbf{x}_n)\right\rangle_{QFT} = \left.\frac{\delta^{\,n}S_\mathrm{b}[\left\{\Xi^{\mathrm{sol}}_\alpha\}_{\alpha\in I}\right]}{\delta\,\Xi^{\mathrm{sol}}_{1,(l)}(t_1,\mathbf{x}_1)\cdots\delta\,\Xi^{\mathrm{sol}}_{n,(l)}(t_n,\mathbf{x}_n)}\right|_{\{\,\Xi^{\mathrm{sol}}_{\alpha,(l)}=0\,\}_{\alpha\in I}}.
\end{equation}
For calculations it will be convenient to state all n-point functions derived from \eqref{eq:n-point_function_GKPW} in momentum space, rather than in spacetime coordinates. In particular, the VEV of a particular operator in Fourier representation $\widehat{\mathcal{O}}_a(\omega,\mathbf{k}) = \int\frac{\mathrm{d}t\mathrm{d}^{d}\mathbf{x}}{(2\pi)^{d+1}}e^{i\omega t-i\mathbf{k}\cdot\mathbf{x}}\widehat{\mathcal{O}}_a(t,\mathbf{x})$ is given by:
\begin{equation}\label{eq:VEV_Fourier_space}
\left.\left\langle\widehat{\mathcal{O}}_a(\omega,\mathbf{k})\right\rangle_{QFT}=\frac{\delta\,S_\mathrm{b}[\left\{\Xi^{\mathrm{sol}}_\alpha\}_{\alpha\in I}\right]}{\delta\,\Xi_{a,(l)}^{\mathrm{sol}}(-\omega,-\mathbf{k})}\right|_{\{\,\Xi^{\mathrm{sol}}_{\alpha,(l)}=0\,\}_{\alpha\in I}}
\end{equation}

Eqs.~\eqref{eq:n-point_function}-\eqref{eq:VEV_Fourier_space} are calculations of correlation functions when the boundary QFT is at equilibrium. The GKPW formula can be adapted for the calculation of transport coefficients by means of linear response theory. If the Hamiltonian of the boundary QFT is perturbed to linear order by a time-dependent source $J_a$ coupled to an operator $\widehat{\mathcal{O}}_a$ as in:
\begin{equation}
    \widehat{H}_{QFT}\longmapsto \widehat{H}_{QFT}'=\widehat{H}_{QFT}+\int\!\mathrm{d}t\mathrm{d}^d\mathbf{x}\, J_a(t,\mathbf{x})\widehat{\mathcal{O}}_a(t,\mathbf{x}),
\end{equation}
the VEV of an operator $\widehat{\mathcal{O}}_b$ in the modified QFT is perturbed by the following quantity:
\begin{equation}\label{eq:linear_response}
\delta\langle\widehat{\mathcal{O}}_b(t,\mathbf{x})\rangle_{QFT} = \int\!\mathrm{d}t'\mathrm{d}^d\mathbf{x}'\;G_{ab}^R(t'-t,\mathbf{x}'-\mathbf{x})J_a(t',\mathbf{x}'),
\end{equation}
where $G^R(t,\mathbf{x})$ is the retarded Green's function of the operator $\widehat{\mathcal{O}}_b$ with respect to the perturbation operator $\widehat{\mathcal{O}}_a$\footnote{In all bulk systems relevant to this work, translation symmetry is always present, which is why all correlators depend on the relative difference of spacetime coordinates.}:
\begin{equation}\label{eq:retarded_correlator}
    G^R_{ab}(t'-t,\mathbf{x}'-\mathbf{x})=-i\,\theta(t'-t)\left\langle\left[\widehat{\mathcal{O}}_b(t,\mathbf{x}),\widehat{\mathcal{O}}_a(t',\mathbf{x}')\right]\right\rangle_{QFT},
\end{equation}
with the VEV in the right-hand side of \eqref{eq:retarded_correlator} taken with respect to the QFT at equilibrium. By applying the GKPW formula \eqref{eq:VEV_Fourier_space} to the one-point function on the left-hand side of eq.~\eqref{eq:linear_response} in Fourier space, the Fourier transform of the retarded Green's function can be calculated from the on-shell bulk action as:
\begin{equation}\label{eq:retarded_correlator_GKPW}
\left.\frac{\delta^2\,S_\mathrm{b}\left[\left\{\Xi_\alpha^{\mathrm{sol}}\right\}_{\alpha\in I}\right]}{\delta\,\Xi_{a,(l)}^{\mathrm{sol}}(\omega,\mathbf{k})\delta\,\Xi_{b,(l)}^{\mathrm{sol}}(-\omega,-\mathbf{k})}\right|_{\left\{\Xi_{\alpha,(l)}^{\mathrm{sol}}\right\}_{\alpha\neq a}}=G_{ab}^R(\omega,\mathbf{k}).
\end{equation}

\section{Bulk action and equations of motion}\label{sec:bulk_action}
In this section we build the bulk action for our strongly coupled, $2+1$-dimensional CMT boundary system. The theory to be dualized is a QFT at finite temperature (see Appendix \ref{app:AdS/CFT_finite_temp}), which has a global $U(2)$ symmetry broken down to $U(1)$ by an $SU(2)$ operator source, as well as another source that couples to a scalar field in the adjoint representation of $SU(2)$. These boundary operator sources will be labeled $\Delta_1$ for the gauge sector, and $\Delta_2$ for the scalar sector, taking inspiration from the effective action in \eqref{eq:effective_toy_action}. In this work we will solve the background EOMs in both $T=0$ and $T>0$ regimes separately, to measure different quantities.

The holographic dictionary states that global symmetries in the boundary theory must be made local (i.e: gauged) in the bulk \cite{Meert2022:PhDthesis}, which requires the introduction of a covariant derivative for all matter fields. As such, the bulk action of our model is given by \cite{Bahamondes2024:JHEP}:
 \begin{gather}
S_{\mathrm{b}}= \int\!\mathrm{d}^4x\,\sqrt{-g}\left[\frac{1}{2\kappa^2}\left(R+\frac{6}{L^2}\right)-\mathrm{Tr}\left(\left(D^\mu\Phi\right)^\dagger\left(D_\mu\Phi\right)\right)-m^2\mathrm{Tr}\left(\Phi^\dagger\Phi\right)\right.\nonumber\\\left.-\frac{\lambda}{4}\left(\mathrm{Tr}(\Phi^\dagger\Phi)\right)^2-
\frac{1}{4}\mathrm{Tr}\left(G_{\mu\nu}G^{\mu\nu}\right)\right],\label{eq:background_action}
\end{gather}
where $\Phi = \Phi_j\sigma_j$ is a scalar field in the adjoint representation of $SU(2)$, and $G_{\mu\nu} = \partial_\mu B_\nu-\partial_\nu B_\mu +iq[B_\mu,B_\nu]$ is the stress tensor constructed from a non-abelian gauge connection $B = B_{\mu,j}\sigma_j\mathrm{d}x^\mu$. The parameter $m^2$ is the square-mass of the scalar field, and $\lambda$ is a $\phi^4$ coupling constant introduced for stability when the $T=0$ limit is eventually taken. The gauge covariant derivative $D_\mu$ is built from the connection field $B$, and acts on $\Phi$ as:
\begin{equation}\label{eq:adjoint_cov_dev}
D_\mu\Phi=\nabla_\mu\Phi +i q\left[\sigma_j B_{\mu,j},\Phi\right],
\end{equation}
where $q$ is the color charge associated to the scalar field, and $\nabla_\mu$ is the standard metric covariant derivative.

Next, we calculate the EOMs of the bulk fields from the saddle point of \eqref{eq:background_action}. These translate into the Einstein equations for the metric field, the curved Klein-Gordon equation for the scalar field, and the Yang-Mills equation for the gauge field sourced by the scalar:
\begin{align}
    R_{\mu\nu}-\frac{1}{2}Rg_{\mu\nu}-\frac{3}{L^2}g_{\mu\nu}&= \kappa^2\left(T_{\mu\nu}^{\mathrm{B}}+T_{\mu\nu}^{\mathrm{\Phi}}\right)\label{eq:einstein_eqs}\\
    (D_\mu D^\mu-m^2)\Phi &= \frac{\lambda}{2}\mathrm{Tr}\left(\Phi^\dagger\Phi\right)\Phi \label{eq:kleingordon_eq}\\
    D_\mu G^{\mu\nu} &= iq\left(\left[\Phi^\dagger,D^\nu\Phi\right]-\left[\Phi^\dagger,D^\nu\Phi\right]^\dagger\right),\label{eq:yangmills_eq}
\end{align}
with $T_{\mu\nu}^B$ and $T_{\mu\nu}^\Phi$ being the stress-energy tensors of the gauge and scalar sectors of the bulk, respectively:
\begin{align}
    T_{\mu\nu}^B &= \mathrm{Tr}\left(G_\mu^{\;\alpha}G_{\nu\alpha}\right)-\frac{1}{4}g_{\mu\nu}\mathrm{Tr}\left(G_{\alpha\beta}G^{\alpha\beta}\right)\label{eq:gauge_stress_energy_tensor}\\
    T_{\mu\nu}^\Phi&= 2\mathrm{Tr}\left[\left(D_\mu\Phi\right)^\dagger\left(D_\nu\Phi\right)\right]-g_{\mu\nu}\left\{\mathrm{Tr}\left[\left(D_\alpha\Phi\right)^\dagger\left(D^\alpha\Phi\right)\right]+m^2\mathrm{Tr}\left(\Phi^\dagger\Phi\right)+\frac{\lambda}{4}\left[\mathrm{Tr}\left(\Phi^\dagger\Phi\right)\right]^2\right\}.\label{eq:scalar_stress_energy_tensor}
\end{align}
Eqs.~\eqref{eq:einstein_eqs}-\eqref{eq:kleingordon_eq} will be refered to as the background EOMs.

Next, we state appropriate \textit{ansatze} for the matter fields, based on the symmetries of the boundary theory we wish the bulk to reflect. Since the boundary system is translationally invariant, the matter fields $\Phi$ and $B$ must reflect this by being independent of the boundary coordinates. Also, the scalar source in the boundary is along the direction of $\sigma_3$ in the $SU(2)$ algebra, and the gauge source along $\sigma_1$, and has nontrivial spacetime coordinates only in the $\mathrm{d}x$ 1-form.
As such, we will take these two fields to be of the following shape in the bulk:
\begin{equation}\label{eq:matter_fields_ansatz}
    \Phi \equiv \Phi(r)= \phi(r)\sigma_3 \;,\;
    B\equiv B(r)= B(r)\sigma_1dx\,.
\end{equation}
As for the bulk geometry itself, since the boundary will be anisotropic, we use the following metric \textit{ansatz}:
 \begin{equation}\label{eq:metric_ansatz}
     \mathrm{d}s^2 = \frac{L^2}{r^2}\left(-f(r)N(r)^2\mathrm{d}t^2+\frac{\mathrm{d}r^2}{f(r)}+h(r)^2\mathrm{d}x^2+\frac{1}{h(r)^2}\mathrm{d}y^2\right).
 \end{equation}
A non-trivial profile of the function $h(r)$ explicitly breaks $SO(2)$ invariance along the boundary coordinates. As is explained in Appendix \ref{app:AdS/CFT_finite_temp}, the emblackening factor $f(r)$ will allow for black brane solutions when we demand that $f(r_h)=0$ at some finite radial coordinate value $r_h>0$. On the other hand, the function $N(r)$ will allow for a Lifshitz-type anisotropic scaling of the time coordinate when $T=0$ solutions are found in chapter \ref{chap:results}. The only ingredient that remains is suplying all fields with UV boundary conditions. As was explained in section \ref{sec:elements_from_adscft}, the leading coefficients of $\phi$ and $B$ correspond to the sources that couple to the boundary scalar and gauge operators, while the leading terms of the geometry fields must be such that the $r\to0$ limit of \eqref{eq:metric_ansatz} is $\mathrm{AdS}_4$. This is accomplished by the following asymptotic expansion near the conformal boundary:
\begin{align}
     \phi(r\to 0) &= r^{\Delta_\phi}\Delta_2+\phi_{(s)}r^{3-\Delta_\phi}+\cdots\label{eq:boundary_expansion_scalar}\\
     B(r\to 0) &= \Delta_1+B_{(s)}r+\cdots\label{eq:boundary_expansion_gauge}\\
     f(r\to 0) &= 1 +\cdots +f_3 r^3+\cdots\label{eq:boundary_expansion_f}\\
     h(r\to 0) &= 1+\cdots + h_3 r^3+\cdots\label{eq:boundary_expansion_h}\\
     N(r\to 0) &= 1+\cdots + N_3 r^3+\cdots \label{eq:boundary_expansion_N},
 \end{align}
where $\phi_{(s)}$, $B_{(s)}$, $f_3$, $h_3$ and $N_3$ are the subleading coefficients of the solutions to the background EOMs, 
while $\Delta_\phi$ is the dual scalar operator's scaling dimension, set by $\Delta_\phi(\Delta_\phi-3)=m^2L^2$. Finally, since we are looking for black brane solutions to the bulk fields, boundary conditions at the event horizon $r=r_h$ must also be supplied. We simply impose regularity on $r=r_h$ by demanding the fields acquire a power series expansion near the horizon:
 \begin{align}
     \varphi(r\to r_h) &= a_0+a_1(r_h-r)+a_2(r_h-r)^2+a_3(r_h-r)^3+\cdots \label{eq:horizon_scalar}\\
     b(r\to r_h) &= b_0+b_1(r_h-r)+b_2(r_h-r)^2+b_3(r_h-r)^3+\cdots \label{eq:horizon_gauge}\\
     f(r\to r_h) &= f_1(r_h-r)+f_2(r_h-r)^2+f_3(r_h-r)^3+\cdots \label{eq:horizon_f}\\
     h(r\to r_h)&= h_0+h_1(r_h-r)+h_2(r_h-r)^2+h_3(r_h-r)^3+\cdots \label{eq:horizon_h}\\
     N(r\to r_h)&= N_0+N_1(r_h-r)+N_2(r_h+r)^2+N_3(r_h-r)^3+\cdots \label{eq:horizon_N}.
 \end{align}

Notice that eqs.~\eqref{eq:einstein_eqs}-\eqref{eq:yangmills_eq} have the symmetry $B\longrightarrow qB$, $\Phi\longrightarrow q\Phi$, $\kappa^2\longrightarrow \kappa^2/q$, which allows to measure the coupling of the gravity and gauge sectors simultaneously in terms of the single coupling constant $\kappa^2/q$. This scaling symmetry allows for solving the background EOMs in two different regimes: the probe limit ($q\to\infty$) and the backreacted regime ($q$ finite). 

In the probe limit the dynamics of the scalar and Yang-Mills sectors of the bulk decouple from the metric dynamics, since the right-hand side of eq.~\eqref{eq:einstein_eqs} becomes zero, and we are left with the vacuum Einstein's equations for the metric. The probe limit is sufficient for locating the phase transition at finite temperature as long as the scalar and gauge fields are small enough in units of $L$ so as to make the effect of the stress-energy tensors of both fields in Einstein's equations negligible. Indeed, the phase transition in the toy model in section \ref{sec:toy_model} was driven by the gauge and scalar fields, so it should be expected that such mechanism should be captured in the bulk entirely by the dynamics of $\phi$ and $B$. The main limitation of the probe limit is the fact that, to dualize a thermal field theory, the only possible solution to the metric that allows for such dualization as per the holographic dictionary is an $\mathrm{AdS}_4$-Schwarzschild black brane. This restricts the dualization of the boundary theory only to very high temperatures, while limits where the IR geometry would differ significantly from the Schwarzschild solution, such as the $T\to0$ or $T=0$ regimes, are beyond reach. As such, to compute transport coefficients of the boundary theory associated to the background fields, and to find $T=0$ solutions, backreaction needs to be taken into account in the background EOMs. Indeed, the probe limit decouples the gravitational degrees of freedom from the flavor currents in the boundary. Such couplings are needed if one is to consider the effects of the emergent anisotropy of the boundary theory on transport by means of linear response theory.

Given all of the above, the background EOMs will be solved in both the probe and backreaction limits. The probe limit will be used for locating the phase transition, and corroborating it corresponds to a transition between a semimetal towards a band insulator through a semi-Dirac point/phase. This will be done by coupling probe fermions to the background bulk in such probe limit, and measuring the spectrum of the former in the boundary through the fermionic retarded Green's function for quasinormal modes in each phase. Once this is done, and the $\Delta_1-\Delta_2$ phase diagram of the theory is obtained, we will focus on solutions to the backreacted background EOMs and use linear response theory to measure the shear viscosity of the boundary theory, and to find $T=0$ solutions. This last step will be done to corroborate that the semi-Dirac phase at finite $T$ emerges from a Quantum Critical Point (QCP) at $T=0$.

In all subsequent calculations and results, all dimensionfull quantities will be measured in units of $L$. Therefore, we immediately set $L=1$. In this choice of units we also make $m^2=-2$ in order to comply with the boundary Breitenlohner-Friedmann bound \cite{Bahamondes2022:BachelorThesis}.
 This results in a dual scaling dimension for the scalar field of $\Delta_\phi = 1$.

\section{Fermions coupled to the probe-limit background}\label{sec:fermions_coupled_to_background}
In this section we show how we couple probe fermions to the background bulk\footnote{In this section we use the exact same notation found in \cite{Bahamondes2024:JHEP}, although citation is, of course, appropriately given in each step where it is required.}. First we set the background EOMs in the probe limit by setting $\kappa = 0$ and $q=1$. The $\phi^4$ term in \eqref{eq:generic_bulk_action} is only relevant when taking the $T\to 0$ limit, since it provides stability to our system given that the choice of $m^2$ is negative. Since we won't be interested in such regime when calculating boundary fermionic correlators, we ignore it by setting $\lambda = 0$ only for this case, to make calculations simpler in the probe limit. With this in mind, the metric functions $f,h$ and $N$ are given by $f(r)=1-(r/r_h)^3$ and $h=N\equiv 1$. We make the background EOMs dimensionless by re-scaling the radial coordinate $r\mapsto r/r_h$, so that for all calculations $r_h = 1$. The Klein-Gordon and Yang-Mills equations for the matter fields \eqref{eq:matter_fields_ansatz} are \cite{Bahamondes2024:JHEP}:
\begin{align}
(4r^2B(r)^2-2)\phi(r)-r\left[r\frac{\mathrm{d}f}{\mathrm{d}r}\frac{\mathrm{d}\phi}{\mathrm{d}r}+f(r)\left(r\frac{\mathrm{d}^2\phi}{\mathrm{d}r^2}-2\frac{\mathrm{d}\phi}{\mathrm{d}r}\right)\right]&=0\label{eq:scalar_eom}\\
8B(r)\phi(r)^2-r^2\left(\frac{\mathrm{d}B}{\mathrm{d}r}\frac{\mathrm{d}f}{\mathrm{d}r}+f(r)\frac{\mathrm{d}^2B}{\mathrm{d}r^2}\right)&=0.\label{eq:gauge_eom}
\end{align}
We also define a set of \textit{vielbeins} for this geometry, given by \cite{Bahamondes2024:JHEP}:
\begin{equation}\label{eq:vielbeins}
    e_{\underline{0}}\,:=\frac{r}{\sqrt{f(r)}}\partial_t\;,\;e_{\underline{1}}:= r\,\partial_{x}\;,\;e_{\underline{2}}\colon = r\,\partial_y\;,\;e_{\underline{3}}:=r\sqrt{f(r)}\partial_r\;,
\end{equation}
where quantities with underlined indices $\underline{\mu} \in \{0,1,2,3\}$ represent tensor coordinates in the \textit{vielbein} basis.

The Dirac bulk action for probe fermions is inspired by the standard AdS/CFT literature when implementing holographic bulk fermions \cite{Giordano2017:JHEP,Plantz2018:JHEP,Grandi2022:PhysRevD}, and is given by \cite{Bahamondes2024:JHEP}:
\begin{equation}\label{eq:fermionic_bulk_action}
    S_f = i\int\!\mathrm{d}^4x\;\left(\bar\Psi\slashed{D}\Psi-g_Y\bar\Psi\Phi\Psi\right).
\end{equation}
Notice we explicitly introduce a Yukawa-type coupling between the fermions and the scalar field through a coupling constant $g_Y$. This is done because the insulating phase in the boundary can only occur if the boundary fermions acquire an effective mass to induce the formation of a gap between the conduction and valence bands \cite{Plantz2018:JHEP}. Since boundary fermions are massless in the sense that they don't possess a physical mass that could be implemented through a bulk fermion mass parameter $M_f$, the best way to achieve the formation of a gap is by making the scalar field take that role \cite{Plantz2018:JHEP}.

The $D_\mu$ operator in \eqref{eq:fermionic_bulk_action} is the spinor covariant derivative given by \cite{Bahamondes2024:JHEP}:
\begin{equation}\label{eq:spinor_cov_dev}
    D_\mu = 1_{2\times 2}\otimes(\,\nabla_\mu 1_{4\times 4}\,)+ 1_{2\times 2}\otimes\Gamma_\mu+\left(iq_fB_{\mu,j}\sigma_j\right)\otimes 1_{4\times 4},
\end{equation}
where $q_f$ is the color charge of the spinor $\Psi$, and $\Gamma_\mu$ are the affine connections for the background geometry. A Dirac spinor in a $3+1$-dimensional spacetime necesarily is a 4-tuple, while the coupling of the gauge sector to the spinor introduces an aditional flavor index to the bulk fermion field, resulting in a spinor $\Psi$ that is an 8-tuple of the form $\Psi = \begin{bmatrix}
    \psi_1\\\psi_2
\end{bmatrix}$. The construction of this fermionic covariant derivative in curved spacetime is standard literature, and can be found in works like \cite{Giordano2017:JHEP,Plantz2018:JHEP,Grandi2022:PhysRevD}, and in books like \cite{CurvedDiracEquationBook}. The affine connections $\Gamma^\mu$ are given by $\Gamma_\mu = \frac{1}{8}\omega_{\underline{\alpha} \,\underline{\sigma}\,\mu}\left[\gamma^{\underline{\alpha}},\gamma^{\underline{\beta}}\right]$, where $\omega_{\underline{\alpha}\,\underline{\sigma}\,\mu}$ are the spin connections built from the \textit{vielbeins} in \eqref{eq:vielbeins} (see \cite{Grandi:2021bsp} for explicit details). Finally, the "flat" bulk gamma matrices $\gamma^{\underline{a}}$ are chosen in the following representation:
\begin{equation}\label{eq:gamma_matrices}
    \gamma^{\underline{0}} = \begin{bmatrix}
        0 & i\sigma_2\\
        i\sigma_2&0
    \end{bmatrix} \quad , \quad \gamma^{\underline{1}} = \begin{bmatrix}
        0 & \sigma_1\\
        \sigma_1 & 0
    \end{bmatrix}
    \quad , \quad \gamma^{\underline{2}} = \begin{bmatrix}
        0 & \sigma_3\\
        \sigma_3&0
    \end{bmatrix}
    \quad , \quad 
    \gamma^{\underline{3}} = \begin{bmatrix}
        -I_{2\times 2} & 0\\
        0 & I_{2\times 2}.
    \end{bmatrix}
\end{equation}
The curved gamma matrices used in the $\slashed{D} \equiv \gamma^\mu D_\mu$ Dirac notation are given by $\gamma^\mu = e^{\mu}_{\underline{a}}\gamma^{\underline{a}}$, and they clearly satisfy the curved Clifford algebra $\{\gamma^\mu,\gamma^\nu\} = 2g^{\mu\nu}1_{4\times 4}$.

The Dirac equation that results from \eqref{eq:fermionic_bulk_action} is given by:
\begin{equation}\label{eq:Dirac_equation}
    (\slashed{D} - g_Y\Phi)\Psi = 0.
\end{equation}
It is evident from \eqref{eq:Dirac_equation} that the scalar field acts as an effective mass for the bulk fermion field. Recall that since these are probe fermions, their dynamics don't backreact on the background matter fields; the latter are solutions to eqs.~\eqref{eq:scalar_eom}-\eqref{eq:gauge_eom} subject to the boundary conditions \eqref{eq:boundary_expansion_scalar}-\eqref{eq:boundary_expansion_gauge}. Rather, we solve Dirac's equation on top of this background, and measure its effect on the spectrum of the boundary fermion operator.

 As is explained in Appendix \ref{app:fermions_in_AdS/CFT}, the AdS/CFT correspondence requires we project the bulk 8-tuple spinor onto the eigenspace of the flat radial $\gamma$ matrix, and separate each component as a 4-touple spinor. We label each projection $\psi_\pm$, and re-scale them as $\psi_\pm = r^{3/2}f(r)^{-1/4}\zeta_{\pm}$ (see Appendix \ref{app:fermions_in_AdS/CFT} for precise details on this construction). Finally, we Fourier decompose these spinors along the boundary coordinates: $\zeta_{\pm}(t,\mathbf{x},r)=e^{-i\omega t+i\mathbf{k}\cdot\mathbf{x}}\zeta_{\pm}(\omega,\mathbf{k},r)$, and plug everything into Dirac's equation \eqref{eq:Dirac_equation}. The resulting system of coupled differential equations in Fourier space is:
 \begin{align}
    \frac{\mathrm{d}\zeta_+}{\mathrm{d}r}+\frac{i}{\sqrt{f(r)}}U\zeta_-=-\frac{g_Y}{r}\frac{\phi(r)}{\sqrt{f(r)}}\gamma^{\underline{3}}\zeta_+ \label{eq:Dirac_eq_1}\\
    \frac{\mathrm{d}\zeta_-}{\mathrm{d}r}-\frac{i}{\sqrt{f(r)}}U\zeta_+=\frac{g_Y}{r}\frac{\phi(r)}{\sqrt{f(r)}}\gamma^{\underline{3}}\zeta_-\label{eq:Dirac_eq_2}, 
\end{align}
where the matrix $U$ has been defined as:
\begin{equation}\label{eq:operator_u}
    U(r;\omega,k_x,k_y)=\left[\begin{array}{cccc}
        k_y & k_x-\frac{\omega}{\sqrt{f(r)}} & 0 & q_fB(r) \\
        k_x + \frac{\omega}{\sqrt{f(r)}} & -k_y & q_fB(r) & 0 \\
        0 & q_fB(r) & k_y & k_x-\frac{\omega}{\sqrt{f(r)}} \\
        q_fB(r) & 0 & k_x+\frac{\omega}{\sqrt{f(r)}} & -k_y 
    \end{array}\right].
\end{equation}
Finally, we impose infalling boundary conditions at the event horizon $r=1$ for the fermion fields:
\begin{equation}\label{eq:infalling_fermions}
    \zeta_\pm^{\mathrm{IR}}(r) = (1-r)^{-i\omega/4\pi T}\zeta_{\pm,(l)}^{\mathrm{IR}}\,,
\end{equation}
where $T = \frac{3}{4\pi}$ is the dimensionless Hawking temperature of the black brane. Plugging \eqref{eq:infalling_fermions} into \eqref{eq:Dirac_eq_1}-\eqref{eq:Dirac_eq_2} and solving order by order in a series expansion around the event horizon yields the relation $\zeta_{-,(l)}^{\mathrm{IR}}=-i(1_{2\times 2}\otimes\sigma_2)\zeta_{+,(l)}^{\mathrm{IR}}$. This indicates there are only four independent initial conditions to be set at $r=1$ for the fermion fields, which we take to be the four entries of the tuple $\zeta_{+,(l)}^{\mathrm{IR}}$.

The leading part of the spinor $\zeta_+$ in the UV, $\zeta_{+,(l)}^{UV}\equiv\zeta_{+,(l)}^{UV}(\omega,\mathbf{k})$, is interpreted as the source that couples to the boundary fermionic operator. Using eq.~\eqref{eq:VEV_Fourier_space} the holographic dictionary implies that the corresponding leading term of $\zeta_-$ is the operator's VEV. These two are related through a fermionic correlation matrix $S\equiv S(\omega,\mathbf{k}\,;\,\Delta_1,\Delta_2)$ \cite{Liu2011:PhysRevD,Giordano2017:JHEP}, defined by:
\begin{equation}\label{eq:fermion_correlator}
\zeta_{-,(l)}^\mathrm{UV}(\omega,\mathbf{k})=S(\omega,\mathbf{k};\Delta_1,\Delta_2)\zeta_{+,(l)}^\mathrm{UV}(\omega,\mathbf{k}).
\end{equation}
From this definition of $S$, it can be deduced that it relates to the retarded Green's function of boundary fermions through the relation $S(\omega,\mathbf{k},;\Delta_1,\Delta_2) = \gamma^{\underline{0}}G_R(\omega,\mathbf{k}\,;\Delta_1,\Delta_2)$ \cite{Iqbal2009:FortPhys,Amon2010:JHEP}. Therefore, the lowest-lying poles of $S$ in complex $\omega$-space in the retarded prescription correspond to the quasi-normal frequencies of fermionic collective excitations in the boundary theory, and their dependence on momentum $\mathbf{k}$ for different values of $\Delta_{1,2}$ will correspond to the different band structures we expect to find in the different phases of the boundary.

Since the Dirac equation is linear, a linear relation holds between the $\zeta_\pm$ fields in the deep IR and in the UV in the form $\zeta_{\pm,(l)}^{\mathrm{UV}} = M_{\pm}\zeta_{-,(l)}^\mathrm{
IR}$. The relation $\zeta_{-,(l)}^{\mathrm{IR}}=-i(1_{2\times 2}\otimes\sigma_2)\zeta_{+,(l)}^{\mathrm{IR}}$ seen above imples the following:
\begin{eqnarray}\label{eq:numerical_S_matrix}
S=iM_- (1_{2\times 2}\otimes\sigma_2)M_+^{-1}.    
\end{eqnarray}
Therefore we can use the determinant method \cite{Amado2009:JHEP,Gubser2010:JHEP,Grandi2022:PhysRevD} to calculate the poles of $S$ in $\omega$-complex plane from the zeroes of the determinant of $M_+$. 

Finally, notice that after appropriate rescaling of eqs.\eqref{eq:Dirac_eq_1} and \eqref{eq:Dirac_eq_2}, the only free parameters remaining in the whole system are $q_f$ and $g_Y$. Both of these parameters determine the coupling strength of the fermions to the bosonic fields in the background. We set $g_Y =q_f = 1/2$ for all numerical calculations carried out in chapter \ref{chap:results}.

\chapter{Results}\label{chap:results}
In this chapter we outline the main results of this work. In the first section we show numerical results regarding the location of the phase transition at finite temperature, and the reconstruction of the full band structure of boundary probe fermions in each phase of $\Delta_1-\Delta_2$ parameter space. We also expose out-of-bounds hydrodynamic behavior around the semi-Dirac region, where a power-law scaling dependence of the shear viscosity-entropy density ratio with respect to temperature is located (see Appendix \ref{app:hydrodynamics} for theoretical background). In the second section we show both analytical and numerical results regarding $T=0$ solutions to the bulk theory, where we locate the semi-Dirac QCP between the semimetallic and insulating phases at zero temperature, and corroborate its Lifshitz nature through a non-relativistic dynamical critical exponent $z$.

\section{Finite temperature}\label{sec:finite_temperature}
\subsection{Phase diagram and band structure of fermions}\label{subsec:phase_diagram_band_structure}
\begin{figure}[!htb]
    \centering
    \includegraphics[width=\linewidth]{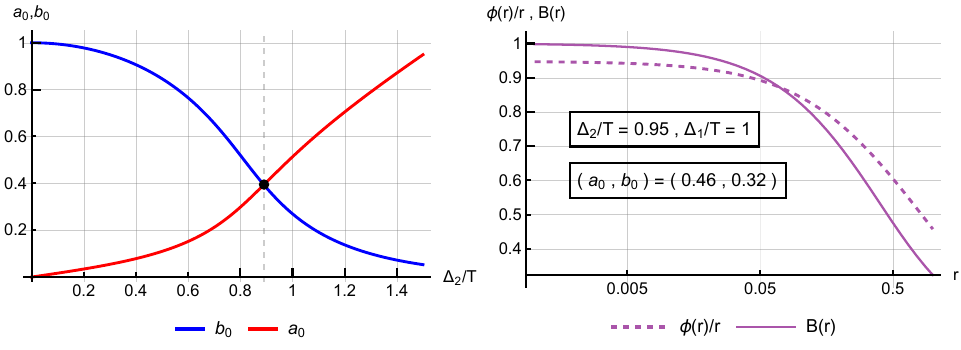}
    \caption{Left plot: Evolution of the scalar and gauge shooting parameters $a_0$ and $b_0$ with respect to $\Delta_2/T$ for $\Delta_1/T = 1$.The point where $a_0=b_0$ is located at the critical value $\Delta_2 \approx 0.883$. Right plot: Profile of matter fields obtained through solving eqs.~\eqref{eq:scalar_eom}-\eqref{eq:gauge_eom} for a sample value of $\Delta_{1,2}/T$, and their corresponding IR shooting parameters $(a_0,b_0)$. The choice of plotting $\phi(r)/r$ is made so as to make the value of $\Delta_2$ clear from the intersection of the dashed line with the vertical axis. The radial coordinate is set in logarithmic scale.}
    \label{fig:background_probe}
\end{figure}

First we numerically solve the background EOMs in the probe limit: eqs.~\eqref{eq:scalar_eom}-\eqref{eq:gauge_eom}. All numerical solutions to the background EOMs will be characterized in terms of the dimensionless parameters $\Delta_1/T$ and $\Delta_2/T$. We do this by implementing a shooting method, using the free event horizon parameters $a_0$ and $b_0$, as defined in eqs.~\eqref{eq:horizon_scalar}-\eqref{eq:horizon_gauge}, to shoot from the IR towards the UV boundary conditions of $\phi$ and $B$; i.e: expansions \eqref{eq:boundary_expansion_scalar}-\eqref{eq:boundary_expansion_gauge}. Each choice of $\Delta_1$ and $\Delta_2$ in the UV yields a unique pair $(a_0,b_0)$ in the IR. In Figure \ref{fig:background_probe} we show the resulting behavior of $a_0$ and $b_0$ as function of $\Delta_2/T$ for fixed $\Delta_1/T = 1$, as well as a typical profile of the matter fields $\phi$ and $B$ for a sample value of the UV boundary conditions.
Recall from section \ref{sec:elements_from_adscft} that the radial bulk coordinate is interpreted as the energy scale of the RG flow of the boundary theory. In the Poincaré patch coordinates used here, the theory's IR is at the event horizon and the UV in the bulk's conformal boundary. Using this scheme we interpret the values of $a_0$
 and $b_0$ as the IR renormalized values of $\Delta_1$ and $\Delta_2$, which would correspond to the bare value of the dual flavor currents \cite{Bahamondes2024:JHEP}. Figure \ref{fig:background_probe} leads to the naive expectation that the critical value of $\Delta_2/T$ where the transition takes place should be located at $(\Delta_2/T)_c \approx 0.883$ (when $\Delta_1/T = 1$), since the QPT at $T=0$ in the toy model of section \ref{sec:toy_model} occured when $\Delta_1 = \Delta_2$. To confirm this expectation we need the explicit band structure of probe fermions coupled to this background.

\begin{figure}[!htb]
    \centering
    \includegraphics[width=\linewidth]{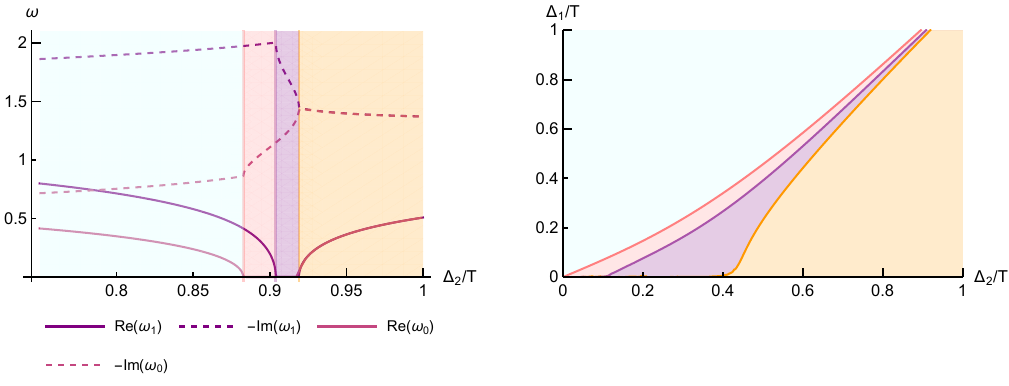}
    \caption{Left plot: Evolution, with respect to $\Delta_2/T$, of the real and imaginary parts of the lowest-lying poles of the retarded fermionic correlator at $\mathbf{k}=\mathbf{0}$ (for $\Delta_1/T=1$). Right plot: Phase diagram in $\Delta_1-\Delta_2$ space, with regions colored according to the behavior of the quasi-normal frequencies $\omega_{0,1}$ for arbitrary values of $\Delta_{1,2}/T$. The light-blue region is the semimetallic phase, the yellow region is the insulating one, and the pink and purple ones are the critical regions where anisotropy arises in the dispersion relations of the lowest-lying fermionic quasi-normal modes arises.}
    \label{fig:phase_diagram}
\end{figure}

 Now we solve Dirac's equations \eqref{eq:Dirac_eq_1}-\eqref{eq:Dirac_eq_2} using the same shooting method to numerically read each entry of the matrices $M_\pm$ defined in section \ref{sec:fermions_coupled_to_background}. First we set $\mathbf{k}=\mathbf{0}$, and locate the lowest-lying zeroes of $\mathrm{det}(M_+)$ on the complex $\omega$-plane; i.e: those with the least negative imaginary part. We locate two such modes for all values of $\Delta_{1,2}/T$, which we call $\omega_0$ and $\omega_1$. The lowest-lying pole (the one interpreted as a one-particle state in the spectrum of the dual fermionic operator) is $\omega_0$, while $\omega_1$ consistently has a greater, or equal, absolute value of its imaginary part. This pole may could be interpreted as a two-particle or many-particle state, yet it will turn out that the mode that exhibits the band structure that we expect is $\omega_0$. Once again, for $\Delta_1/T=1$ we plot the evolution of $\omega_{0,1}$ as a function of $\Delta_{2}/T$, and we plot the results in Figure \ref{fig:phase_diagram}.
We see that, at zero spatial momentum and for $\Delta_1/T = 1$, the real part of $\omega_0$ becomes zero at precisely $\left(\Delta_2/T\right)_c$. The region where $\mathrm{Re}(\omega_0)$ is non-zero at $\mathbf{k}=\mathbf{0}$ for $0\leq\Delta_2/T\leq \left(\Delta_2/T\right)_c$ is shown in light blue in Figure \ref{fig:phase_diagram}. Unlike the naive expectation from our toy model, where we would expect $\mathrm{Re}(\omega_0)$ to bounce back immediately towards a gapped band structure, the system remains gapless for a finite range of $\Delta_2/T$; this region is shown in pink in Figure \ref{fig:phase_diagram}. Mode $\omega_1$ becomes gapless at a different value of $\Delta_2/T$, and remains gapless for a finite range of $\Delta_2/T$ (the region shown in purple in Figure \ref{fig:phase_diagram}). Finally, both modes meet at another critical value $(\Delta_2/T)'_{c} \approx 0.918$, and become massive for all larger values of $\Delta_2/T$ (yellow region in Figure \ref{fig:phase_diagram}). 

\begin{figure}[!htb]
    \centering
    \begin{subfigure}{\textwidth}
    \centering
    \includegraphics[width=\textwidth]{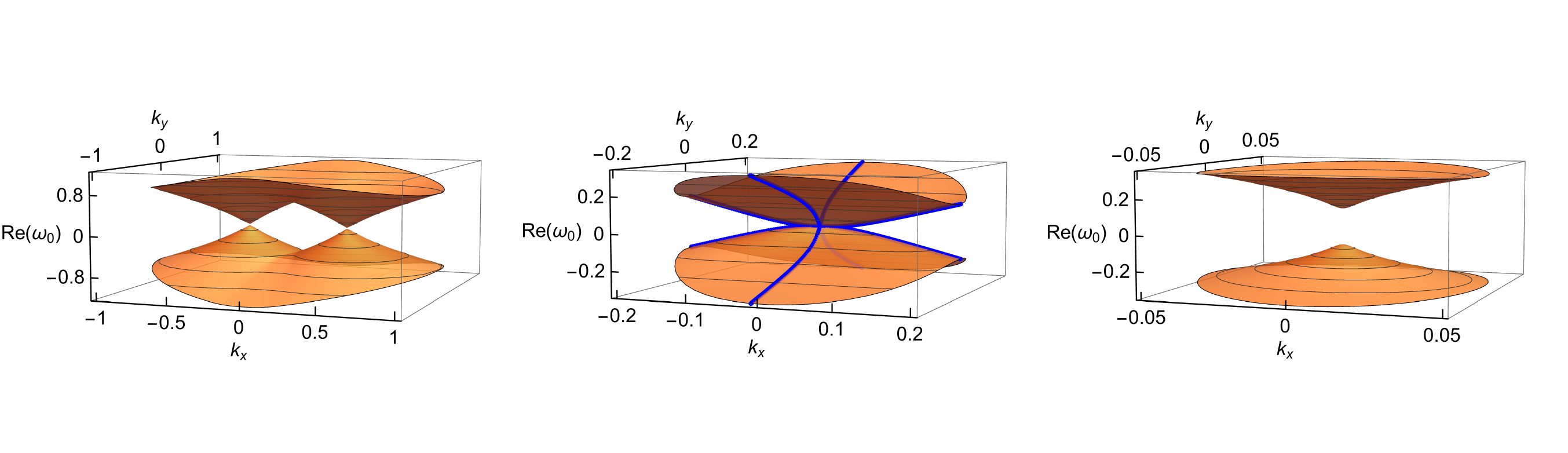}
    \vspace*{-1.8cm}
    \end{subfigure}
    \begin{subfigure}{\textwidth}
    \centering
        \includegraphics[width=\textwidth]{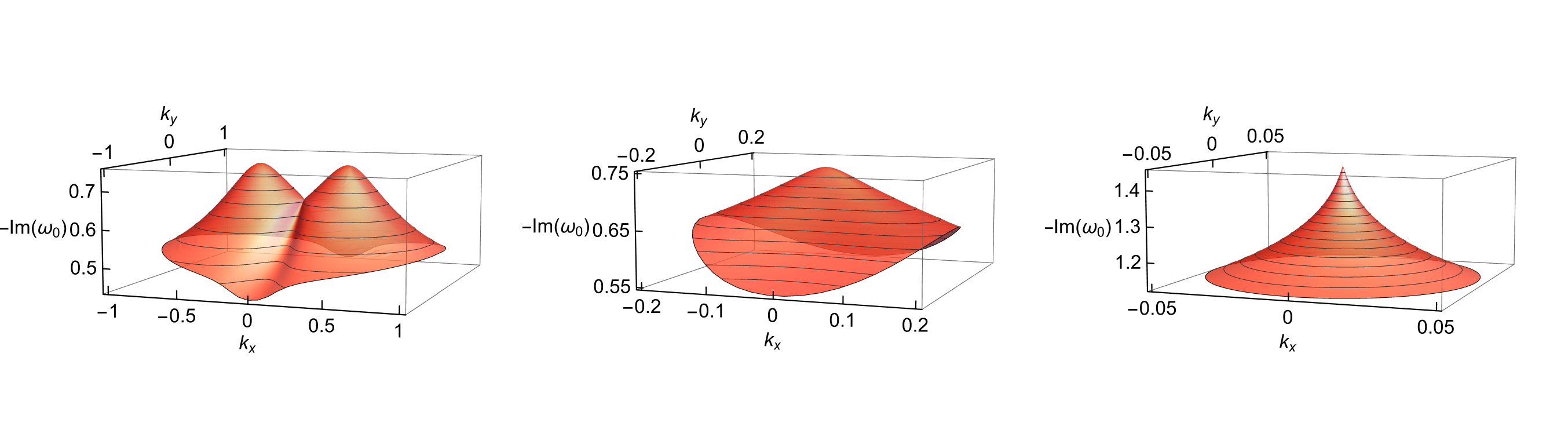}
    \end{subfigure}
    \caption{Dispersion relation at finite $\mathbf{k}$ for the lowest-lying quasi-normal frequency $\omega_0$, as calculated in sample points of the light-blue, pink, and yellow regions of the phase diagram in Figure \ref{fig:phase_diagram} (left, center, and right plots, respectively).}
    \label{fig:3D_plots}
\end{figure}

\begin{figure}[!htb]
    \centering
    \begin{subfigure}{\textwidth}
    \centering
 \includegraphics[width=\textwidth]{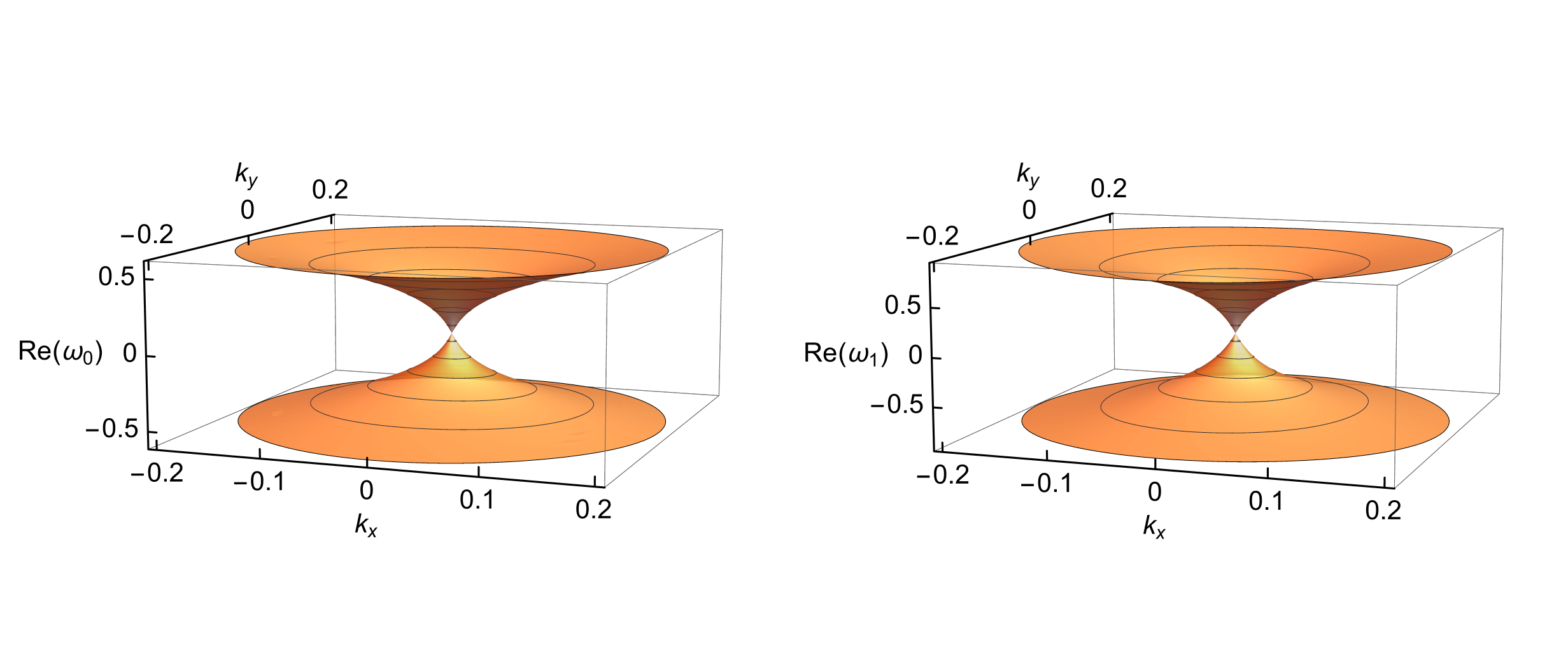}
    \vspace*{-1.6cm}
   \end{subfigure}
   \begin{subfigure}{\textwidth}
   \centering
       \includegraphics[width=\textwidth]{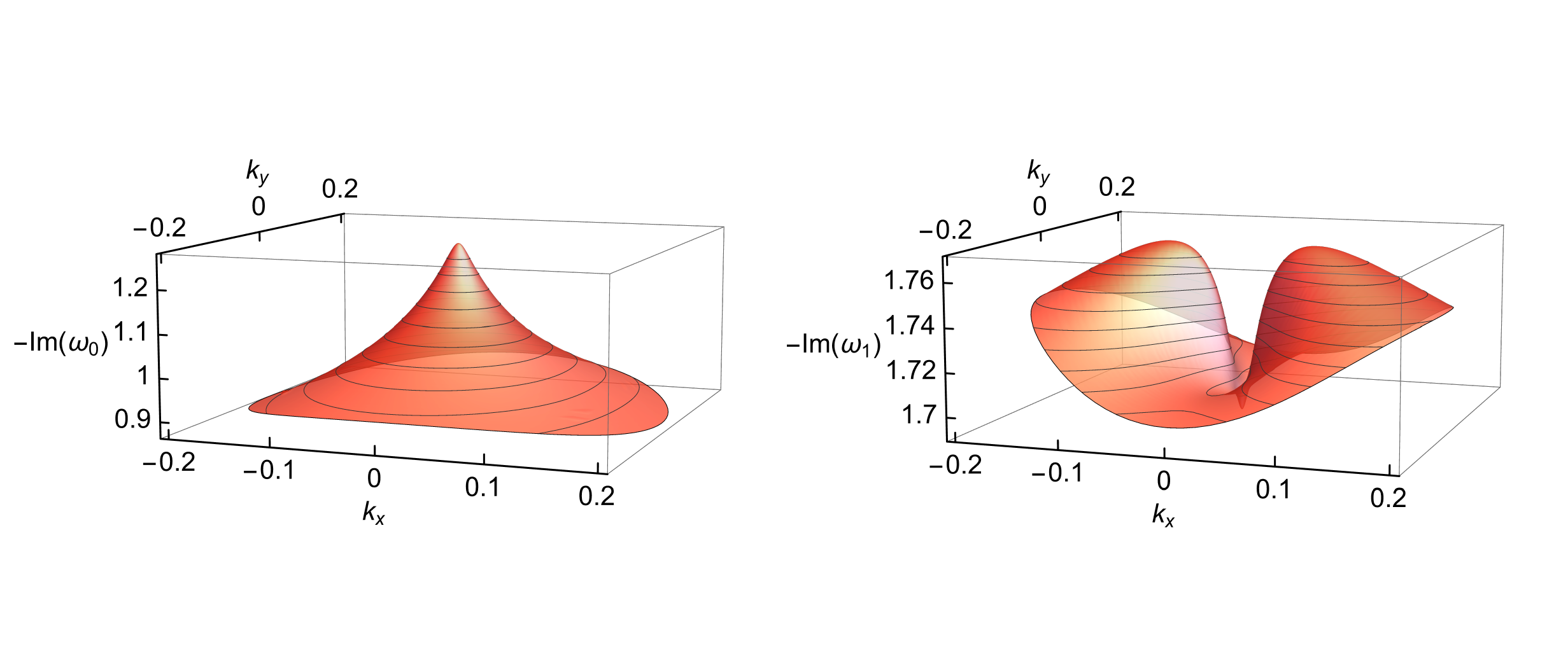}
   \end{subfigure}
     \caption{Dispersion relations of $\omega_0$ (left plots) and $\omega_1$ 
     (right plots) for $\Delta_1/T = 1$ and $\Delta_2/T \approx 0.909$.
     This corresponds to a sample point in the purple region of the $\Delta_1-\Delta_2$
     phase diagram (see Figure \ref{fig:phase_diagram}).}
    \label{fig:final_3d_plots}
\end{figure}

The regions in the plots of Figure \ref{fig:phase_diagram} are coloured distinct from each other because 
they correspond to the expected semimetallic, insulating, and semi-Dirac phases that we expected the boundary 
fermions to showcase in their spectrum; at least for the lowest mode $\omega_0$. This is shown in 
Figure \ref{fig:3D_plots}. The dispersion relation for $\omega_0$ at finite $\mathbf{k}$ corresponds to a double Dirac cone separated along the $k_x$ 
direction for fermions in the light-blue region of the phase diagram in \ref{fig:phase_diagram}. This 
characterizes this region as the semimetallic phase of the theory.
 Dispersion relations for those same modes 
in the yellow region of the phase diaram feature a band gap, which makes this region of $\Delta_1-\Delta_2$ space an insulating phase.
 Finally, points in the pink region of the phase diagram in Figure \ref{fig:phase_diagram} feature the expected 
 anisotropic semi-Dirac dispersion relation, with a vanishing Fermi velocity along the $k_x$ direction:
 \begin{equation}\label{eq:anisotropic_dispersion}
    \omega_0(k_x,0) = -i\alpha+(\pm\beta-i\gamma)k_x^2+\cdots\;,\;\omega_0(0,k_y)=-i\alpha\pm v_fk_y+\cdots,
 \end{equation}
 where the parameteres $\alpha,\,\beta,\,\gamma$ and $v_f$
 must be fitted to the numerical data. We also note that the $\omega_0$ dispersion relation remains anisotropic in the purple region; however as soon as the 
 first excited state $\omega_1$ becomes massless such anisotropy ceases to be semi-Dirac, and becomes linear along both 
 spatial directions, with different Fermi velocities (see Figure \ref{fig:final_3d_plots}).

The previous results indicate that when the gauge field dominates over the scalar field in the
deep IR of the theory, the system is driven towards its semimetallic phase (see, for instance, \cite{Grandi2022:PhysRevD} for similar conclusions), while the scalar
field drives the system towards an effective band gap when $a_0 \gg b_0$. This fact gives
confirmation to the fact that the Yukawa coupling between the fermions and scalar field
in action \eqref{eq:fermionic_bulk_action} gives the fermions mass, as was also the case
in the holographic Weyl fermions built by \cite{Plantz2018:JHEP}.

The fact that the critical anisotropic phases persist for a finite range of $\Delta_2/T$ 
can be attributed directly to finite temperature in the
boundary system. Indeed, the phase transition of the toy model was a QPT with a definite
QCP for the order parameter $\Delta_2/\Delta_1$ because it was a theory at $T=0$, unlike
this probe-limit bulk theory. We can interpret that thermal fluctuations in the boundary
canonical ensemble dominate over quantum fluctuations of the underlying Hamiltonian, which
results in the spreading-out of the $T=0$ QCP into a finite critical region; we expect this 
region to be a quantum critical region as defined in \cite{Sachdev:BookQPTs,Sondhi1997:RevModPhys}. The turning of a QPT 
into a thermal phase transition is expected from the theory of quantum 
criticality \cite{Sondhi1997:RevModPhys}, and has also been seen 
in holographic CMT models, like holographic 
topological semimetals \cite{Landsteiner2019:SciChinaPhysMechAstron}. The fact 
that the anisotropic phases of the phase diagram in Figure \ref{fig:phase_diagram} 
constitue a Quantum Critical Region 
is a bold claim. Such a region is defined as a phase of the dual system 
where scaling of observables and correlation functions scale non-trivialy 
with respect to temperature, and whose associated critical exponents 
are determined by the QCP at $T=0$ when $T$ 
is sufficiently low \cite{Frerot2019:NatCommun}.
 To confirm if the anisotropic phases in the phase diagram are, indeed,
 a Quantum Critical Region that comes from a QCP at $T=0$ we need 
 solutions to the background EOMs at $T=0$. We do this in section
 \ref{sec:zero_temperature_results}. Furthermore, we need an observable whose 
 scaling behavior with respect to temperature we can measure in the $T\to0$ limit, to 
 determine if its critical exponent with respect to temperature in the anisotropic 
 phase is related to the $T=0$ critical poit. This observable 
 will be the shear viscosity-entropy density ratio, and we measure it in subsection \ref{subsec:viscosity}.

\subsection{Backreacted background and shear viscosity}\label{subsec:viscosity}
In this subsection, we show the results of solving the background 
EOMs with backreaction ($q <\infty$), and the $\phi^4$ self-interacting
term in the bulk Lagrangian turned on. In all subsequent calculations 
in this chapter, we take $q=1$, $\lambda = 1$ and $\kappa^2=1$. First, we plug the \textit{anstaze}
for the background fields, eqs.~\eqref{eq:matter_fields_ansatz} and \eqref{eq:metric_ansatz}
into the background EOMs, and look for black brane solutions, whose dimensionless horizon
is at $r_h = 1$. To solve the resulting five ODEs, we use the four IR initial conditions of 
each background field as shooting parameters: $a_0$, $b_0$, $h_0$ and $N_0$ (see 
eqs.~\eqref{eq:horizon_scalar}-\eqref{eq:horizon_N}).
\begin{figure}[!htb]
    \centering
    \includegraphics[width=\linewidth]{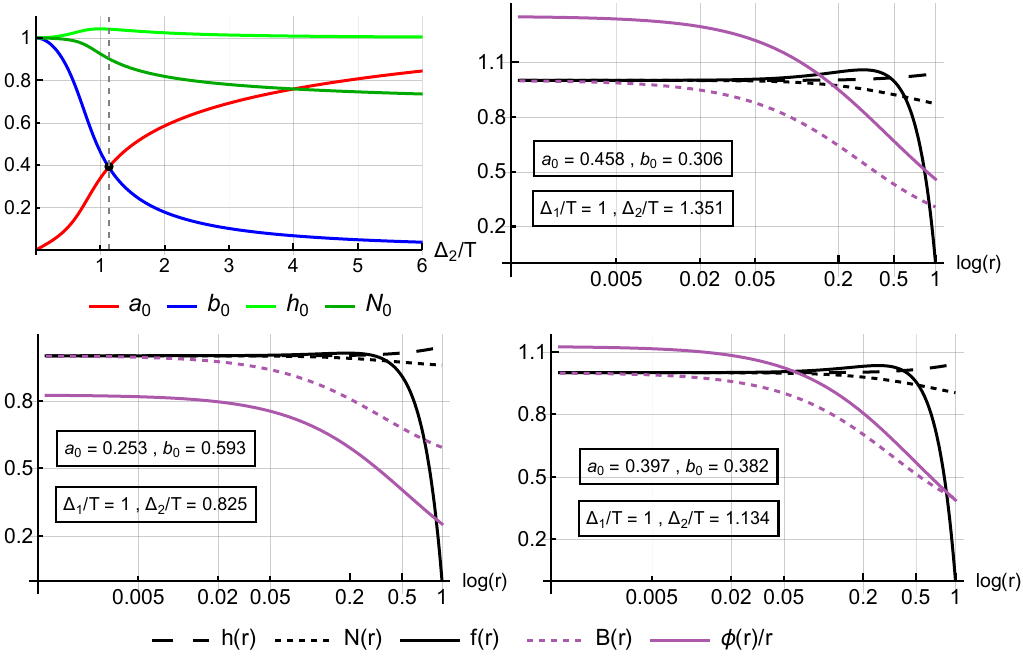}
    \caption{Top left plot: Shooting parameters for background fields
     as a function of $\Delta_2/T$ for fixed $\Delta_1/T = 1$.  The qualitative behavior 
     of the renormalized $\Delta_1/T$ and $\Delta_2/T$ values near the phase transition 
     is the same as that obtained in the probe limit. Also shown are the 
     numerical profiles of said background fields for a sample values of shooting 
     parameters, in logarithmic scale, for each phase of the model: insulating (top right
      plot), semi-Dirac (bottom right plot), and semimetallic (bottom left plot).}
    \label{fig:background_profiles_backreaction}
\end{figure}
We must impose five UV boundary 
conditions on the background fields (eqs.\eqref{eq:boundary_expansion_f}-\eqref{eq:boundary_expansion_N}).
Apart from the boundary sources $\Delta_{1,2}$ that act as boundary conditions for $\phi$
and $B$, now we also require asymptotically $\mathrm{AdS}_4$ geometry in the $r\to 0$ limit.
At first glance the system looks overdefined, since we only have four shooting parameters
in the IR to shoot towards five boundary conditions in the UV. Fortuntately the EOMs 
impose automatically the condition $f\to 1$ to leading order when expanding the system of
ODEs in series around the $r=0$ boundary, leaving only four boundary conditions to be fixed
by the shooting parameters. 

The numerically obtained behavior of the shooting parameters with respect to $\Delta_2/T$
for fixed $\Delta_1/T$ is displayed in Figure \ref{fig:background_profiles_backreaction},
 in the same fashion as in the previous subsection, where we include the metric field shooting
 parameters and profiles along the radial coordinate. Again, the point of crossing of $a_0$ and
 $b_0$ gives a naive indication of where the transition from the semimetallic phase towards the
 anisotropic one takes place.

  \begin{figure}[!htb]
    \centering
    \includegraphics[width=0.8\textwidth]{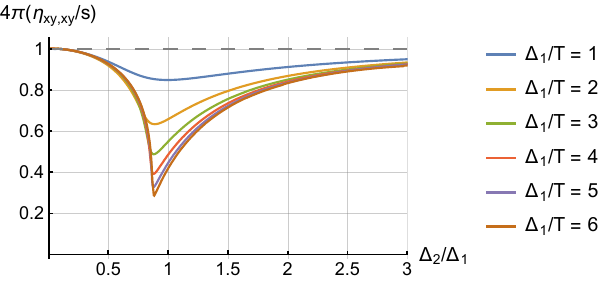}
    \caption{Plot of the $\eta/s$ ratio for the boundary theory as a function of $\Delta_1/\Delta_2$ 
    for a set of different fixed values of $\Delta_1/T$. It can be deduced that in the range 
    of values of $\Delta_2/\Delta_1$ where we expect to find the anisotropic phase of the
    dual theory there is a monotonous scaling of $\eta/s$ with respect to temperature. As was 
    expected, the value of $\eta/s$ is found to be below the KSS bound near the critical region,
    while for the $\Delta_2/T = 0$ and $\Delta_2/T\to\infty$ limits $\eta/s$ returns to the universal 
    value of $1/4\pi$.}
    \label{fig:viscosity_plot}
\end{figure}
 
 Having the background numerically solved in the backreacted regime allows for the implementation
 of linear response theory for the calculation of any transport coefficient asociated to 
 the dual operators. We are specifically interested in the shear viscosity-entropy density 
 ratio, $\eta/s$, since several holographic models of strongly coupled fluids predict that explicit breaking of
 $SO(2)$ symmetry along a given plane of the boundary theory results in violations of the 
 Kovtun-Sons-Starinets (KSS) bound \cite{Kovtun2004:PhysRevLett}:
 \begin{equation}\label{eq:KSS_bound}
    \frac{\eta}{s}\geq \frac{1}{4\pi}.
 \end{equation}
The aforementioned systems range from holographic Weyl semimetals 
\cite{Landsteiner2016Odd:PhysRevLett} to holographic anisotropic 
plasmas \cite{Rebhan2012:PhysRevLett,Critelli2014:PhysRevD}. The violation of \eqref{eq:KSS_bound}
is expected in strongly coupled anisotropic fluids in holography, since the original conjecture
of \eqref{eq:KSS_bound} as a universal lower bound for $\eta/s$ relied heavily on the use of 
the full Lorentz group for the calculation of the shear viscosity \cite{Kovtun2004:PhysRevLett}.

As is shown in Appendix \ref{app:hydrodynamics}, the shear viscosity in a holographic fluid 
can be calculated from the standard theory of hydrodynamics through Kubo's formula:
\begin{equation}\label{eq:Kubo_formula}
    \eta = -\frac{1}{\omega}\lim_{\omega\to0}\mathrm{Im}\left(G_{xy,xy}^R(\omega,\mathbf{k}=\mathbf{0})\right),
\end{equation}
where $G_{xy,xy}^R$ is the retarded Green's function of the shear entry of the boundary energy-
momentum tensor, $T_{xy}$, with respect to itself in the context
of linear response theory (i.e: the retarded correlator as defined in eq.~\eqref{eq:retarded_correlator}). Using equation \eqref{eq:retarded_correlator_GKPW}
with the on-shell action of our model,
we numerically calculate the shear viscosity through Kubo's formula as an implicit
function of $\Delta_1$ and $\Delta_2$ (see Appendix \ref{app:holographic_renormalization} for details). Numerical results are shown in Figure \ref{fig:viscosity_plot}.
\begin{figure}[!htb]
    \centering
    \includegraphics[width=0.7\linewidth]{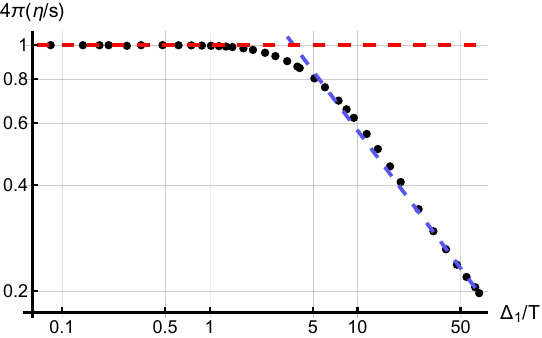}
    \caption{Log-log plot of $\eta/s$ as a function of $\Delta_1/T$ for $\Delta_2/\Delta_1 \approx 0.879$.
    In the limit of high temperature, the $\eta/s$ ratio behaves universaly, as thermal fluctuations
    supress the scaling with $T$ that takes place at the QCP. When $T\to 0$ a power-law
    scaling of the form $\eta/s \sim T^\nu$ appears. The numerical fitting of the data poitns
    results in a value of this scaling exponent of $\nu\approx 0.561$.}
    \label{fig:scaling_viscosity}
\end{figure}

We see from Figure \ref{fig:viscosity_plot} that the minimum of $\eta/s$ is achieved at a
specific critical value of $\Delta_2/\Delta_1$, where scaling of $\eta/s$ with 
temperature $T$ takes place. This is the first hint at the anisotropic region of the $\Delta_1-\Delta_2$ diagram 
being Quantum Critical. We locate this critical value of $\Delta_2/\Delta_1$ approximately
at $(\Delta_2/\Delta_1)_c\approx 0.879$. This value turns out to be the critical point 
in $\Delta_2/\Delta_1$ space where a semi-Dirac QPT takes place in the $T=0$ boundary theory,
as will be shown in section \ref{sec:zero_temperature_results}. When fixing $\Delta_2/\Delta_1 = (\Delta_2/\Delta_1)_c$
and lowering the value of $T/\Delta_1$, we find an interpolation between universal 
$\eta/s = 1/4\pi$ behavior at high temperatures and a monotone scaling $\eta/s\sim T^\nu$
at very low temperatures. We make a numerical fit to the log-log data $(\eta/s,\Delta_1/T)$
for a wide range of values of $\Delta_1/T$, and calculate $\nu\approx 0.561$, as shown 
in Figure \ref{fig:scaling_viscosity}. We will see how $\nu$ is related to the underlying
$T=0$ quantum critical bulk geometry in section \ref{sec:zero_temperature_results}. This will give 
further confirmation that the anisotropic region is quantum critical, since it will show that 
the physics at finite, yet low temperature are ruled by the characteristics of the QCP.

\section{Zero temperature}\label{sec:zero_temperature_results}

In this section we find explicit solutions to the background EOMs that do not feature a black brane 
in the deep IR; i.e: solutions with zero temperature for the background fields\footnote{In 
this whole section we use notation and concepts introduced in \cite{Bahamondes2025:arXiv}.
Of course, we give appropriate citation when called for.}. 
Depending on the phase of the dual theory we are in, the geometry of the bulk and 
profile of the matter fields will be different. This means that we will have to propose 
three different families of IR boundary conditions. Each IR solution we postulate will 
not necessarily be valid as a dual geometry for our theory, since they will not 
generically  fulfill the UV boundary conditions in 
eqs.~\eqref{eq:boundary_expansion_scalar}-\eqref{eq:boundary_expansion_N}. following
the procedure outlined in \cite{Bahamondes2025:arXiv}, we will impose
the UV boundary conditions required by AdS/CFT by modifying these IR solutions through 
irrelevant perturbations and using the associated free parameters to shoot towards the
boundary expansions in eqs.~\eqref{eq:boundary_expansion_scalar}-\eqref{eq:boundary_expansion_N}.
The resulting numerical profiles are domain walls that interpolate between the theory's UV and 
IR \cite{Landsteiner2016Odd:PhysRevLett,Grandi:2021bsp}.

\subsection{Insulating phase}\label{subsec:insulating_phase}

In this subsection we build domain wall solutions that dualize the zero temperature insulating
phase of the boundary theory. The deep IR geometry in this phase is an exact solution to 
the backgroud EOMs, and they correspond to constant geometry and matter fields, realizing an 
$\mathrm{AdS}_4$ geometry with a constant scalar field\footnote{We recover the variables
$\kappa$, $\lambda$, $L$ and $q$ only to make clear that the $T=0$ solutions depend
on the choice of such values. When running any remaining numerics, the same previous values
for these parameters are used.}:
\begin{equation}\label{eq:insulating_phase_zero_T}
    \phi_0(r) \equiv \frac{1}{L}\sqrt{\frac{-m^2L^2}{\lambda}}\;,\; f_0(r)\equiv 1+\frac{(m^2L^2)^2\kappa^2}{3L^2\lambda}\;,\; N_0(r)\equiv N_0\;,\;h_0(r)\equiv h_0\;,\;B_0(r)\equiv 0\;,
\end{equation}
where $h_0$ and $N_0$ are free, positive parameters. At this point we see the necesity of the
including the $\phi^4$ term in our bulk system, since the profiles in \eqref{eq:insulating_phase_zero_T}
would not be well defined if $\lambda = 0$. Again, this solution is only valid 
in the deep IR ($r\to\infty$) because it does not satisfy the appropriate asymptotically 
AdS boundary conditions in the UV ($r\to 0$). We remedy this by perturbing the fields in 
\eqref{eq:insulating_phase_zero_T} by the following irrelevant perturbations:
\begin{align}
\phi(r)&=\phi_0(r)+\delta\phi(r)\;,\;f(r)=f_0(r)+\delta f(r)\;,\;N(r)=N_0(r)+\delta N(r)\;,\nonumber\\
\;h(r)&=h_0(r)+\delta h(r)\;,\;B(r)=B_0(r)+\delta B(r)\label{eq:perturbed_insulating_phase_zero_T}.
\end{align}
Up to linear order the scalar and gauge EOMs decouple from the Einstein equations, 
as we show in \cite{Bahamondes2025:arXiv}, resulting in:
\begin{align}
    m^2L^2\delta\phi(r)-r\left(1+\frac{\kappa^2(m^2L^2)^2}{3L^2\lambda}\right)\delta\phi'(r)+\frac{r^2}{2}\left(1+\frac{\kappa^2(m^2L^2)^2}{3L^2\lambda}\right)\delta\phi''(r)&=0\label{eq:linearized_KG}\\
    \frac{8m^2L^2q^2}{\lambda}\delta B(r)+r^2\left(1+\frac{(m^2L^2)^2\kappa^2}{3L^2\lambda}\right)\delta B''(r)&=0\label{eq:linearized_YM}.
\end{align}
Equation \eqref{eq:linearized_KG} is the Klein-Gordon equation in pure $\mathrm{AdS}_4$, 
with an effective modified mass of 
$M^2L^2 \equiv \frac{m^2L^2}{1+\kappa^2(m^2L^2)^2/3L^2\lambda}$, and 
\eqref{eq:linearized_YM} is the EOM of a non-Abelian gauge field coupled to 
the same negative geometry and massive scalar. The solutions to these 
perturbation equations are \cite{Bahamondes2025:arXiv}:
\begin{align}
    \delta\phi(r)& =\phi_0r^{\Delta_-^{(s)}}+\phi_1r^{\Delta_+^{(s)}}\,,\;\Delta_{\pm}^{(s)} = \frac{3}{2}\pm\frac{1}{2}\sqrt{9-\frac{24L^2(m^2L^2)}{3L^2\lambda+(m^2L^2)^2\kappa^2}}\label{eq:perturbation_sol_insulating_scalar}\\
    \delta B(r) &=B_0r^{\Delta_-^{(g)}}+B_1r^{\Delta_+^{(g)}}\,,\;\Delta_{\pm}^{(g)}=\frac{1}{2}\pm\frac{1}{2}\sqrt{1-\frac{96L^2(m^2L^2)}{3L^2\lambda+(m^2L^2)^2\kappa^2}}\label{eq:perturbation_sol_insulating_gauge}.
\end{align}

 Finally, retaining only those solutions that vanish when $r\to \infty$ (which is 
 what makes the perturbations irrelevant) the full solutions that interpolate between the 
 UV $\mathrm{AdS}_4$ and the IR $\mathrm{AdS}_4$ are the following, as shown in \cite{Bahamondes2025:arXiv}:
 \begin{align}
     f(r) & =1+\frac{(m^2L^2)^2\kappa^2}{3L^2\lambda}+\cdots\label{eq:insulating_f_perturbed}\\
     N(r) &= N_0+\cdots\label{eq:insulating_N_perturbed}\\
     h(r)&=h_0+\cdots\label{eq:insulating_h_perturbed}\\
     \phi(r)&= \frac{1}{L}\sqrt{\frac
     {-m^2L^2}{\lambda}}+\phi_0r^{\Delta_-^{(s)}}+\cdots\label{eq:insulating_scalar_perturbed}\\
     B(r)&=B_0r^{\Delta_-^{(g)}}+\cdots\label{eq:insulating_gauge_perturbed},
 \end{align}
 with higher order corrections implicitly contained in the $\cdots$. We use $(h_0,N_0,\phi_0,B_0)$ 
 as shooting parameters to numerically solve for a set of fields that asymptote, to leading order, 
 to $h,N\xrightarrow[r\to 0]{} 1$ and $B\xrightarrow[r\to 0]{}\Delta_1$, 
 $\phi\xrightarrow[r\to 0]{}\Delta_2$.

\begin{figure}[!htb]
     \centering
     \begin{subfigure}{\textwidth}
     \centering
    \includegraphics[width=\textwidth]{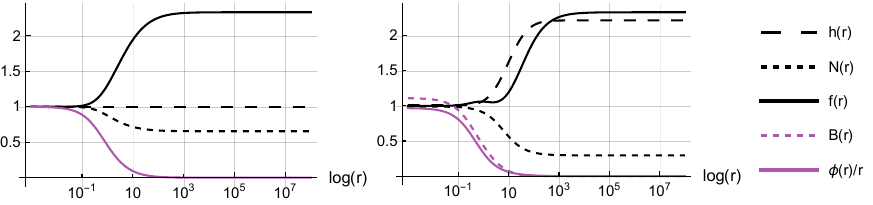}
     \end{subfigure}
     \begin{subfigure}{\textwidth}
     \centering
     \includegraphics[width=\textwidth]{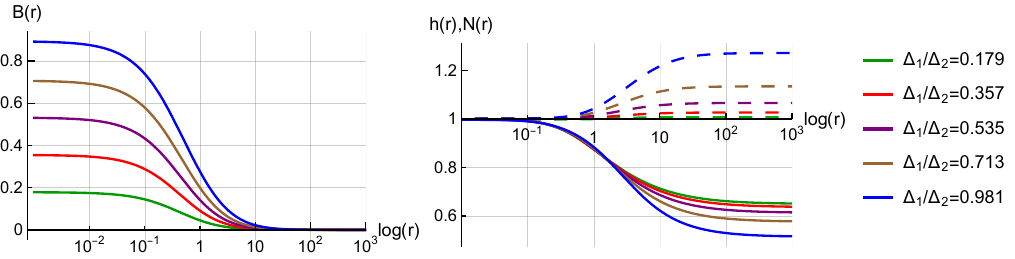}
     \end{subfigure}
     \caption{Top panel: Background fields at $T=0$ in the insulating phase of the 
     boundary system for a sample value of $\Delta_1 = 0$ (left plot) and 
     $\Delta_2/\Delta_1 = 0.879\ldots$ (right plot). When $\Delta_1=0$ the solution
     correponds to the deep insulating phase, where there is no gauge field competing
     with the non-trivial profile of $\phi$. Bottom panel: Profiles of the background 
     gauge field (left plot) for increasing values of $\Delta_1/\Delta_2$. The evident anisotropy that the gauge 
     induces can be seen from the appearance of a similar non trivial interpolation of 
     the $h(r)$ function (dashed line of right plot) between the IR and UV, 
     and $N(r)$ (continuous line of right plot).}
\label{fig:zero_T_fields_insulating_sample}
 \end{figure}

 Since this phase does not feature a black brane horizon, the temperature can not act a 
 reference scale for making the EOMs dimensionless, like we did when setting
 $r_h=1$ with the finite temperature solutions. This means that the boundary conditions 
 imposed on the matter fields $\phi$ and $B$ will be slightly different than those 
 for finite temperature. Near the UV boundary, the gauge and scalar fields still must 
 behave as was stated in eqs.~\eqref{eq:matter_fields_ansatz}. We rescale the radial coordinate
 by $\Delta_2$ as: $r\to \Delta_2 r$, which translates into the following UV 
 boundary conditions for the appropriately rescaled, dimensionless, matter fields:
 \begin{align}
     B(r\to 0)=\Delta_1+B_{(s)}r+\cdots&\longmapsto B(r\to0)=\frac{\Delta_1}{\Delta_2}+\frac{B_{(s)}}{\Delta_1^2}r+\cdots\label{eq:rescaled_gauge_insulating_UV}\\
     \phi(r\to 0)=\Delta_2r+\phi_{(s)}r^2+\cdots &\longmapsto \phi(r\to 0) = r+\frac{\phi_{(s)}}{\Delta_2^2}r^2+\cdots\label{eq:rescaled_scalar_insulating}.
 \end{align}
 The boundary conditions \eqref{eq:rescaled_gauge_insulating_UV} 
 and \eqref{eq:rescaled_scalar_insulating} are the solutions that we shoot towards from 
 the IR using the shooting parameters described above, choosing different values of the 
 ratio $\Delta_1/\Delta_2$ from $0$ up to a critical value 
 $(\Delta_1/\Delta_2)_c=1.137\ldots$, above which the background solution in 
 eqs.~\eqref{eq:insulating_f_perturbed}-\eqref{eq:insulating_gauge_perturbed}
 is no longer valid for the numerical method used for solving the EOMs. We note that, to avoid confusion, from the next subsection on, we describe all 
 solutions in terms of the ratio $\Delta_2/\Delta_1$ instead of $\Delta_1/\Delta_2$ (that means 
 the critical value for which insulating solutions cease to exist is 
 $(\Delta_2/\Delta_1)_c=0.879\ldots$). The field profiles obtained by this method are 
 shown in Figure \ref{fig:zero_T_fields_insulating_sample}. As it will be seen in 
 subsection \ref{subsec:Lifshitz_phase}, when approaching the critical point 
 from the inside this phase (i.e: for decreasing values of $\Delta_2/\Delta_1$ 
 above $(\Delta_2/\Delta_1)_c$) the shooting parameter $B_0$ diverges close to the 
 transition. The critical value $(\Delta_2/\Delta_1)_c$ corresponds to the quantum 
 critical point, below which the system should enter the semimetallic phase.

\subsection{Semimetallic phase}
Now we present the domain wall solutions corresponding to the dual 
geometries to the semimetallic phase. The IR geometry is, again, an exact $\mathrm{AdS}_4$ 
background, now with a constant gauge field on top of it:
\begin{equation}\label{eq:semimetalic_phase_zero_T}
    \phi_0(r)\equiv 0\;,\;f_0(r)\equiv 1\;,\;h_0(r)=h_0\;,\;N_0(r)\equiv N_0\;,\;B_0(r)\equiv B_0
\end{equation}
Perturbation of the background \eqref{eq:semimetalic_phase_zero_T} results in an exponential 
series expansion \cite{Bahamondes2025:arXiv}. Again, retaining only the IR regular solutions to the 
perturbations of the background EOMs, the full $T=0$ solution of the semimetallic region is given by 
the fields shown in \cite{Bahamondes2025:arXiv}:
\begin{align}
    f(r) &=1+\frac{2\kappa^2(h_0-2B_0r)\phi_1}{h_0}r^2e^{-\frac{4B_0r}{h_0}}+\cdots\label{eq:semimetalic_zeroT_f}\\
    h(r) &=h_0-\frac{\kappa^2h_0(8B_0^2r^2+4B_0h_0r+h_0^2)\phi_1^2}{16B_0^2}e^{-\frac{4B_0r}{h_0}}+\cdots\label{eq:semimetalic_zeroT_h}\\
    N(r)&=N_0+\frac{\kappa^2N_0(32B_0^2r^3-8B_0^2h_0r^2+4B_0h_0^2r+h_0^3)}{16B_0^2h_0}e^{-\frac{4B_0r}{h_0}}+\cdots\label{eq:semimetalic_zeroT_N}\\
    \phi(r)&=\phi_0re^{-\frac{4B_0r}{h_0}}+\cdots\label{eq:semimetalic_zeroT_scalar}\\
    B(r)&= B_0+\frac{h_0^2\phi_1^2}{2B_0}e^{-\frac{4B_0r}{h_0}}+\cdots\,.\label{eq:semimetallic_zeroT_gauge}
\end{align}
In this case, the integration constants $(h_0,N_0,\phi_0,B_0)$ are taken as shooting parameters to 
numerically solve the EOMs with the appropriate boundary conditions. By scaling the $r$-coordinate by $r\mapsto \Delta_1 r$ for numerical convenience, the 
following re-scaling of the UV boundary conditions of the matter fields is used for the
background fields in this $T=0$ solution:
\begin{align}
     B(r\to 0)=\Delta_1+B_{(s)}r+\cdots&\longmapsto B(r\to0)=1+\frac{B_{(s)}}{\Delta_1^2}r+\cdots\label{eq:rescaled_gauge_semimetallic_UV}\\
     \phi(r\to 0)=\Delta_2r+\phi_{(s)}r^2+\cdots &\longmapsto \phi(r\to 0) = \frac{\Delta_2}{\Delta_1}r+\frac{\phi_{(s)}}{\Delta_1^2}r^2+\cdots\label{eq:rescaled_semimetalic_insulating}.
 \end{align}

\begin{figure}[!htb]
      \centering
      \includegraphics[width=\textwidth]{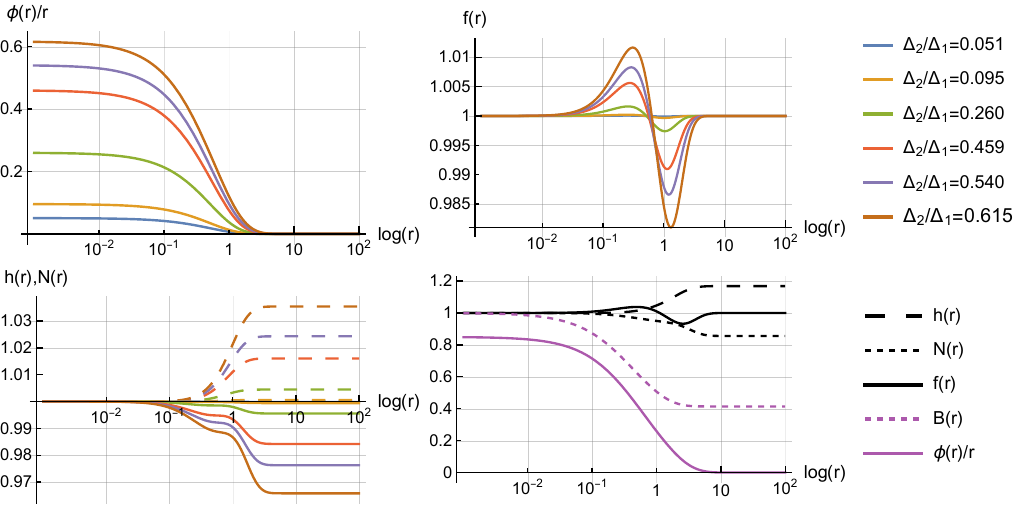}
      \caption{Top panel and left plot of bottom panel: Background fields at $T=0$ in the semimetallic phase of boundary theory, 
      for increasing values of $\Delta_2/\Delta_1$ below the critical point $(\Delta_2/\Delta_1)_c=0.879\ldots$ that separates it from the insulating phase. 
      The dashed lines of the bottom left plot are the $h(r)$ function, and the continous 
      line is $N(r)$. Bottom right plot: Background fields at $T=0$ for $(\Delta_2/\Delta_1)_c = 0.879\ldots$ as numerically obtained 
      from the shooting procedure towards the boundary conditions \eqref{eq:rescaled_gauge_semimetallic_UV}-\eqref{eq:rescaled_gauge_semimetallic_UV}.}
    \label{fig:zero_T_fields_semimetallic_sample}
\end{figure}

 We solve the background EOMs for increasing 
 values of $\Delta_2/\Delta_1$, from $\Delta_2/\Delta_1=0$ up until a critical value 
 $(\Delta_2/\Delta_1)_c$ for which solutions for the EOMs with the IR boundary conditions of 
 eqs.~\eqref{eq:semimetalic_zeroT_f}-\eqref{eq:semimetallic_zeroT_gauge} cease to exist. This
 value is precisely $(\Delta_2/\Delta_1)_c\approx 0.879$; the same critical value
 obtained in the insulating phase, and the same critical point in $\Delta_2/\Delta_1$ parameter space 
 for which there was monotone scaling $\eta/s\sim (T/\Delta_1)^\nu$ in subsection 
 \ref{subsec:viscosity}. The case of $\Delta_2/\Delta_1 = 0$ corresponds to the deep region of the semimetallic 
  phase of the theory, for which there is no scalar field, and the full geometry of the 
  bulk is just a trivial $\mathrm{AdS}_4$ spacetime (i.e: $B=f=h=N\equiv 1$ and 
  $\phi\equiv 0$). As $\Delta_2/\Delta_1$ is increased the scalar field acquires a 
  non-trivial profile in the $r$ coordinate (see Figure 
  \ref{fig:zero_T_fields_semimetallic_sample}).

\subsection{Lifshitz phase}\label{subsec:Lifshitz_phase}
\begin{figure}[!htb]
    \centering
    \includegraphics[width=\textwidth]{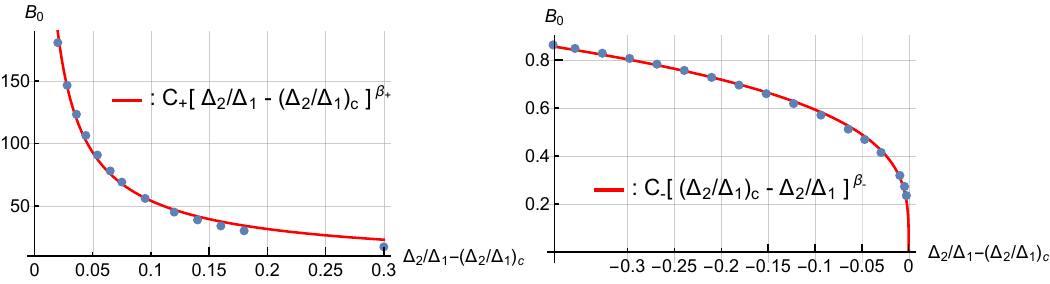}
    \caption{Critical behavior of the $B_0$ shooting 
    parameter in both the insulating (left plot) and 
    semimetallic (right plot) phases. In either side of the 
    critical point a non-linear fit was performed so as to 
    determine the critical exponent of the shooting 
    parameter in both phases.}
    \label{fig:phase_transition}
\end{figure}

The two previous phases meet at the critical point 
$(\Delta_2/\Delta_1)_c = 0.879\ldots$, where the anisotropic transition at zero 
temperature of the boundary theory takes place. We are confident in calling this change of 
bulk solutions a QPT because, as can be seen in Figure 
\ref{fig:phase_transition}, the shooting parameter associated to the gauge field 
$B(r)$ (i.e: the shooting parameter in the IR in either phase for the gauge field, as 
shown in eqs.~\eqref{eq:insulating_gauge_perturbed} and \eqref{eq:semimetallic_zeroT_gauge}) 
features a well-defined critical exponent $\beta_\pm$ for values of 
$\Delta_2/\Delta1$ that are very close to $(\Delta_2/\Delta_1)_c$. This means 
$B_0\sim \left|\frac{\Delta_2}{\Delta_1}-\left(\frac{\Delta_2}{\Delta_1}\right)_c\right|^{\beta_\pm}$, 
with $\beta_+$ corresponding to the insulating phase and $\beta_-$ to the semimetallic 
phase. The numerical values 
of both critical exponentes are  $\beta_+ = -0.776\ldots$ and $\beta_-=0.275\ldots$ \cite{Bahamondes2025:arXiv}.

\begin{figure}[!htb]
    \centering
    \includegraphics[width=\textwidth]{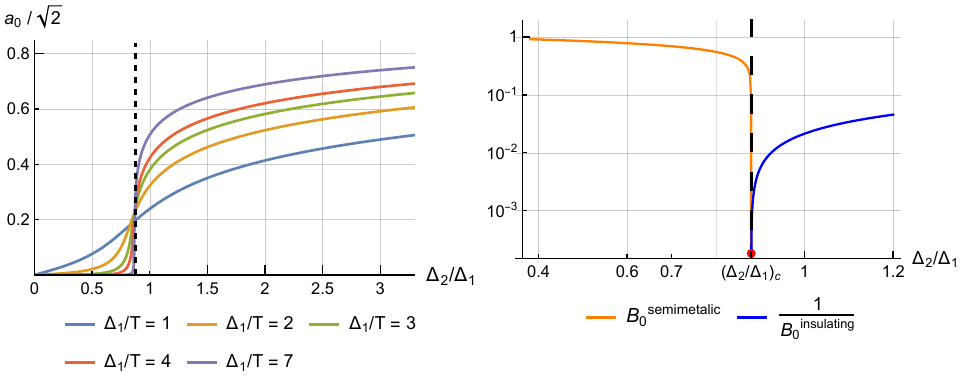}
    \caption{Left plot: Plot of the behavior of the finite 
    $T$ event horizon value of $\phi(r)$ for decreasing 
    temperature. The dashed line is located at the critical 
    point $(\Delta_2/\Delta_1)_c$, where it can be seen that 
    the behavior of $a_0$ tends towards a discontinous jump 
    from $a_0=0$ to a finite value, which approaches 
    $\sqrt2 = \sqrt{-m^2/\lambda}$ as $T/\Delta_1$ decreases. 
    Right plot: Behavior of the nonlinear fits (see Figure 
    \ref{fig:phase_transition}) for the gauge field shooting 
    parameters at $T=0$ as a function of 
    $\Delta_2/\Delta_1$. Both curves meet at the critical 
    point, where the QPT takes place (vertical axis is set 
    in logarithmic scale for better visualization).}
    \label{fig:lowering_temp_a0}
\end{figure}

We can also see that the numerical data for the matter fields that was obtained for the 
finite temperature case tend towards the $T=0$ solutions. To see this, remember that as 
the fixed value of $\Delta_1/T$ increases by lowering the temperature, the bulk black 
brane becomes colder, and its event horizon recedes ever deeper into the $r\to\infty$ region. That 
means that, in the $T\to 0$ limit, the value of $a_0$ (using the notation from section 
\ref{sec:finite_temperature}) should tend towards the deep IR value of $\phi(r)$ in both the semimetallic and 
insulating phases. As we can see from the $T=0$ solutions in \eqref{eq:insulating_phase_zero_T} and \eqref{eq:semimetalic_phase_zero_T}, 
the transition from the semimetallic towards the insulating phase results in a 
discontinous change in the value of $\phi(r\to\infty)$, from $\phi\equiv 0$ to 
$\phi\equiv (1/L)\sqrt{-m^2L^2/\lambda}=\sqrt{2}$. The change in $a_0$ as a function of 
$\Delta_2/\Delta_1$ as $T\to 0$ tends towards this discontinous transition, which 
is what is shown in Figure \ref{fig:lowering_temp_a0}. 

The geometry of the fields in the critical point $(\Delta_2/\Delta_1)_c$ is given by the 
following Lifshitz-type geometry in the deep IR, which corresponds to an exact solution 
to the background EOMs \cite{Bahamondes2025:arXiv}:
\begin{equation}\label{eq:critical_phase_zero_T}
    \phi_0(r)\equiv\phi_{0,c}\;,\;f_0(r)\equiv f_{0,c}\;,\;h_0(r)=\frac{r^{-\alpha}}{N_0\sqrt{f_{0,c}}}\;,\;N_0(r)=N_0r^{\alpha}\;,\;B_0(r)=\frac{B_{0,c}}{N_0}r^{-1-\alpha},
\end{equation}
where $\phi_{0,c}$, $f_{0,c}$, $B_{0,c}$ and $\alpha$ are solutions to a trascendental 
equation that solves the EOMs at zeroth order ($N_0$ is a free parameter). For 
$\kappa=\lambda=1$ the numerical values of these parameters are $(\phi_{0,c}\,,\,f_{0,c}\,,\,B_{0,c}\,,\,\alpha)\approx(0.455,0.919,0.698,-0.309)$. 
It can readily be seen when plugging this \textit{ansatz} into the metric in 
eq.~\eqref{eq:metric_ansatz} that the IR features the scaling symmetry 
$(r,t,x,y)\mapsto (\lambda^{\frac{1}{1-\alpha}} r,\lambda t,\lambda^{\frac{1+\alpha}{1-\alpha}} x,\lambda y)$. 
This justifies the classification of this phase as a Lifshitz-type phase, as defined in 
previous works like \cite{Grandi:2021bsp} or \cite{Grandi2022:PhysRevD}, since one of the spatial 
coordinates scales differently with respect to $t$ than the other. We define the 
dynamical-critical exponent $z\equiv \frac{1-\alpha}{1+\alpha}$, which means that if 
$t\mapsto \lambda t$ then $y\mapsto \lambda y$ and $x\mapsto \lambda^{1/z} x$.

 Using the numerical value of $\alpha$ the dynamical exponent is approximately 
 $z\approx 1.896$. This indicates that $t$ roughly scales quadratically with distance in 
 the $x$-direction, yet linearly in the $y$-direction. Furthermore, runing the numerics shows 
 that tuning the shooting parameter $N_0$ of the geometry fields so that 
 $h\xrightarrow[r\to 0]{} 1$ to leading order also makes 
 $N\xrightarrow[r\to 0]{} 1$ to leading order, which means there is no need to include 
 perturbations to the background shown in \eqref{eq:critical_phase_zero_T} to connect it 
 to the $\mathrm{AdS}$ UV. Numerically reading off the leading terms of $B(r)$ and 
 $\phi(r)$ results in $\Delta_2/\Delta_1 = (\Delta_2/\Delta_1)_c$, confirming that 
 the geometry in \eqref{eq:critical_phase_zero_T} actually corresponds to the critical 
 point that separates the two phases previously found to collide at 
 $(\Delta_2/\Delta_1)_c$.

 Finally, as was anticipated at the end of subsection \ref{subsec:viscosity}, the dynamical
 critical exponent $z$ is related to the scaling of $\eta/s$ with respect to $T$ in the low
 temperature limit. Indeed, solely from dimensional analysis (see, for instance, \cite{Hartnoll2016bumpy:JHEP,Landsteiner2016Odd:PhysRevLett,Ling2016:JHEP},
 for detailed calculations) it can be seen that, for a theory with the scaling symmetry 
 of the Lifshitz critical spacetime, we have $\eta/s\sim T^{1/z}$. Using te numerical value
 of $z$ outlined above, the numerical value of the power-law scaling exponent $\nu \approx 0.561$
 agrees with the analytical prediction within a 6.4\% margin of error \cite{Bahamondes2025:arXiv}.

\begin{figure}[!htb]
    \centering
    \includegraphics[width=0.6\textwidth]{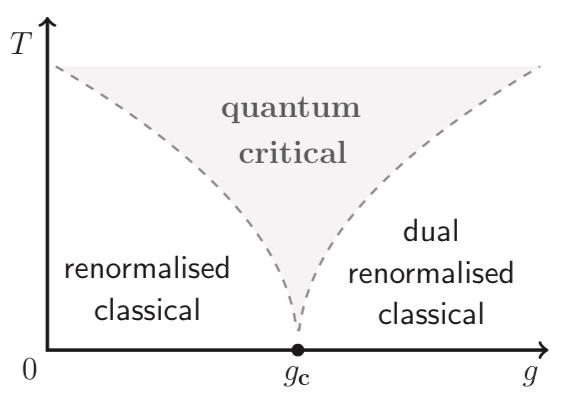}
    \caption{Cartoon of a quantum critical $T-g_c$ phase diagram condensed matter system, where $g_c$ 
    plays the role of $\Delta_2/\Delta_1$ in this work's model. The quantum critical region is equivalent to the 
    anisotropic phases of the $\Delta_2-\Delta_1$ phase diagram in Figure \ref{fig:phase_diagram} at sufficiently low
    temperature. The dual renormalized classical regions are the semimetalic and insulating phases of the model. Image 
    obtained from \cite{Zaanen2015:BookChptr1}.}
    \label{fig:final_cartton}
\end{figure}

These results give confidence for identifying the anisotropic regions of the phase diagram in 
Figure \ref{fig:phase_diagram} as a Quantum Critical Region, since observables like 
$\eta/s$ scale monotonously with temperature with exponents determined by the physics at the $T=0$ QCP. The 
qualitative shape of the $T-\Delta_2/\Delta_1$ phase diagram of the model is sketched as in Figure \ref{fig:final_cartton}. 
This $T-\Delta_2/\Delta_1$ phase diagram is quite general for systems that feature a QPT when they are set at finite temperature (see 
\cite{Landsteiner2016Odd:PhysRevLett,Sachdev2023:book}). A QCP that sits at the $T\to0$ endpoint of a quantum critical region is 
usually called a quantum critical wedge \cite{Zaanen2015:BookChptr1}.

\begin{figure}[!htb]
    \centering
    \includegraphics[width=0.8\textwidth]{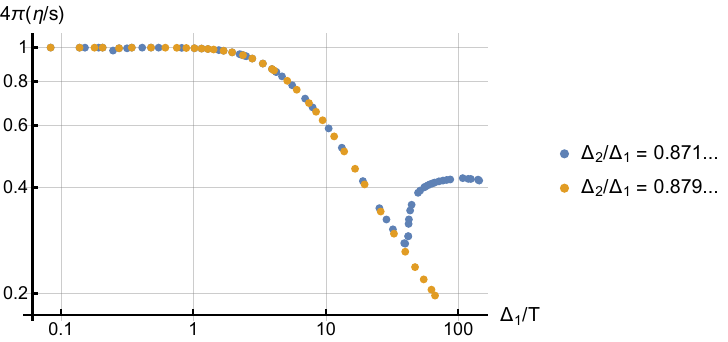}
    \caption{Log-log plot of the $\eta/s$ numerical data as a function of $\Delta_1/T$ for $\Delta_2/\Delta_1$ 
    at the critical value (same data as shown in Figure \ref{fig:scaling_viscosity}), and at a slightly lesser value. It 
    can be seen that for the slightly smaller value of the $\Delta_2/\Delta_1$ parameter, the monotnous scaling of $\eta/s$ 
    ceases at sufficiently low temperatures, giving further indication that the anisotropic phase is, indeed, a quantum critical region.}
    \label{fig:final_victory}
\end{figure}
If the anisotropic phase of our model is a quantum critical region that stems from a quantum critical wedge, this 
should be evident when lowering the system's temperature when $\Delta_2/\Delta_1$ is slightly driven away from 
$(\Delta_2/\Delta_1)_c$. In such a case, lowering the temperature should result in two different behaviors 
for the $\eta/s$ ratio. At very high temperatures we should see $\eta/s = 1/4\pi$, just as it was 
seen for $\Delta_2/\Delta_1=(\Delta_2/\Delta_1)_c$ in Figure \ref{fig:viscosity_plot}. Then the system should enter the quantum 
critical region, and so we should see $\eta/s\sim (T/\Delta_1)^{1/z}$. Finally, the system would reach a critical temperature 
where it leaves the critical phase and enters either the insulating or semimetallic phases. This would imply that the scaling of 
$\eta/s$ with temperature disappears, returning to $\eta/s\sim T^0$ behavior. This is precisely what is seen in Figure \ref{fig:final_victory}, which 
lends further confidence to the statement that the anisotropic region is, in fact, a Quantum Critical Region.
\chapter{Conclusions}\label{chap:conclusions}
In this work we built a bottom-up holographic model for a strongly coupled, thermal QFT in $2+1$-dimensions, 
that features both thermal and quantum phase transitions from a semimetalic towards an insulating phase. The critical 
phase/point turned out to feature anisotropic semi-Dirac nature, which was imprinted both in 
the dispersion relation of quasi-normal fermionic modes and the $T=0$ critical geometry through a non-trivial 
dynamical critical exponent $z$. These results serve to contribute yet another condensed matter model at strong coupling 
whose physics are capable of being probed through the AdS/CFT holographic correspondence. In essence, this construction 
relied on introducing relevant deformations to a scalar and gauge fields, which induces a non-trivial RG flow 
from a relativistic fixed UV towards a non-trivial IR, where all the physics of the dual theory are embeded in the 
dynamics of the matter content and geometry of the spacetime. This work also shows that semi-Dirac anisotropy is, in the context 
of AdS/CFT, a strongly emergent feature of the IR physics of QFTs that are covered by the universality class of theories this 
bottom-up construction models. As such, it is not unique to a specific underlying Hamiltonian of a solid-state system (like 
the toy model in section \ref{sec:toy_model} would suggest at first glace). This has been also confirmed in other works, 
where explicit semi-Dirac anisotropy is detected using standard QFT methods in weakly coupled theories (see \cite{Link:2017ora}); 
the fact that this feature has been reproduced in the context of AdS/CFT is a further sanity check 
to the validity of the conjecture itself.

 Another sanity check that is obtained from this work for AdS/CFT regards non-universality of the $\eta/s$ ratio. Indeed, we managed to 
 show that the quantum critical point at $T=0$ induces a monotonous scaling of $\eta/s$ with respect to temperature in the $T\to0$ limit, as 
 was shown to be the case in other holographic setups (see \cite{Landsteiner2016Odd:PhysRevLett}). This result serves to give 
 further evidence for the fact that rotational invariance is key for saturating the KSS bound. Eventhough this is not an equivalence, since 
 some particular holographic models that feature anisotropy have been shown to preserve the KSS bound \cite{Baggioli2023:JHEP}, it neverteless 
 gives a trustworthy recipee for creating a whole variety of condensed matter models that feature low viscosity. For the case of 
 semi-Dirac metals, the recent discovery of semi-Dirac excitations in experimental settings in works like \cite{Yinming2024:PhysRevX} makes the 
 results of this work all the more relevant for probing the qualitative/thermodynamical properties of these 
 types of materials, and gives a starting ground for searching for more physically relevant properties. Also, numerical evidence was given 
 for characterizing the anisotropic region of the boundary theory as a Quantum Critical Region, where the finite, low-$T$ physics are ruled by the 
 quantum critical point at $T=0$, specifically regarding the scaling of observables with temperature. Further research into this region should be 
 developed in order to determine with certainty that it is a Quantum Critical Region. The results obtained in this work give, nevertheless, strong evidence 
 to support such hypothesis.

 Potential future research on the line of results shown in this work lies mostly in the context of AdS/CMT. For example, the 
 question of spontaneous symmetry breaking in this system remains, to the author's knowledge, unexplored. By turning on a $U(1)$ 
 gauge field in the bulk, we could probe the boundary theory at finite density through a chemical potential. By adding more bulk fields 
 that dualize more matter content in the boundary, spontaneous symmetry breaking like the kind seen in holographic superconductivity \cite{Hartnoll2008:PhysRevLett,Giordano2017:JHEP} 
 and holographic flat bands \cite{Grandi:2021bsp,Grandi2024:JHEP} can be studied. This would allow to potentially enrich the phase diagram of the model 
 by studying the possibility of coexisting or competing phases in the semimetalic, insulating and anisotropic regions found in the main body of this work.

 \subsection*{Acknowledgements}
This work was supported by  Fondecyt (Chile) Grant No. 1241033 (R.S.-G. and S.B.). I.S.L. thanks ICTP and Universidad Católica de Chile for hospitality during early stages of this project. I.S.L would like to acknowledge support from the ICTP through the Associates Programme (2023-
2028). 

I would like to thank my friends and colleagues who supported me both academically and emotionally throughout my master's degree, who include 
but are not limited to: Nicolás Cáceres, Mayrim Busniego, Rogelio Albarracín, Sol Covacic, Ignacio Rojas, Melanie Martínez, Juan Pablo Esparza, 
Francisco Jara, Loreto Osorio, Sofia Nash, Javiera Mellado, Sebastián Salvadores, and Denisse Ortiz. A most special thanks to my 
parents, who have supported me continually and consistently throught my academic life, and to my advisor Rodrigo Soto-Garrido, for his 
invaluable guidance throughout the seven years I have known him.
\chapter*{Appendix}\label{chap:appendix}
\appendix

{\let\clearpage\relax \chapter{AdS/CFT at finite temperature}\label{app:AdS/CFT_finite_temp}}
In this appendix we explain how to dualize QFTs with temperature holographically. The formalism that was outlined in 
section \ref{sec:elements_from_adscft} for the calculation of correlation functions in the boundary from the dynamics of 
fields in the bulk was outlined in the context of regular QFT at zero temperature. The starting point of AdS/CFT, the GKPW rule,
can be extended to include theories with temperature; in such a case the GKPW rule in the large-$N$, strongly coupled limit, of AdS/CFT 
is stated as:
\begin{equation}\label{eq:thermal_GKPW}
    e^{iS_\mathrm{b}\left[\left\{\Xi^{\mathrm{sol}}_\alpha\right\}_{\alpha\in I}\right]}=\mathcal{Z}\left[\left\{J_\alpha = \Xi^{\mathrm{sol}}_{\alpha,(l)}(r\to0)\right\}_{\alpha\in I}\right],
\end{equation}
where $\mathcal{Z}$ corresponds to the thermal generating functional of correlation functions, defined as the thermal average of the exponential of currents:
\begin{equation}
    \mathcal{Z}[\left\{J_\alpha\right\}_{\alpha\in I}] = \mathcal{Z}_0\left\langle \exp{\left[-\int\!\mathrm{d}\tau\mathrm{d}^d\mathbf{x}\,J_\alpha(\tau,\mathbf{x})\phi(\tau,\mathbf{x})\right]}\right\rangle,
\end{equation}
where $\langle\rangle$ is the thermal average taken with respect to the partition function $\mathcal{Z}_0$. Notice that the boundary QFT is considered to be in the 
Euclidean time formalism, and as such the bulk must also have Euclidean signature when making sense of eq.~\eqref{eq:thermal_GKPW}. This means, most importantly, that the time dimension 
is compactified in a circle of period $\beta = 1/T$. This means that the background geometry of the boundary changes from being Minkowski $\mathbb{R}^{d,1}$ to $\mathbb{R}^d\times\mathbb{S}^1$.

In order to dualize a thermal QFT in the bulk we need a clear way to embed the compactification of the time dimension in the bulk metric. Clearly the pure $\mathrm{AdS}_{d+2}$ 
geometry is not enough, so the geometry of an $\mathrm{AdS}_{d+2}$-black brane with a generic emblackening factor is proposed. Throughout this chapter the choice of 
Poincaré patch coordinates shown in the metric \eqref{eq:pure_AdS_metric} will be used. As is outlined in \cite{Zaanen2015:BookChptr6}, such a metric is of the form:
\begin{equation}\label{eq:generic_euclidean_black_brane}
    \mathrm{d}s_E^2=g_{t\,t}\mathrm{d}\tau^2+\delta^{jk}g_{jk}\mathrm{d}x^j\mathrm{d}x^k+\frac{1}{g^{r\,r}}\mathrm{d}r^2.
\end{equation}
For the metric of this work's model, the metric components are $g_{tt}=\frac{r^2}{L^2}f(r)N(r)^2$, $g^{rr}=\frac{r^2}{L^2}f(r)$, $g_{xx} = h(r)^2$ and $g_{yy}=1/h(r)^2$. In 
general, the metric components are assumed to depend only on $r$, and not on the boundary coordinates. We 
will now argue why this geometry is appropriate for dualizing the finite temperature on the boundary. The defining feature 
of a black brane is the presence of a horizon at some finite value of the bulk radial coordinate; i.e: $g_{tt}(r=r_h)=g_{rr}(r=r_h)=0$. 
That means that, near the horizon, the following series expansion around $r_h$ can be taken:
\begin{align}
    g_{tt}(r)&=\left.\frac{\mathrm{d}g_{tt}}{\mathrm{d}r}\right|_{r=r_0}(r-r_0)+\frac{1}{2!}\left.\frac{\mathrm{d}^2g_{tt}}{\mathrm{d}r^2}\right|_{r=r_0}(r-r_0)^2+\cdots\nonumber\\
    g^{rr}(r)&=\left.\frac{\mathrm{d}g^{rr}}{\mathrm{d}r}\right|_{r=r_0}(r-r_0)+\frac{1}{2!}\left.\frac{\mathrm{d}^2g^{rr}}{\mathrm{d}r^2}\right|_{r=r_0}(r-r_0)^2+\cdots.\label{eq:emblackening_series_expansion}
\end{align}
Plugging this expansion into the line element \eqref{eq:generic_euclidean_black_brane} and series expanding around $r_h$ again, 
the following expression for the metric near the event horizon results \cite{Zaanen2015:BookChptr6}:
\begin{equation}\label{eq:euclidean_black_brane_expansion}
    \mathrm{d}s_E^2=\left.\frac{\mathrm{d}g_{tt}}{\mathrm{d}r}\right|_{r=r_0}(r-r_0)\mathrm{d}\tau^2+\frac{\mathrm{d}r^2}{\left.\frac{\mathrm{d}g^{rr}}{\mathrm{d}r}\right|_{r=r_0}(r-r_0)}+\delta^{jk}g_{jk}(r_0)\mathrm{d}x^j\mathrm{d}x^k+\cdots,
\end{equation}
where we have assumed that $g_{jk}(r_0)\neq 0$, and put all order two, or higher, terms of $r-r_h$ in the $\cdots$. Thus the $r$-dependence of 
the expansion \eqref{eq:euclidean_black_brane_expansion} is contained only in the $r-r_h$ terms in the $\tau$ and $r$ coordinates. We further assume 
that the event horizon is a single zero of the black brane, and therefore $\left.\frac{\mathrm{d}g_{tt}}{\mathrm{d}r}\right|_{r=r_h},\left.\frac{\mathrm{d}g_{rr}}{\mathrm{d}r}\right|_{r=r_h}\neq 0$\footnote{Explicit 
black brane/hole solutions that are asymptotically $\mathrm{AdS}_{d+2}$ have a double zero at the event horizon, such as the case of the extremal 
Reissner-Nördstromm solution. In the context of this work, such extreme cases will not be of interest.}. As outlined in \cite{Zaanen2015:BookChptr6}, now we make a change of coordinates given 
by $\rho = 2\sqrt{r-r_0}/\sqrt{\left.\frac{\mathrm{d}g^{rr}}{\mathrm{d}r}\right|_{r=r_0}}$, which turns the metric into:
\begin{equation}
    \mathrm{d}s^2_E = \frac{1}{4}\left(\left.\frac{\mathrm{d}g^{rr}}{\mathrm{d}r}\right|_{r=r_0}\cdot\left.\frac{\mathrm{d}g_{tt}}{\mathrm{d}r}\right|_{r=r_0}\right)\rho^2\mathrm{d}\tau^2+\mathrm{d}\rho^2+\cdots,
\end{equation}
where the boundary coordinate terms of $\mathrm{d}s_E^2$ have been included in the irrelevant terms contained in $\cdots$. Calling the constant 
prefactor of the Euclidean time coordinate $C_{r_h}^2 :=\frac{1}{4}\left(\left.\frac{\mathrm{d}g^{rr}}{\mathrm{d}r}\right|_{r=r_0}\cdot\left.\frac{\mathrm{d}g_{tt}}{\mathrm{d}r}\right|_{r=r_0}\right)$, 
we notice that we are left with the metric of a circle in $\mathbb{R}^2$ in polar coordinates: $\mathrm{d}s_E^2 = C_{r_h}^2\rho^2\mathrm{d}\tau^2+\mathrm{d}\rho^2+\cdots$, 
with $\tau$ acting as the angular variable. Now it is clear why the geometry \eqref{eq:generic_euclidean_black_brane} is appropriate to dualize the compactification of the Euclidean time dimension
in the bulk: to have a smooth metric in the near-horizon region of the bulk in Euclidean signature, the $\tau$ dimension must be periodic. The necessity for smoothness of spacetime in Euclidean signature thus 
enforces the periodicity of Euclidean time, while in the boundary theory the Wick rotation of the many-body propagator in the functional integral enforced periodicity. 
\begin{figure}[!htb]
    \centering
    \includegraphics[width=0.7\textwidth]{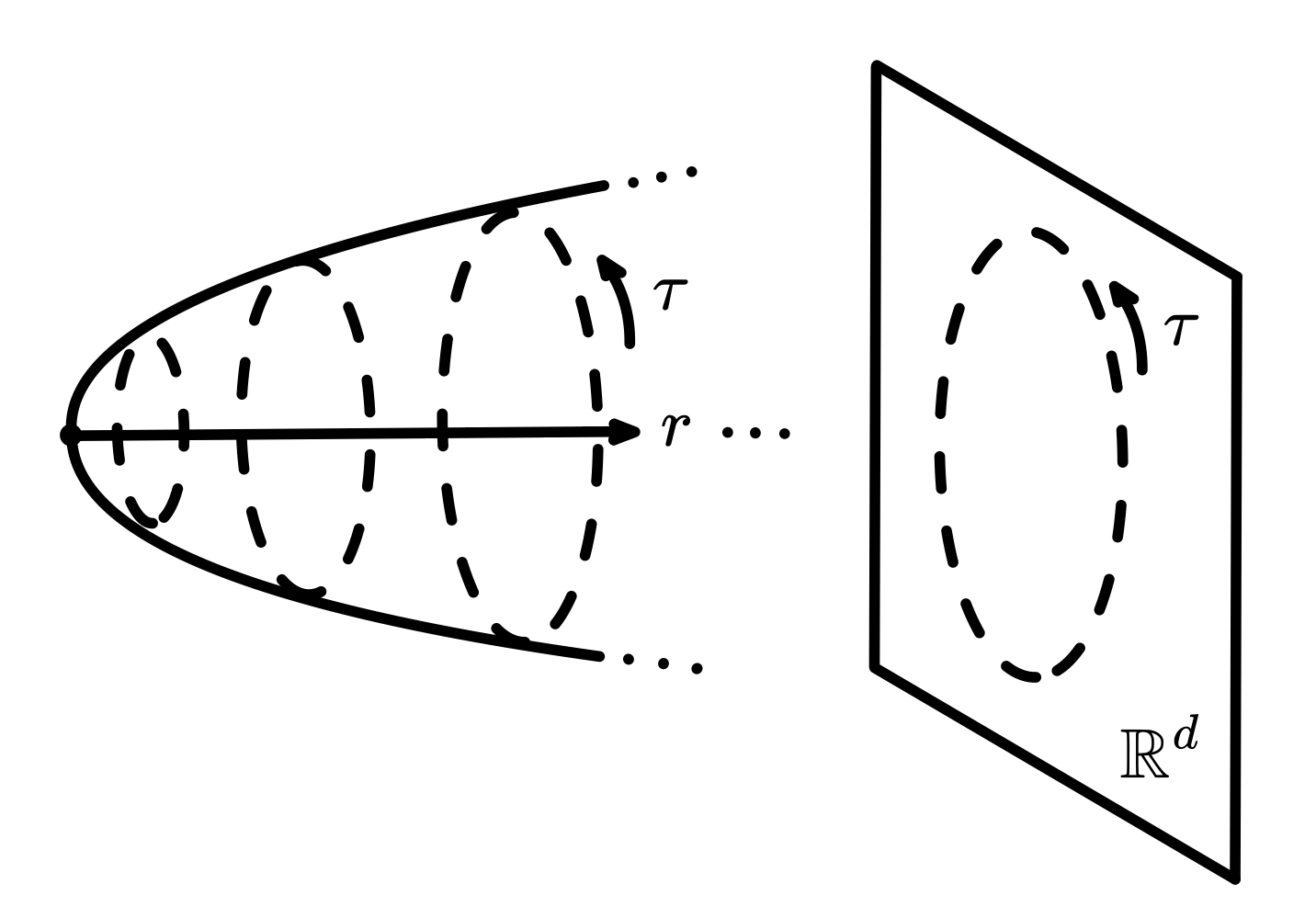}
    \caption{Cartoon of the matching between the Euclidean time period in the bulk and the boundary. In the conformal boundary of the 
    negatively curved spacetime, the geometry of the black brane can be thought of like a cigar-like shape, where the radius of the Euclidean 
    circle becomes infinitely large in the limit of $r\to\infty$ (the conformal boundary). Up to an $r$ factor, we "match" the Euclidean time circle 
    of the boundary to that in the bulk, and conclude that the periodicity of both must be the same.}
    \label{fig:cigar}
\end{figure}
The exact period of $\tau$ in the Euclidean bulk can be deduced from a final coordinate transformation: $\theta = C_{r_h}\tau$, giving the exact form of the $\mathbb{S}^1$ metric:
\begin{equation}\label{eq:circle_metric}
    \mathrm{d}s_E^2=\rho^2\mathrm{d}\theta^2+\mathrm{d}\rho^2+\cdots.
\end{equation}
The period of the $\theta$ coordinate is thus $2\pi$, which means that the period of the $\tau$ coordinate 
is $\frac{2\pi}{C_{rh}} = 4\pi/\sqrt{\frac{\mathrm{d}g_{tt}}{\mathrm{d}r}|_{r=r_h}\frac{\mathrm{d}g^{rr}}{\mathrm{d}r}|_{r=r_h}}$. By matching the period of the boundary 
Euclidean time coordinate to the period it has in the bulk, as is sketched in Figure \ref{fig:cigar}, we can conclude that the period of the bulk Euclidean coordinate must be $\beta = 1/T$. Therefore:
\begin{gather}\label{eq:Hawking_temp}
    \beta = \frac{4\pi}{\sqrt{\left.\frac{\mathrm{d}g_{tt}}{\mathrm{d}r}\right|_{r=r_h}\left.\frac{\mathrm{d}g^{rr}}{\mathrm{d}r}\right|_{r=r_h}}}\Longrightarrow T = \frac{1}{4\pi}\sqrt{\left.\frac{\mathrm{d}g_{tt}}{\mathrm{d}r}\right|_{r=r_h}\left.\frac{\mathrm{d}g^{rr}}{\mathrm{d}r}\right|_{r=r_h}}.
\end{gather}
The left-hand side of eq.~\eqref{eq:Hawking_temp} is the temperature of the boundary QFT, while the right-hand side is the expression for the black brane's 
Hawking temperature \cite{Hartnoll2009:ClassQuantGrav_lectures}. This establishes the holographic dictionary entry that states that 
a QFT at finite temperature is holographically dualized by a black hole/brane bulk geometry, where the brane/hole's Hawking temperature is equal 
to the QFT's temperature. 

 For the model of this work, we plug the metric coordinates of \eqref{eq:metric_ansatz} into \eqref{eq:Hawking_temp}, 
 which results in the temperature formula:
 \begin{equation}\label{eq:my_Hawking_temp}
    T = \frac{r_h^2}{4\pi L^2}N(r_h)\left.\frac{\mathrm{d}f(r)}{\mathrm{d}r}\right|_{r=r_h}.
 \end{equation}
 Eq.~\eqref{eq:Hawking_temp} implies that thermal effects are an essentially IR effect, since the QFT's temperature 
 is expressed exclusively in terms of quantities at the event horizon of the bulk, which is in the deep IR of the RG flow. Eq.~\eqref{eq:my_Hawking_temp} 
 is the formula used in the numerical and analyical calculations of all quantities in chapters \ref{chap:theoretical_background} and \ref{chap:results} that 
 involve the temperature $T$ of the boundary theory. In particular, when working in the probe limit where $f(r) = 1-(r_h/r)^3$ and $N\equiv 1$ 
 for this choice of Poincaré coordinates, the theory's temperature reduces to:
 \begin{equation}
    T = \frac{3}{4\pi L^2}r_h.
 \end{equation}
 When working in dimensionless coordinates by scaling the event horizon to $r_h=1$, the Hawking temperature reduces to $T = 3/4\pi$, as was 
 mentioned in section \ref{sec:fermions_coupled_to_background}. Also, notice from equation \eqref{eq:my_Hawking_temp} 
 that even in the backreacted setting, where explicit formulas for $N$ and $f$ are not known, the Hawking temperature scales 
 as $r_h$ (again, in the choice of Poincaré coordinates used in this chapter). Lowering the black brane temperature would means
 reducing the value of $r_h$, which means that the event horizon recedes ever deeper into the bulk when $T\to 0$ is taken, justifying the 
 interpretation of the value of $a_0$ (in the notation of chapter \ref{chap:results}) tending into the $T=0$ deep IR value of $\phi(r)$. This 
 is the contrary to what happens in flat spacetime, where black branes/holes become hotter as their horizon becomes smaller.

 One final detail worth mentioning is the mechanism of extracting real-time, Lorentzian signature, information in thermal AdS/CFT when all quantities are 
 defined in terms of Euclidean signature spacetime. As is the case in many-body quantum statistical mechanics, all observables in a many-body system at thermodynamic 
 equilibrium can be obtained from thermal, Euclidean signature $n$-point functions \cite{NegeleOrland:Book}:
 \begin{equation}\label{eq:thermal_n_point_function_def}
    \mathcal{G}(\tau_1\mathbf{x}_1\,,\ldots ,\,\tau_n\mathbf{x}_n):=\langle\widehat{\mathcal{O}}_1(\tau_1,\mathbf{x}_1)\cdots\widehat{\mathcal{O}}_n(\tau_n,\mathbf{x}_n)\rangle,
 \end{equation}
where Euclidean time-ordering is assumed in this definition. Eventhough these $n$-point functions can be perfectly calculated 
using the GKPW rule by taking functional derivatives of eq.~\eqref{eq:thermal_GKPW} in the same fashion as is done in the $T=0$ 
case with Minkowski signature spacetime, it is not enough for the purposes of real-time linear response theory (recall section \ref{sec:elements_from_adscft}). Indeed, physical observables 
are measured in real time, and transport coefficients such as the shear viscosity only make sense in Lorentzian signature where causality is manifest. This requires 
calculating the real time $n$-point functions of eq.~\eqref{eq:n-point_function}, rather than the thermal $n$-point functions defined in eq.~\eqref{eq:thermal_n_point_function_def}.

 Let us restrict ourselves to the link between thermal and real time Green's function for the 
 case of 2-point functions between a bosonic operator $\mathcal{O}$ with respect to itself 
 at the event $t=\mathbf{x}=0$. The complication of recovering the retarded Green's function relevant for linear response, from $\mathcal{G}$ lies 
 in the fact that the latter is only defined for the discrete set Matsubara frequencies \cite{Son2002:JHEP}. The exact relation between the former 
 and the latter, in momentum space, is given by \cite{Son2002:JHEP}:
 \begin{equation}\label{eq:undoing_Wick_rotation}
    G^R(i\omega_n,\mathbf{k})=-\mathcal{G}(\omega_n,\mathbf{k})\;,\;\omega_n=2\pi n\;,\,(n=0,1,2,\ldots),
 \end{equation}
 where $\omega_n$ are the bosonic Matsubara frequencies. In order to recover the full function $G_R$ in 
 momentum space, one should perform analytical continuation of eq.~\eqref{eq:undoing_Wick_rotation} from the discrete set 
 of Matsubara frequencies towards the entire complex $\omega$-plane. This is technically unfeasable in holographic bottom-up models 
 like the one used in this work, since it requires knowing $\mathcal{G}$ in all Matsubara frencies. Since holography is concerned in the 
 in working with Fourier modes of bulk fields with low energy and momentum (for example, hydrodynamics is an effective theory for 
 describing the boundary in the long wavelength regime), this won't usually be the case.

How to circunvent the above mentioned limitation was a very dynamic field of research in the early years of AdS/CFT 
\cite{Balasubramanian1999:PhysRevD,Balasubramanian1999two:PhysRevD,Son2002:JHEP,Herzog2003:JHEP}, since it was very desirable 
to be able to compute real-time correlators at finite temperature simply working in Lorentzian signature in the bulk from the start, 
avoiding the Euclidean signature alltogether. Such a goal would allow to compute (at least numerically) retarded correltaros in real time 
at finite temperature, without performing analytic continuation of the thermal Green's function \eqref{eq:thermal_n_point_function_def}, by using 
the GKPW formula in real time as a starting point even when the boundary theory is thermal (which is what is done in this work 
in subsection \ref{subsec:viscosity}, and further explained in Appendix \ref{app:hydrodynamics} for the computation of the shear viscosity). The conjecture of applying real-time GKPW calculations for real-time correlators 
at finite temperature has shown to be true for AdS/CFT, and a full treatise on the subject can be found in 
\cite{Zaanen2015:BookChptr6}, and also in earlier research like \cite{Herzog2003:JHEP}, so we refer the reader 
to such material for further insight on the subject. 

{\let\clearpage\relax \chapter{Hydrodynamics in holography}\label{app:hydrodynamics}}
In this appendix we outline the basic ingredients of hydrodynamic theory that are needed to understand 
the relevance of calculating the $\eta/s$ ratio in bottom-up AdS/CFT holographic models. Hydrodynamics 
is the theory that describes the collective behavior of systems in the regime of long distances and long times, where 
translational symmetry is conserved. Most condensed matter bulk-type systems (i.e: those that are thermodynamically large)
don't break translational symmetry, unless an external factor, like an electric or magnetic field, explicitly breaks it. This 
makes hydrodynamic properties of systems universal, since they don't depend on the microscopic details of a Hamiltonian or lattice, given 
that they rule the dynamics of the collective, low energy degrees of freedom of the theory of interest. Indeed, according to \cite{Zaanen2015:BookChptr7},
hydrodynamical laws refer to equations of state for conserved extensive quantities, such as energy and momentum, which are based on 
thermodynamic principles. In the regime of long times and distances, microscopic quantum effects should be averaged out, leading 
to the classical formulation of hydrodynamics for any such system. This is even true in strongly coupled QFTs with no quasiparticles, where hydrodynamic flow can not 
be understood as a consequence of Boltzmann-like kinetics between particles that collide with each other \cite{Zaanen2015:BookChptr7}, 
even though flow still exists.

With the above background established, it is clear that hydrodynamical laws should be tractable through AdS/CFT for 
strongly coupled condensed matter systems near criticality. This is indeed the case, and has provided one of the only 
experimental tests for predictions related to condensed matter systems from holographic principles; namely, the 
shear viscosity to entropy density ratio, $\eta/s$, of the quark-gluon plasma (QGP) \cite{Demir2009:PhysRevLett}. Indeed, 
heavy-ion collisions conducted at the RHIC laboratory at Brookhaven produced an experimental 
realization of the QGP whose $\eta/s$ ratio was anomalously low \cite{Shuryak2005:NucPhysA}, being very close to the exact holographic 
lower bound for such ratio deduced by \cite{Kovtun2004:PhysRevLett}: eq.~\eqref{eq:KSS_bound}. Eversince, the bottom-up approach 
has been extensively used to create new models that seem to violate this bound, mainly using explicit rotational symmetry breaking
in the boundary theory by deforming the bulk appropriately \cite{Rebhan2012:PhysRevLett,Critelli2014:PhysRevD,Landsteiner2016Odd:PhysRevLett,Hartnoll2016bumpy:JHEP}. 
As it was seen in secion \ref{subsec:viscosity}, anisotropic Dirac semimetals are another such system where
the KSS bound is violated.

In the following calculations we show how to set hydrodynamical calculations in the holographic context, using a 
fully relativistic boundary theory as a toy model. Given a boundary theory in $d+1$-dimensions at equilibrium, with $SO(d,1)$ symmetry,
 the most fundamental hydrodynamical law is the conservation of energy:
 \begin{equation}\label{eq:continuity_eq}
    \partial_a\,\langle\widehat{T}^{ab}\rangle_{QFT}=0.
 \end{equation}
Hydrodynamics is generally established in terms of equations of motion, rather than an action principle, due to the 
presence of dissipation. Dissipation of energy would be introduced, for example, by perturbing the QFT out of equilibrium 
through linear response theory. As indicated in \cite{Son2007:review}, eq.~\eqref{eq:continuity_eq} needs to be supplied with a set of constitutive relatuions for the 
energy-momentum tensor. This is done by defining $\langle\widehat{T}^{ab}\rangle_{QFT}$ in terms of the local $d+1$-velocity field 
of the fluid: $u^a = (u^0,u^1,\ldots,u^d)\equiv u^a(x)$, as well as the fluids temperature field: $T\equiv T(x)$\footnote{In this chapter 
we refer to $x$ as the boundary theory coordinate; i.e: not including any aditional radial coordinate.}. Since $u^2 =-1$, eq.~\eqref{eq:continuity_eq} 
defines a set of $d+1$-equations for $d$ variables, making the profile of $\langle\widehat{T}^{ab}\rangle_{QFT}$ solvable in principle.

When dissipation is present the energy-momentum tensor $T^{ab}$ is expanded in a gradient expansion of the form:
\begin{equation}\label{eq:gradient_expansion}
    \widehat{T}_{ab}=\widehat{T}_{ab}^{(0)}+\widehat{\tau}_{ab}+\cdots,
\end{equation}
where $\widehat{\tau}_{ab}$ is the dissipative part of the energy-momemtun tensor that contains the least amount 
of derivatives of the constitutive fields $u$ and $T$ \cite{Heller2020:PhysRevD}. The term $\widehat{T}_{ab}^{(0)}$ is the 
conserved part of the energy-momentum tensor, and satisfies eq.~\eqref{eq:continuity_eq}.

\begin{figure}[!htb]
    \centering
    \includegraphics[width=0.8\linewidth]{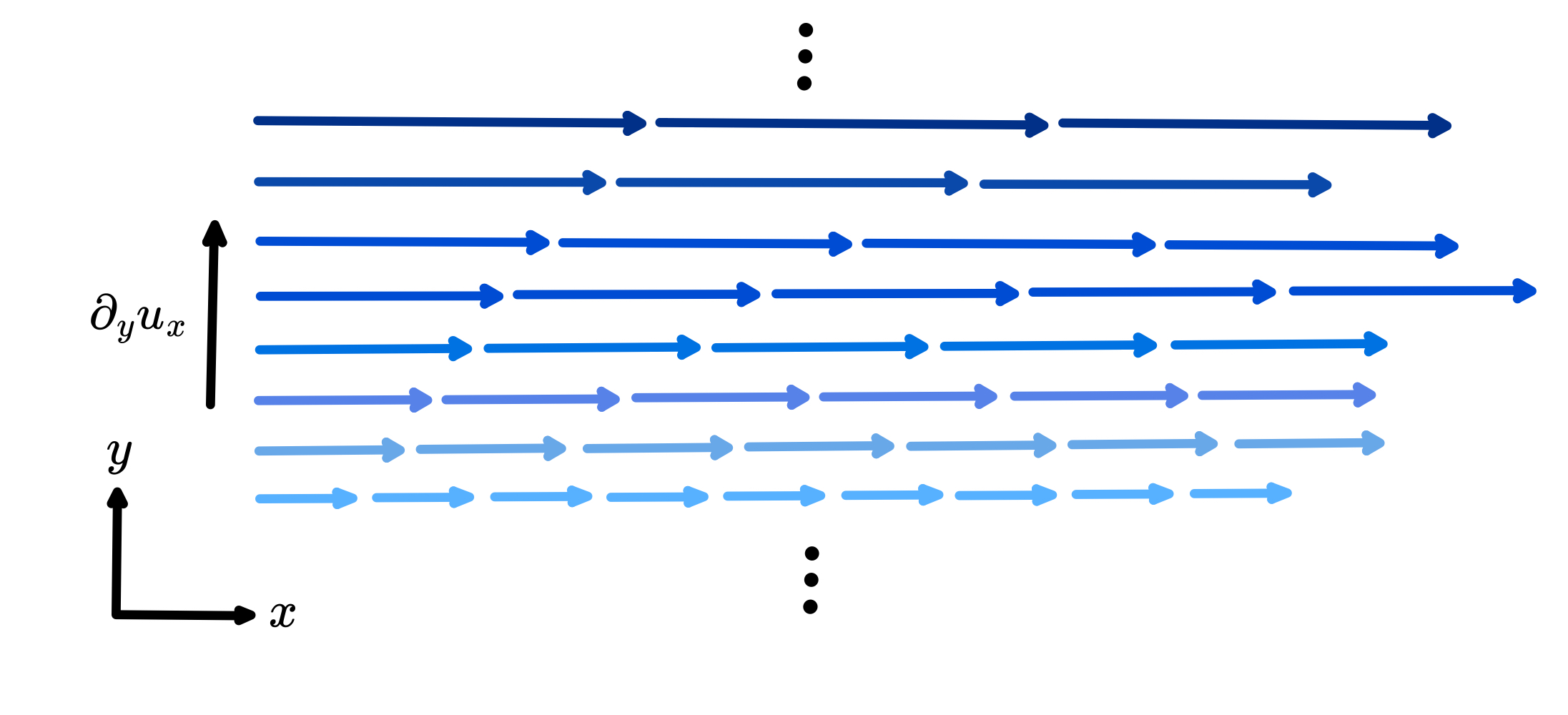}
    \caption{Cartoon of a two-dimensional fluid where disipation is present by the 
    shear stress caused by a gradient in the $u_x$ component of the velocity along the $y$ direction.}
    \label{fig:fluid_cartoon}
\end{figure}

A precise covariant expression for each term of the gradient expansion \eqref{eq:gradient_expansion} is usually achieved 
by symmetry principles. In the case of our relativistic $d+1$-dimensional theory, Lorentz invariance implies that:
\begin{equation}
    \langle\widehat{T}_0^{ab}\rangle_{QFT}=(\epsilon+P)u^au^b+P\eta^{ab},
\end{equation}
where $P$ and $\epsilon$ are the fluids local pressure and energy densities in the center-of-mass reference frame \cite{Son2007:review}. 
Lorentz invariance also restricts the shape of $\sigma_{ab}$, and is given by:
\begin{equation}\label{eq:almost_there}
    \langle\widehat{\tau}^{ab}\rangle_{QFT}=\Pi^{ac}\Pi^{bd}\left[\eta\left(\partial_{c}u_{d}+\partial_{d}u_{c}-\frac{2}{d}\eta_{cd}(\partial\cdot u)\right)+\zeta\eta_{cd}(\partial\cdot u)\right]\;
\end{equation} 
where $\Pi^{cd}=\eta^{cd}-u^cu^d$ is the projector transverse to the $d+1$-velocity field. The terms $\eta$ and $\zeta$ 
appear as proportionality constants throughout the construction of \eqref{eq:almost_there} using rotational invariance (see \cite{Heller2020:PhysRevD} 
for a complete derivation), and they correspond to the fluid's shear and bulk viscosity, respectively \cite{Son2007:review}. Assuming the
fluid to be incompressible (i.e: $\zeta = 0$), the disipative part of the energy-momentum tensor only involves 
the shear viscosity $\eta$. As a hydrodynamic quantity, $\eta$ represents the response of a fluid to the shear 
stress caused by an induced gradient of velocity along its perpendicular direction (see Figure \ref{fig:fluid_cartoon}).

The generation of a gradient in four-velocity can be understood to be an external source introduced into the
QFT, which perturbs the VEV of the energy momentum tensor. Using the formalism of linear response theory, the
viscosity tensor $\eta_{abcd}$ can be defined by:
\begin{equation}\label{eq:linear_response_energy_momentum}
    \langle\widehat{\tau}^{ab}\rangle_{QFT}=\eta_{abcd}\,\partial^{(c}u^{d)}.
\end{equation}
This equation is simply the statement of the linear-response relation in eq.\eqref{eq:linear_response} for 
a hydrodynamics, with the external perturbation corresponding, in this case, to a gradient in the $d+1$-velocity. This 
means that we should be able to relate the viscosity tensor $\eta_{abcd}$ to the retarded Green's function of the 
energy-momentum tensor.

We proceed to make this relation explicit in the case of the Lorentz-invariant fluid. We know that 
the perturbations that couple to the energy-momentum tensor are fluctuations in the spacetime metric. Following the 
recipe of \cite{Son2007:review}, we perturb the Minkwoski metric as in $g_{ab}=\eta_{ab}+h_{ab}$, focusing 
only in the case where the perturbations are uniform in space, yet explicitly dependent on time, for simplicity. By moving to the 
fluid's rest frame, we can choose a gauge where only the spatial indices of $h_{ab}$ are non-trivial:
\begin{equation}
    h_{00}(t)=0\;,\;h_{0j}(t)=0.
\end{equation}
Now we compute the entries of $\langle\widehat{\tau}_{ab}\rangle$. Knowing that in the fluid's rest frame $u^a=(1,0,\ldots,0)$, we
use \eqref{eq:almost_there} with the $d+1$-velocity now coupled to a curved background given by $g_{ab}$. This turns the partial derivatives 
into covariant derivatives: $\partial_au^b\mapsto\nabla_au^b = \partial_au^b+\Gamma_{ac}^bu^c$. This results in:
\begin{equation}
    \left\langle\widehat{\tau}_{ab}(t,\mathbf{x})\right\rangle_{QFT} = \begin{bmatrix}
        0 & 0 & 0 \\
        0 & Ph_{xx}(t)-\frac{1}{2}\eta\left(\frac{\mathrm{d}h_{xx}}{\mathrm{d}t}-\frac{\mathrm{d}h_{yy}}{\mathrm{d}t}\right) & Ph_{xy}(t)+\eta\frac{\mathrm{d}h_{xy}}{\mathrm{d}t} \\
        0 & Ph_{xy}(t)+\eta\frac{\mathrm{d}h_{xy}}{\mathrm{d}t} & Ph_{yy}(t)-\frac{1}{2}\eta\left(\frac{\mathrm{d}h_{xx}}{\mathrm{d}t}+\frac{\mathrm{d}h_{yy}}{\mathrm{d}t}\right) 
    \end{bmatrix}.
\end{equation}
Knowing that $\delta\langle\widehat{T}_{ab}\rangle_{QFT} = \langle\widehat{\tau}_{ab}\rangle_{QFT}$ in the context of 
linear response theory, and going into momentum space, we have:
\begin{equation}\label{eq:terrible_equation}
    \delta\left\langle\widehat{T}_{ab}(\omega,\mathbf{k}=\mathbf{0})\right\rangle_{QFT}=\begin{bmatrix}
                                                                                        0 & 0 & 0 \\
                                                                                        0 & \left(P+i\omega\frac{\eta}{2}\right)h_{xx}(\omega)-i\omega\frac{\eta}{2}h_{yy}(\omega) & \left(P+i\omega\eta\right)h_{xy}(\omega) \\
                                                                                        0 & \left(P+i\omega\eta\right)h_{xy}(\omega) & \left(P+i\omega\frac{\eta}{2}\right)h_{yy}(\omega)-i\omega\frac{\eta}{2}h_{xx}(\omega)
                                                                                        \end{bmatrix}+\mathcal{O}[\omega^2]
\end{equation}
By comparing each entry of \eqref{eq:terrible_equation} to the momentum space version of \eqref{eq:linear_response} 
for the energy-momentum tensor:
\begin{equation}\label{eq:momentum_linear_response}
    \delta\langle\widehat{T}_{ab}(-\omega,-\mathbf{k})\rangle_{QFT}=G_{ab,cd}^R(\omega,\mathbf{k})h^{cd}(-\omega,-\mathbf{k}),
\end{equation}
we have the following relations for our toy relativistic fluid:
\begin{align}
    G_{xx,xx}^R(\omega,\mathbf{k}=\mathbf{0})=P+i\omega\frac{\eta}{2}+\mathcal{O}[\omega^2] \;&,\; G_{xx,yy}^R(\omega,\mathbf{k}=\mathbf{0})=-i\omega\frac{\eta}{2}+\mathcal{O}[\omega^2] \nonumber\\
    G_{yy,yy}^R(\omega,\mathbf{k}=\mathbf{0})=P+i\omega\frac{\eta}{2}+\mathcal{O}[\omega^2] \;&,\; G_{xy,xy}^R(\omega,\mathbf{k}=\mathbf{0})=P-i\omega\eta+\mathcal{O}[\omega^2]. \label{eq:free_retarded_correlators}
\end{align}
In particular, we deduce Kubo's formula for the shear viscosity (eq.\eqref{eq:Kubo_formula}) from the last entry of 
equation \eqref{eq:free_retarded_correlators}.

Notice that the exact correspondence between the shear viscosity $\eta$ as a hydrodynamic coefficient and Kubo's formula 
has been established only for a Lorentz invariant fluid. The QFT featured in this work explicitly breaks rotational invariance,
which means eqs.\eqref{eq:almost_there} is not valid as a covariant expression for the disipative part of the energy-momentum tensor. For these cases,
the "shear viscosity" that is calculated through Kubo's formula is not necessarily the hydrodynamic coefficient $\eta$ that appears in the
hydrodynamical equations of motion; rather, it is the $xy-xy$ entry of the viscosity tensor $\eta_{abcd}$ as defined through eq.\eqref{eq:linear_response_energy_momentum}. 
In the case of a Lorentz-invariant fluid, this entry coincides precisely with the hydrodynamic shear viscosity $\eta$. In any other case, this might not be true. For example, 
$\eta_{xy,xy}$ might be some linear combination of the bulk and shear viscosities, or any other hydrodynamic parameter of the theory. From the perspective of linear response theory 
for a generic, strongly coupled field theory, eq.~\eqref{eq:Kubo_formula} acts as a definition of the shear viscosity, and is related to the linear response of the energy-momentum tensor
by being the shear entry of the viscosity tensor $\eta_{abcd}$, as it is defined through \eqref{eq:linear_response_energy_momentum}. In other works where anisotropic holographic fluids are treated, Kubo's formula is taken as the definition of the shear viscosity of the 
dual field theory (see for example \cite{Erdmenger2012:JHEP,Jain2015:JHEP,Ling2016:JHEP}). The KSS-bound violations reported in such works, alongside 
all proposed improved bounds (see \cite{Hartnoll2016bumpy:JHEP}), are all probed in terms of Kubo's formula as a definition of the shear viscosity.

{\let\clearpage\relax \chapter{Holographic renormalization, GKPW calculations and linear response}\label{app:holographic_renormalization}}
In chapter \ref{chap:theoretical_background} it was explained how to calculate n-point correlation
functions in the boundary theory from the on-shell bulk action. This implies evaluating
the action on the field configuration that satisfy the action's Euler-Lagrange equations 
of motion, which results, in general, in an ill-defined quantity. Indeed, the on-shell bulk 
action is usually divergent, and it is associated to the infinite volume of AdS spacetime
that needs to be integrated over for evaluating it \cite{PapadimitriouLectures}. This patological aspect 
of AdS/CFT can be thought of as the dual manifestiation of the UV divergences of the boundary QFT
when calculating correlation functions, and they are fixed in the same way as it is in standard QFT; 
namely by adding counterterms to the on-shell action. To have a finite value for $S_{\mathrm{b}}^{\mathrm{on-shell}}$, counterterms are added to the bulk action
which do not affect the EOMs of the fields, while, at the same time, the integration over the radial
coordinate is evaluated at a regularized cut-off $r=\varepsilon>0$ \cite{deHaro2001:CommunMathPhys,Bianchi2002:NuclPhysB,Skenderis2002:ClassQuantGrav}. 
The counter-terms to be added depend on the specific matter content and dimensionality of the bulk, and they 
cancel the divergences that arise from the bare, on-shell action when taking the $\varepsilon\to 0$ limit. The 
specific nuance and details on how to build a holographically renormalized bulk theory can be found 
in standard literature like \cite{deHaro2001:CommunMathPhys}, or \cite{Skenderis2002:ClassQuantGrav}. For bulk model
presented in this work, we simply state the counterterms that cancel the divergences that arise for the
bulk action \eqref{eq:background_action} when evaluated on-shell. The fully renormalized action
for our model is:
\begin{align}\label{eq:renormalized_action}
      S_{\mathrm{b}}^{\,\mathrm{ren}} = \int_{r=\varepsilon}\!\mathrm{d}^4x\,\sqrt{-g}&\left[\frac{1}{2\kappa^2}\left(R+\frac{6}{L^2}\right)-\mathrm{Tr}\left(\left(D^\mu\Phi\right)^\dagger\left(D_\mu\Phi\right)\right)-m^2\mathrm{Tr}\left(\Phi^\dagger\Phi\right)-\frac{\lambda}{4}\left(\mathrm{Tr}(\Phi^\dagger\Phi)\right)^2\right.\nonumber \\\left.-
\frac{1}{4}\mathrm{Tr}\left(G_{\mu\nu}G^{\mu\nu}\right)\right] 
+&\frac{1}{2\kappa^2}\int_{r=\varepsilon}\!\mathrm{d}^2\mathbf{x}\,\mathrm{d}t\,\sqrt{-\gamma}\left(4+R[\gamma]+2K\right)+\int_{r=\varepsilon}\mathrm{d}^2\mathbf{x}\,\mathrm{d}t\,\mathrm{Tr}\left(\Phi^\dagger\Phi\right).
  \end{align}
Here, $\gamma$ is the induced metric on the hypersurface created at the $r=\varepsilon$ cut-off, $R[\gamma]$ is the unduced Ricci scalar, and $2K$ is the Gibbons-Hawinkg-York (GHY)
counterterm, which is introduced to have a well-defined variational problem when ultimately taking variations with respect to metric components. Notice that 
all counterterms are boundary quantities that depend only upon the boundary conditions imposed on the background fields, which doesn't affect the solution to the
variational equation $\delta\,S_{\mathrm{b}}=0$ in the bulk. 

The choice of counterterms that renormalize this bulk action were taken from previous work in holographic 
fluids that have a $SU(2)$ gauge sector in, like \cite{Landsteiner2016Odd:PhysRevLett} and \cite{Ammon2010:PhysRevLett,Arias:2012py}. For the 
purposes of this work, we are interested in using equation \eqref{eq:VEV_Fourier_space} for
calculating the VEVs of the operators dual to $\phi$ and $B$ in the bulk. To do this, we take variations of $S_{\mathrm{b}}^{\mathrm{ren}}$ it with respecto to 
the bulk fields, evaluate it on-shell, take the limit $\epsilon\to 0$, and take functional derivatives with respect to the dual field sources. We execute 
each of these steps one at a time.

First, we take variations of $S_{\mathrm{b}}^{\mathrm{ren}}$ with respect to the bulk fields on general (without evaluating them on-shell), 
which results in:
\begin{align}\label{eq:covariant_variation_renormalized_action}
\delta\,S_\mathrm{b}^{\mathrm{ren}}&=\frac{1}{2\kappa^2}\int_{r=\varepsilon}\!\mathrm{d}t\mathrm{d}^d\mathbf{x}\,\sqrt{-\gamma}\,n^\mu\left(\nabla^\nu\delta\,g_{\mu\nu}-g^{\sigma\rho}\nabla_\mu\delta\,g_{\sigma\rho}\right)-\int\!\mathrm{d}t\mathrm{d}^d\mathbf{x}\,\sqrt{-\gamma}\,n^\mu\mathrm{Tr}\left[\left(D_\mu\Phi\right)\delta\,\Phi^\dagger+\right.\nonumber\\
&\left.\left(D_\mu\Phi\right)^\dagger\delta\,\Phi\right]+\int_{r=\varepsilon}\!\mathrm{d}t\mathrm{d}^d\mathbf{x}\,\sqrt{-\gamma}\,n_\mu\mathrm{Tr}\left(G^{\mu\nu}\delta\,B_\nu\right)+\delta\,S_{\mathrm{b}},
\end{align}
where $\delta\,S_{\mathrm{b}}$ is the variation of the bare bulk action, and is proportional to the background EOMs.

 Now we evaluate \eqref{eq:covariant_variation_renormalized_action} on-shell. Notice that by doing this, $\delta\,S_{\mathrm{b}^{\mathrm{on-shell}}}=0$, and $\delta\,S_{\mathrm{b}}^{\mathrm{ren},\mathrm{on-shell}}$ 
 only consists of boundary terms at $r=\varepsilon$. That means that we can substitute the background fields in the on-shell evaluation of \eqref{eq:covariant_variation_renormalized_action} 
 by their asymptotic expansion near $r=0$, instead of their full bulk solution. Notice, however, that the 
 variation \eqref{eq:covariant_variation_renormalized_action} includes terms proportional to both the bulk on-shell fields (whose boundary 
 expansions are \eqref{eq:boundary_expansion_scalar}-\eqref{eq:boundary_expansion_N}), and their variations. 
 On the boundnary at $r=\varepsilon$, the variations $\delta\,g_{\mu\nu}$, $\delta\,\Phi$ and $\delta\,B$ are realized 
 by taking variations of the leading and subleading coefficients of the corresponding solutions to the EOMs. This translates 
 into the following asymptotic expansion at $r=\varepsilon$ for the variations:
\begin{align}
     \delta\,\phi_j(r=\varepsilon\,;\,\omega,\mathbf{k}) &= r\delta\,\varphi_{j,(l)}(\omega,\mathbf{k})+\delta\,\varphi_{j,(s)}(\omega,\mathbf{k})r^2+\cdots\;(j=1,2,3)\label{eqap:boundary_expansion_scalar}\\
     \delta\,B_{\mu,j}(r=\varepsilon\,;\,\omega,\mathbf{k}) &= b_{\mu,j,(l)}(\omega,\mathbf{k})+b_{\mu,j,(s)}(\omega,\mathbf{k})r+\cdots\;(j=1,2,3)\label{eqap:boundary_expansion_gauge}\\
     \delta\,h_{\mu\nu}(r=\varepsilon\,;\,\omega,\mathbf{k}) &= \frac{1}{r^2}\left[h_{\mu\nu,(l)}(\omega,\mathbf{k})+\cdots+r^3h_{\mu\nu,(s)}(\omega,\mathbf{k})+\cdots\right]\label{eqap:boundary_expansion_metric}
\end{align}
Plugging \eqref{eq:boundary_expansion_scalar}-\eqref{eq:boundary_expansion_N} and \eqref{eqap:boundary_expansion_scalar}-\eqref{eqap:boundary_expansion_metric} into \eqref{eq:covariant_variation_renormalized_action}, we obtain, 
up to first order in variations of the fields (and immediately taking the limit $\varepsilon\to 0$):
\begin{align}\label{eq:on-shell_full_variation}
    \delta\,S_{\mathrm{b},1}^{\mathrm{ren}\,,\,\mathrm{on-shell}}&=\int\!\mathrm{d}\omega\mathrm{d}^{d}\mathbf{k}\,\frac{2f_3(2h^{\mathrm{UV}}_{tt,(l)}+h_{xx,(l)}+h_{yy,(l)}-6h_3(h_{xx,(l)}-h_{yy,(l)}))}{4\kappa^2}\nonumber\\
    &-\int\!\mathrm{d}\omega\mathrm{d}^{d}\mathbf{k}\,\left(2B_{(s)}\delta\,b_{1,1}(\omega,\mathbf{k})-2\phi_{(s)}\delta\,\varphi_{3,(l)}(\omega,\mathbf{k})\right.\nonumber\\&\left.+\Delta_2(\omega,\mathbf{k})(\delta\,h_{tt,(l)}(\omega,\mathbf{k})+\delta\,h_{xx,(l)}(\omega,\mathbf{k})+\delta,h_{yy,(l)}(\omega,\mathbf{k}))\right),
\end{align}

Finally, we use eq.\eqref{eq:VEV_Fourier_space} to find $\langle\widehat{\mathcal{O}}_B\rangle_{QFT}$ and $\langle\widehat{\mathcal{O}}_\phi\rangle_{QFT}$,
where $\widehat{\mathcal{O}}_B$ and $\widehat{\mathcal{O}}_\phi$ are the dual gauge and scalar operators in the boundary QFT, respectively. This leads to:
\begin{equation}\label{eq:explicit_VEVs}
    \langle\widehat{\mathcal{O}}_B\rangle_{QFT} = 2B_{(s)}\;,\;\langle\widehat{\mathcal{O}}_\phi\rangle_{QFT} = -2\phi_{(s)}.
\end{equation}

The overall factors of $\pm 2$ can be scaled away by a re-scaling of the bulk fields. The main conclusion is the fact that 
both VEVs are proportional to the sub-leading coefficients of the boundary expansions of the bulk fields. Now we want to use \eqref{eq:retarded_correlator_GKPW} to use linear response theory to calculate the retarded 
Green's function of dual operators. These would allow for calculation of a wide range of transport coefficients, 
like conductivities and bulk viscosities (see \cite{Erdmenger2012:JHEP} for procedures on calculating such transport coefficients).
However, we will be only interested in calculating the shear viscosity of our dual, strongly coupled fluid. To do this 
we need to use Kubo's formula (eqs.~\eqref{eq:Kubo_formula}), and to do that we need the retarded Green's function $G_{xy,xy}^R$ 
associated to the shear entry of the boundary energy-momentum tensor: $T_{xy}$. As it was explained 
in chapter \ref{chap:theoretical_background}, to implement linear perturbations on the boundary theory,
we simply have to use linear response theory as usual in the boundary, while implementing the perturbation currents
that take the boundary system out of equilibrium as classical fields, following the standard AdS/CFT procedure \cite{Hartnoll2009:ClassQuantGrav_lectures}.

For calculating the shear viscosity in a regular fluid through linear response, we introduce a linear preturbation
in the boundary that couples to the $\widehat{T}_{xy}$ operator, and determine its VEV after the source 
has been turned on \cite{Son2007:review}. The source that couples to the energy-momentum tensor $\widehat{T}_{\mu\nu}$ is 
nothing more than the metric tensor $g_{\mu\nu}$. Therefore, the classical field that dualizes this perturbation 
is, again, a perturbation of the $xy$ component of the bulk metric field, with its on-shell leading boundary solution 
to the linearized background EOMs identified as the boundary source. We call this fluctuation, which is effectively a 
gravitational wave mode propagating on the background, $h_{xy}\equiv h_{xy}(t,\mathbf{x},r)$.

 Since our model includes also the matter fields $\phi$ and $B$, a perturbation on the gravitational sector of the bulk 
 will backreact into the Klein-Gordon and Yang-Mills equations, inducing fluctuations of the scalar and gauge sectors as well. This 
 means that in order to have a consistent set of linearized EOMs for $h_{xy}$, we need to couple it to a given set of 
 fluctuations of the matter fields. It turns out that the $h_{xy}$ mode of the gravitational sector only couples
 to the $b_{y,2}$ mode of the gauge sector, remaining uncoupled from any other fluctuations on the remaining metric or matter fields.
  Therefore, to implement linear response theory in the holographic bulk system, the following two fluctuations are turned on:
  \begin{align}
    \delta\,g_{\mu\nu}\equiv h_{\mu\nu}(t,\mathbf{x},r)&=\int\!\frac{\mathrm{d}\omega\mathrm{d}^d\mathbf{k}}{(2\pi)^{d+1}}\,e^{-i\omega t+i\mathbf{k}\cdot\mathbf{x}}\begin{bmatrix}0&0&0&0\\
                                                                                                                                          0&0&h_{xy}(\omega,\mathbf{k},r)&0\\
                                                                                                                                          0&h_{xy}(\omega,\mathbf{k},r)&0&0\\
                                                                                                                                          0&0&0&0\end{bmatrix}\label{eq:metric_fluctuation_ansatz}\\
    \delta\,B_{\mu}\equiv b_{\mu}(t,\mathbf{x},r)&=\int\!\frac{\mathbf{d}\omega\mathbf{d}^d\mathbf{k}}{(2\pi)^{d+1}}\,e^{-i\omega+i\mathbf{k}\cdot\mathbf{x}}\begin{bmatrix}
                                                                                                                                                             0\\
                                                                                                                                                             0\\
                                                                                                                                                             b_{y,2}(\omega,\mathbf{k},r)\sigma_1\\
                                                                                                                                                             0\\
                                                                                                                                                             0\end{bmatrix}\label{eq:gauge_fluctuation_ansatz}
  \end{align}

  By using the GKPW formula \eqref{eq:VEV_Fourier_space}, the VEV of the dual energy-momentum tensor
  is obtained by functional diferentiation of \eqref{eq:covariant_variation_renormalized_action} with
  respect to $\delta\,g_{\mu\nu}$, and evaluating on-shell:
  \begin{equation}\label{eq:energy_tensor_vev}
    \langle T^{ab}(\tilde{x})\rangle_{QFT}=\frac{1}{2\kappa^2}\lim_{\varepsilon\to 0}\sqrt{-\gamma}\left\{\gamma^{ab}K-K^{ab}+\frac{1}{2}\gamma^{ab}\left[4+R[\gamma]+2\kappa^2\mathrm{Tr}\left(\Phi^\dagger\Phi\right)\right]\right\}^{\mathrm{(on-shell)}}.
\end{equation}
Using eq.~\eqref{eq:energy_tensor_vev} with the bulk that includes the linear perturbations \eqref{eq:metric_fluctuation_ansatz}-\eqref{eq:gauge_fluctuation_ansatz},
(i.e: the bulk that dualizes the linearly perturbed dual theory) it can be seen that it contains both the background/unperturbed fields, and the fluctuations themselves.
 Again, since \eqref{eq:energy_tensor_vev} is evaluated at $r=\varepsilon\to0$, we can insert expansions \eqref{eq:boundary_expansion_scalar}-\eqref{eq:boundary_expansion_N} 
 (for the background fields) and \eqref{eqap:boundary_expansion_scalar}-\eqref{eqap:boundary_expansion_metric} (for the perturbations) into it for its on-shell evaluation. 
 Transforming into momentum space, this results in:
 \begin{equation}\label{eq:fourier_energy_tensor_vev}
    \langle \widehat{T}_{xy}(-\omega,-\mathbf{k})\rangle_{QFT}=4\phi_{(s)}h_{xy,(l)}(\omega,\mathbf{k})-\frac{3}{2\kappa^2}h_{xy,(s)}(\omega,\mathbf{k})-\frac{1}{2\kappa^2}f_3h_{xy,(l)}(\omega,\mathbf{k})+\cdots,
 \end{equation}
where higher order contact terms that involve $\omega^2$ and $\mathbf{k}^2$ products are contained in $\cdots$. Finally, using eq.~\eqref{eq:retarded_correlator_GKPW} by taking 
the derivative of \eqref{eq:fourier_energy_tensor_vev} with respect to $h_{xy,(l)}$ results in:
\begin{equation}
    \mathrm{Im}\left(G_{xy,xy}^R(\omega,\mathbf{k})\right)=-\frac{3}{2\kappa^2}\mathrm{Im}\left(\frac{\delta\,h_{xy,(s)}(-\omega,-\mathbf{k})}{\delta\,h_{xy,(l)}(-\omega,-\mathbf{k})}\right).
\end{equation}

This analytical formula then allows computation of $\eta\equiv\eta_{xy,xy}$ using Kubo's formula \eqref{eq:Kubo_formula}:
\begin{equation}
    \eta = \frac{3}{2\kappa^2}\lim_{\omega\to 0}\frac{1}{\omega}\mathrm{Im}\left(\frac{\delta\,h_{xy,(s)}(-\omega,\mathbf{k}=\mathbf{0})}{\delta\,h_{xy,(l)}(-\omega,\mathbf{k}=\mathbf{0})}\right),
\end{equation}
where the dependence of the subleading part of $h_{xy}$ with respect to its leading part comes from imposing infalling boundary conditions
on the fluctuating field at the black brane horizon:
\begin{equation}
    h_{xy}(\omega)=(1-r)^{-i\frac{\omega}{4\pi T}}v(r),
\end{equation}
for some function $v(r)$. This formula, even though is exact, requires nowing precisely how the subleading term of $h_{xy}$ depends on the leading one.
For numerical convenience, the results in subsection \ref{subsec:viscosity} (specifically those shown in Figures \ref{fig:viscosity_plot} and \ref{fig:scaling_viscosity}) 
where obtained using equation \eqref{eq:linear_response} as the definition of $G_{xy,xy}^R$, rather than the exact 
holographic formula derived from \eqref{eq:retarded_correlator_GKPW}. Indeed, in momentum space we have:
\begin{equation}\label{eq:final_eq}
    \delta\langle\widehat{T}_{xy}(-\omega,-\mathbf{k})\rangle_{QFT}=G_{xy,xy}^R(\omega,\mathbf{k})h_{xy}(-\omega,-\mathbf{k}).
\end{equation}
By using the boundary condition of $h_{xy}$ and $b_{y,1}$ on the black brane event horizon as a shooting parameter, we numerically solve the 
linearized equations of motion for said fluctuations so that the leading term of $h_{xy}$ at the UV is normalized to 1 for all. Then, using
\eqref{eq:fourier_energy_tensor_vev} to calculate the left-hand side of \eqref{eq:final_eq} for that specific numerical solution, 
we can read off the numerical value of $G_{xy,xy}^R$ in momentum space from \eqref{eq:final_eq}. 

{\let\clearpage\relax \chapter{Fermions in AdS/CFT}\label{app:fermions_in_AdS/CFT}}
The procedure of encoding fermionic operators is a bit more involved that the encodigin of scalar and bosonic gauge fields, 
such as ha been the case so far. This has to do, esentially, with the fact that the Dirac equation is 
a first order differential equation, unlike the Klein-Gordon, Einstien, or Yang-Mills equations of 
the background fields of the bulk system. This, in principle, would reduce the amount of independent 
boundary conditions that can be imposed on the bulk fields from two to only one \cite{Henningson1998:PhysLettB}. In this 
appendix we justify the dictionary entry used to identify the VEV and source of the dual fermionic operator 
of the boundary theory that was used in section \ref{sec:fermions_coupled_to_background} to calculate the 
retarded fermionic Green's function (i.e: where eq.~\eqref{eq:fermion_correlator} comes from).

First, let us restate the Dirac action used to encode probe fermions in the bulk:
\begin{equation}\label{eqap:Dirac_action}
    S_f = i\int\!\mathrm{d}^4x\sqrt{-g}\,\bar\Psi(\slashed{D}-g_Y\Phi)\Psi.
\end{equation}
To associate sources and VEVs in the boundary to the dynamics of the classical fields in the bulk, 
we must use the GKPW formula again. To do that, we take variations of action \eqref{eqap:Dirac_action}
 with respect to the Dirac spinors $\bar\Psi$ and $\Psi$ (not yet evaluating it on-shell). This results in a bulk term and a boundary term:
 \begin{equation}\label{eq:fermion_action_variation}
    \delta\,S_f = i\int\!\mathrm{d}^4x\sqrt{-g}\,\left[(\delta\,\bar\Psi)\left(\slashed{D}-g_Y\Phi\right)\Psi+\overline{\left(\slashed{D}-g_Y\Phi\right)\Psi}(\delta\,\Psi)\right]+\int_{r=\varepsilon}\!\mathrm{d}t\mathrm{d}^2\mathbf{x}\sqrt{-\gamma}n_\mu\bar\Psi\gamma^\mu(\delta\,\Psi).
 \end{equation}
 Now, recall this model dualizes a pair of Dirac spinorial operators in the boundary, each of which is a 2-tuple, which combined 
 form a $SU(2)$ doublet. That means we need four independent boundary conditions at $r\to0$ to dualize the source that couples to this doublet. 
 Now the convenience of the fact Dirac spinors in $3+1$ dimensions have double the components of those in $2+1$ dimensions becomes evident: since 
 $\Psi$ is an 8-tuple (a $SU(2)$ doublet formed by a pair of 4-tuple Dirac spinors), we can use half of the entries of the 8-tuple to dualize these sources. 
 We expect, therefore, that the remaining 4 entries will correspond to the VEV of the boundary fermionic operator. This is the reason 
 why in section \ref{sec:fermions_coupled_to_background} we projected the 8-tuple spinor onto the eigen-space of the 
 radial $\gamma$-matrix: it explicitly separates the entries of $\Psi$ that are interpreted as boundary sources, and those that are not \cite{Henningson1998:PhysLettB}. 
 Using the notation:
 \begin{equation}
    \Psi_\pm = \left(1_{2\times 2}\otimes\frac{1}{2}\left(1_{4\times 4}\pm\gamma^{\underline{3}}\right)\right)\Psi = \begin{bmatrix}\frac{1}{2}(1_{4\times4}\pm\gamma^{\underline{3}})\psi_1\\\frac{1}{2}(1_{4\times4}\pm\gamma^{\underline{3}})\psi_2\end{bmatrix}\equiv\begin{bmatrix}\psi_{1,\pm}\\\psi_{2,\pm}\end{bmatrix},
 \end{equation}
 such that $\gamma^{\underline{3}}\psi_{j,\pm}=\pm\psi_{j,\pm}$, it is clear that $\bar\Psi_\pm\Psi_\pm = 0$, where $\bar\Psi = \Psi^\dagger (1_{2\times2}\otimes\gamma^{\underline{0}})$, 
 simply using the anticommutation relations of the flat $\gamma$ matrices. As such, expanding \eqref{eq:fermion_action_variation} 
 in terms of $\Psi_\pm$ and $\bar\Psi_\pm$, and evaluating on-shell, we have:
 \begin{equation}\label{eq:dirac_action_variation_on_shell}
    \delta\,S_{f}^{\mathrm{on-shell}}=\int_{r=\varepsilon}\!\mathrm{d}t\mathrm{d}^2\mathbf{x}\sqrt{-\gamma}\,\left[\bar{\Psi}_+(\delta\Psi_-)-\Psi_-(\delta\Psi_+)\right],
 \end{equation}
 where we have used that the unit vector normal to the hypersurface at $r=\varepsilon$ is given by $n_\mu = \left(0,0,0,\frac{L}{r\sqrt{f(r)}}\right)$, while $\gamma^{3}=\frac{r\sqrt{f(r)}}{L}\gamma^{\underline{3}}$, 
 so that $n_\mu\gamma^\mu = \gamma^{\underline{3}}$, and $\bar\Psi_\pm\gamma^3 = \pm\bar\Psi_\pm$.

  Eq.~\eqref{eq:dirac_action_variation_on_shell} is still unsuitable for use of the GKPW formula, since 
  it involves variations of both $\Psi_+$ and $\Psi_-$. In order to have a well-defined variational problem 
  in the boundary, variations of the action, on-shell, should be expressed exclusively in terms of the boundary conditions 
  that are interpreted as the sources, so that we do not have to take functional derivatives with respect to the 
  remaining components we expect to correspond to the VEV of the dual operator. To remedy this, we follow the 
  holographic renormalization procedure shown in Appendix \ref{app:holographic_renormalization}, adding to \eqref{eqap:Dirac_action} 
  a boundary term that doesn't spoil the shape of the bulk Dirac equation:
  \begin{equation}\label{eq:renormalized_Dirac_action}
    S_f^{\mathrm{ren}} = i\int\!\mathrm{d}^4x\sqrt{-g}\,\bar\Psi(\slashed{D}-g_Y\Phi)\Psi - i\int_{r=\varepsilon}\!\mathrm{d}t\mathrm{d}^2\mathbf{x}\sqrt{-\gamma}\,\bar\Psi_+\Psi_-.
  \end{equation}
  Variations of the boundary term cancels the boundary portion of $\delta\,S_f^{\mathrm{on-shell}}$ proportional to $\delta\,\Psi_-$\footnote{Adding 
  the counterterm with an opposite relative sign to the bulk action would result in cancelation of the term proportional 
  to $\delta\,\Psi_+$, resulting in the alternative quantization scheme that interprets $\Psi_-$ as the source.}. 
  Therefore:
  \begin{equation}\label{eq:renormalized_dirac_action_variation}
  \delta\,S_f^{\mathrm{ren},\mathrm{on-shell}} = -\int_{r=\varepsilon}\!\mathrm{d}t\mathrm{d}^2\mathbf{x}\sqrt{-\gamma}\,\left[\Psi_-\left(\delta\Psi_+\right)-\left(\delta\bar\Psi_+\right)\Psi_-\right].
  \end{equation}

  Now, we explain the differente between the notation $\psi_\pm$ of section \ref{sec:fermions_coupled_to_background}, and the $\Psi_\pm$ 
  introduced here. Expanding the $\Psi$ Dirac field tuple as a full 8-entry vector column, we have:
  \begin{gather*}
    \Psi = \begin{bmatrix}\psi_1\\\psi_2\\\psi_3\\\psi_4\\\psi_5\\\psi_6\\\psi_7\\\psi_8\end{bmatrix}\Longrightarrow \Psi_+ = \begin{bmatrix}0\\0\\\psi_3\\\psi_4\\0\\0\\\psi_7\\\psi_8\end{bmatrix}\,,\,\Psi_- = \begin{bmatrix}\psi_1\\\psi_2\\0\\0\\\psi_5\\\psi_6\\0\\0\end{bmatrix}\Longrightarrow\psi_+:=\begin{bmatrix}\psi_3\\\psi_4\\\psi_7\\\psi_8\end{bmatrix}\,,\,\psi_-:=\begin{bmatrix}\psi_1\\\psi_2\\\psi_5\\\psi_6\end{bmatrix}.
  \end{gather*}
  These 4-tuples $\psi_\pm$ are those that we call the decomposition of the 8-tuple $\Psi$ onto the eigenspaces of the radial $\gamma$ matrix. Notice that 
  these fields are still $SU(2)$ doublets, since the upper and lower $\mathbb{C}^2$ sectors of $\psi_\pm$ do not mix under $SU(2)$ 
  transformations of the original $\Psi$ doublet. Also, the upper and lower 2-tuples $\psi_{+,1} :=\begin{bmatrix}\psi_3\\\psi_4\end{bmatrix}$, $\psi_{+,2} = \begin{bmatrix}\psi_7\\\psi_8\end{bmatrix}$, 
  $\psi_{-,1}:=\begin{bmatrix}\psi_1\\\psi_2\end{bmatrix}$, $\psi_{-,2}:=\begin{bmatrix}\psi_5\\\psi_6\end{bmatrix}$ are each Dirac spinors. This can be seen from the explicit form of the generators of 
  Lorentz transformations of the boundary spacetime in the spinorial representation, using the $\gamma$ matrices outlined in eq.~\eqref{eq:gamma_matrices}. Recall that these are defined as $\sigma^{\underline{a}\,\underline{b}} = \frac{i}{4}\left[\gamma^{\underline{a}},\gamma^{\underline{b}}\right]$, with 
  $\sigma^{\underline{\mu}\,\underline{0}}$ being the generators of boosts and $\sigma^{\underline{i}\,\underline{j}}$ being the generators of rotations. The choice of $\gamma$ matrix representation used in this 
  work is the same as that of \cite{Giordano2017:JHEP}, and therefore the Lorentz generators are the same as those outlined in that work:
  \begin{equation}\label{eq:Lorentz_boundary_generators}
    \sigma^{\underline{0}\,\underline{1}} = \frac{i}{2}\begin{bmatrix}\sigma_3 & 0 \\
    0 & \sigma_3\end{bmatrix}\;,\;\sigma^{\underline{0},\underline{2}} = \frac{i}{2}\begin{bmatrix}-\sigma_1 & 0 \\ 0 & -\sigma_1\end{bmatrix}\,,\,\sigma^{\underline{1}\,\underline{2}}=\frac{1}{2}\begin{bmatrix}\sigma_2 & 0 \\ 0 & \sigma_2\end{bmatrix}
  \end{equation}
  This explicitly shows, as pointed out in \cite{Giordano2017:JHEP}, that the Lorentz transformations of the bulk theory, restricted to boosts and rotations 
  in the boundary coordinates, decompose into two irreducible spin-$1/2$ sectors. This justifies the apparent artificial construction of the 4-tuples $\psi_\pm$ from 
  $\Psi_\pm$, since $\psi_\pm$ correspond to a $SU(2)$ doublet of Dirac spinors in $2+1$ dimensions. This remains so after the re-scaling of the spinors 
  $\psi_\pm := r^{3/2}f(r)^{-1/4}\zeta_\pm$.
  
  Finally, we turn to analyzing the near-boundary solution to Dirac's equation when expressed in terms of $\zeta_\pm$; i.e: eq.~\eqref{eq:Dirac_eq_1}-\eqref{eq:Dirac_eq_2}. 
  Since $f(r)\to 0$, $\phi(r)\to r\Delta_1$ and $B(r)\to\Delta_2$ to leading order when $r\to0$, Dirac's equatino for $\zeta_\pm$ near the 
  boundary reduce to:
  \begin{align}
    \frac{\mathrm{d}\zeta_+}{\mathrm{d}r}+i\begin{bmatrix}k_y & k_x-\omega & 0 & q_f\Delta_1 \\
    k_x+\omega & -k_y & q_f\Delta_1 & 0 \\
    0 & q_f\Delta_1 & k_y & k_x-\omega \\
    q_f\Delta_1 & 0 & k_x+\omega & -k_y\end{bmatrix}\zeta_- & =-g_Y\Delta_2\gamma^{\underline{3}}\zeta_+\\
        \frac{\mathrm{d}\zeta_-}{\mathrm{d}r}-i\begin{bmatrix}k_y & k_x-\omega & 0 & q_f\Delta_1 \\
    k_x+\omega & -k_y & q_f\Delta_1 & 0 \\
    0 & q_f\Delta_1 & k_y & k_x-\omega \\
    q_f\Delta_1 & 0 & k_x+\omega & -k_y\end{bmatrix}\zeta_+ & =g_Y\Delta_2\gamma^{\underline{3}}\zeta_-.
  \end{align}
  A series expansion around $r=0$ of these two equations results in a leading constant term for $\zeta_+$ and $\zeta_-$:
  $\zeta_\pm(r\to 0)=\zeta_{\pm,(l)}+\mathcal{O}[r]$, where $\zeta_{-,(l)}$ is independent of $r$, but related to $\zeta_+$ 
  through the explicit solution of the above equations. This dependence (which is linear) is implicitly represented 
  through the matrix equation \eqref{eq:fermion_correlator}, which is where the numerical correlation matris $S$ introduced in 
  section \ref{sec:fermions_coupled_to_background} comes from (the exact analyitical shape of this matris is not relevant to the numerics). 

   Finally, identifying explicitly $\zeta_{+,(l)}$ as the source of the dual fermionic operator, 
   we have, from eq.~\eqref{eq:renormalized_dirac_action_variation}:
   \begin{gather*}
    \delta\,S_f^{\mathrm{ren},\mathrm{on-shell}}=\int_{r=\varepsilon}\!\mathrm{d}t\mathrm{d}^2\mathbf{x}\,\sqrt{-\gamma}\left[\bar\psi_-\left(\delta\psi_+\right)-\left(\delta\bar\psi_+\right)\psi_-\right] = \int_{r=\varepsilon}\!\mathrm{d}t\mathrm{d}^2\mathbf{x}\,\frac{\sqrt{f(r)}}{r^3}\left[\left(\delta\bar\psi_+\right)\psi_--\bar\psi_-\left(\delta\psi_+\right)\right]\\
    = \int_{r=\varepsilon}\!\mathrm{d}t\mathrm{d}^2\mathbf{x}\int\!\frac{\mathrm{d}\omega\mathrm{d}^2\mathbf{k}}{(2\pi)^3}\int\!\frac{\mathrm{d}\omega'\mathrm{d}^2\mathbf{k}'}{(2\pi)^3}e^{-i(\omega+\omega')t+i(\mathbf{k}+\mathbf{k}')\cdot\mathbf{x}}\left[\left(\delta\bar\zeta_+(\omega,\mathbf{k},r)\right)\zeta_-(\omega',\mathbf{k}',r)-\bar\zeta_-(\omega',\mathbf{k}',r)\left(\delta\zeta_+(\omega,\mathbf{k},r)\right)\right]\\
    =\int_{r=\varepsilon}\!\frac{\mathrm{d}\omega\mathbf{d}^2\mathbf{k}}{(2\pi)^3}\left[\left(\delta\bar\zeta_+(\omega,\mathbf{k},r)\right)\zeta_-(-\omega,-\mathbf{k},r)-\bar\zeta_-(-\omega,-\mathbf{k},r)\left(\delta\zeta_+(\omega,\mathbf{k},r)\right)\right].
   \end{gather*}
   Taking the $r=\varepsilon\to 0$ limit of this last expression, and functionaly differentiating with respect to $\bar\zeta_{+,(l)}(\omega,\mathbf{k})$ results 
   in the expected result (up to a re-scalable constant factor of $(2\pi)^3$):
   \begin{equation}
    \langle\widehat{\mathcal{O}}_\psi(-\omega,-\mathbf{k})\rangle_{QFT} = \zeta_{-,(l)}(\omega,\mathbf{k}) = \widehat{S}(\omega,\mathbf{k}\,;\,\Delta_1,\Delta_2)\zeta_{+,(l)}
   \end{equation}

\bibliographystyle{apalike}
\bibliography{referencias}
\end{document}